\documentclass[a4paper,fleqn]{mnras}

\usepackage{times}

\usepackage[T1]{fontenc}
\usepackage{ae,aecompl}

\usepackage{graphicx}	
\usepackage{amsmath}	
\usepackage{amssymb}	
\usepackage{multirow}

\title[R136 dissected with HST/STIS. I. FUV spectroscopy]
{The R136 star cluster dissected with Hubble Space Telescope/STIS. I. Far-ultraviolet 
spectroscopic census and the origin of He\,{\sc ii} $\lambda$1640 in young star clusters}

\author[P.A. Crowther et al.]{Paul A. Crowther$^{1}$\thanks{Paul.Crowther@shef.ac.uk},
S.M. Caballero-Nieves$^{1}$, K.A. Bostroem$^{2,3}$, J. Ma\'{i}z Apell\'{a}niz$^{4}$, 
\newauthor F.R.N. Schneider$^{5,6}$, N.R. Walborn$^{2}$, C.R. Angus$^{1, 7}$,
I. Brott$^{8}$, A. Bonanos$^{9}$, \newauthor A. de Koter$^{10,11}$, S.E. de Mink$^{10}$, C.J. 
Evans$^{12}$, G. Gr\"{a}fener$^{13}$, A. Herrero$^{14,15}$, \newauthor I.D. Howarth$^{16}$, N. Langer$^{6}$, 
D.J. Lennon$^{17}$, J. Puls$^{18}$, H. Sana$^{2,11}$, J.S. Vink$^{13}$
\vspace{3mm} \\ 
$^{1}$Department of Physics and Astronomy, University of Sheffield, 
Sheffield S3 7RH, UK\\
$^{2}$Space Telescope Science Institute, 3700 San Martin Drive, Baltimore MD 21218, 
USA\\
$^{3}$Department of Physics, University of California, Davis, 1 Shields Ave, Davis CA 95616, USA\\
$^{4}$Centro de Astrobiologi\'{a}, CSIC/INTA, Campus ESAC, Apartado Postal 78, E-28\,691 Villanueva de la Ca\~nada, Madrid, Spain\\
$^{5}$ Department of Physics, University of Oxford, Denys Wilkinson Building, Keble Road, Oxford, OX1 3RH, UK\\
$^{6}$ Argelanger-Institut f\"{u}r Astronomie der Universit\"{a}t Bonn, Auf dem 
H\"{u}gel 71, D-53121 Bonn, Germany \\
$^{7}$ Department of Physics, University of Warwick, Gibbet Hill Rd, 
Coventry CV4 7AL, UK\\
$^{8}$ Institute for Astrophysics, Tuerkenschanzstr. 17, AT-1180 Vienna, Austria\\
$^{9}$ Institute of Astronomy \& Astrophysics, National Observatory of Athens, I. 
Metaxa \& Vas. Pavlou St, P. Penteli 15236, Greece \\
$^{10}$ Astronomical Institute Anton Pannekoek, University of Amsterdam, Kruislaan 403, 
1098 SJ, Amsterdam, Netherlands\\
$^{11}$ Institute of Astronomy, KU Leuven, Celestijnenlaan 200 D, 3001 Leuven, Belgium\\
$^{12}$ UK Astronomy Technology Centre, Royal Observatory Edinburgh, Blackford Hill, 
Edinburgh, EH9 3HJ, UK\\
$^{13}$ Armagh Observatory, College Hill, Armagh, BT61 9DG, UK\\
$^{14}$ Instituto de Astrof\'{i}sica de Canarias, E-38200 La Laguna, Tenerife, Spain\\
$^{15}$ Departamento de Astrof\'{i}sica, Universidad de La Laguna, E-38205 La Laguna, 
Tenerife, Spain \\
$^{16}$ Dept of Physics \& Astronomy, University College London, Gower St, London, WC1E 
6BT, UK\\
$^{17}$ European Space Astronomy Centre, ESA, Villanueva de la Ca\~{n}ada, Madrid, Spain\\
$^{18}$ Uinversit\"{a}ts-Sternwarte, Scheinerstrasse 1, 81679 M\"{u}nchen, Germany 
}
\date{\today}

\pubyear{2015}
\begin{document}\label{firstpage}
\pagerange{\pageref{firstpage}--\pageref{lastpage}} 
\maketitle

\begin{abstract} 
We introduce a HST/STIS stellar census of R136a, the central ionizing star cluster 
of 30 Doradus. We present low resolution far-ultraviolet STIS/MAMA 
spectroscopy of R136 using 17 contiguous 52$\times$0.2 arcsec slits which 
together provide complete coverage of the central 0.85 parsec (3.4 
arcsec). We provide spectral types of 90\% of the 57 
sources brighter than $m_{\rm F555W}$ = 16.0 mag within a radius of 0.5 parsec of 
R136a1, plus 8 additional nearby sources including R136b (O4\,If/WN8). We measure wind 
velocities for 52 early-type stars from C\,{\sc iv} 
$\lambda\lambda$1548--51, including 16 
O2--3 stars. For the first time we spectroscopically classify all Weigelt 
\& Baier members of R136a, which comprise three WN5 stars (a1-a3), two O 
supergiants (a5-a6) and three early O dwarfs (a4, a7, a8). A complete 
Hertzsprung-Russell diagram for the most massive O stars in R136 is 
provided, from which we obtain a cluster age of 1.5$^{+0.3}_{-0.7}$ Myr.
In addition, we 
discuss the integrated ultraviolet spectrum of R136, and highlight the 
central role played by the most luminous stars in producing the prominent 
He\,{\sc ii} $\lambda$1640 emission line. This
emission is totally dominated by very massive stars with initial masses above 
$\sim 100 M_{\odot}$. The presence of strong He\,{\sc ii} 
$\lambda$1640 emission in the integrated light of very young star clusters (e.g 
A1 in NGC~3125)  favours an initial mass function extending well 
beyond a conventional upper limit of 100 $M_{\odot}$. We include montages of 
ultraviolet spectroscopy for LMC O stars in the 
Appendix. Future studies in this series will focus 
on optical STIS/CCD medium resolution observations.
\end{abstract}


\begin{figure*}
\begin{center}
 \includegraphics[height=2.0\columnwidth,angle=-90]{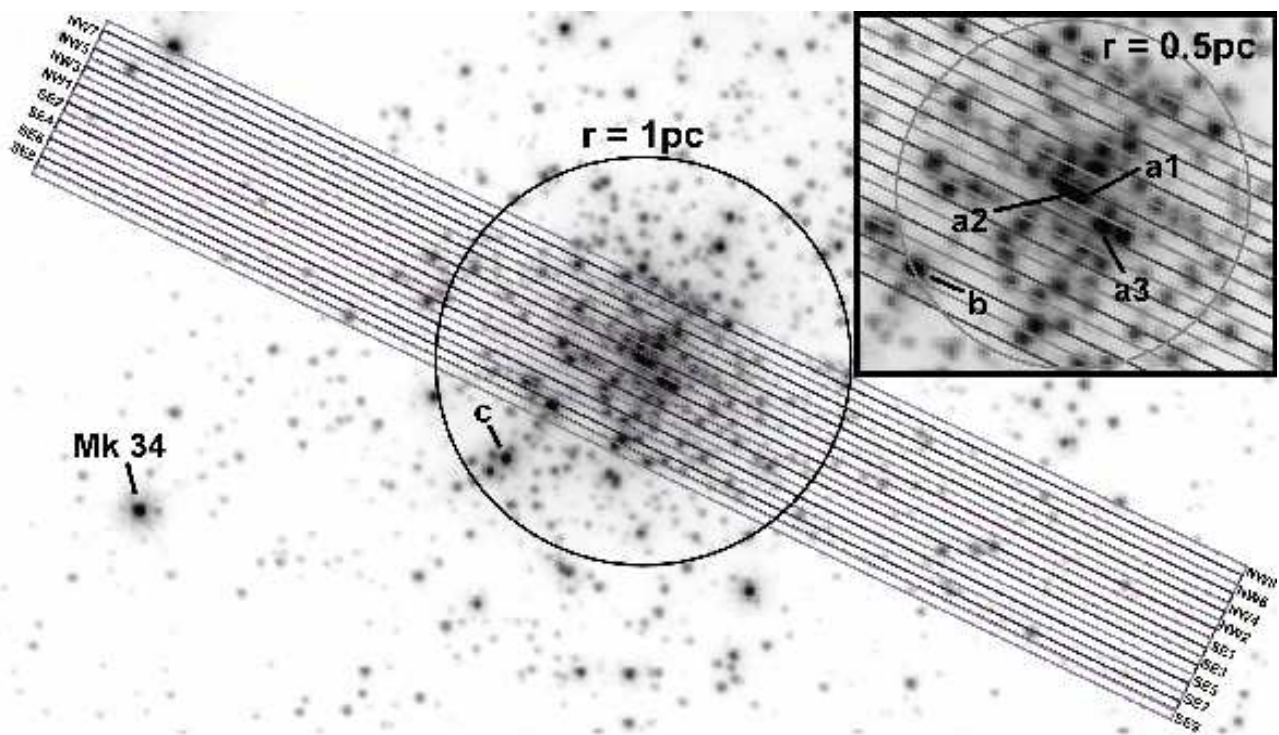} 
 \caption{HST/STIS slits (52$''\times 0''.2$) superimposed 
upon an F336W WFC3/UVIS image of R136 (de Marchi et al. 2011), oriented 
with North up and East to left, together with a circle of radius 4.$''$1 (1 parsec) and identification of the
acquisition star Melnick 34. The active slit length for MAMA observations is the central
25$''$. The zoom highlights the central region, including 
identification of individual slits and the integrated R136a cluster (2.$''$05 radius circle, 
centred upon R136a1 equivalent to 0.5 parsec at the distance of the LMC)}
\label{slits}
\end{center}
\end{figure*}

\section{Introduction}

The formation of massive stars remains an unsolved problem in 
astrophysics. This is especially challenging since very massive stars
 (VMS, $\geq 100 M_{\odot}$) are usually located within young
massive star clusters ($\geq 10^{4} M_{\odot}$) with small core radii
($r_{c} \sim$ 0.1--0.2 pc, Portegies Zwart et al. 
2010).  The short lifetime of such stars (2--3 Myr, Yusof et al. 2013) 
restricts suitable Milky Way star clusters to only a handful, 
including the Arches (Martins et al. 2008) and NGC~3603 (Schnurr et al. 
2008), both of which are highly reddened.
Therefore, observations of young star clusters at ultraviolet
wavelengths -- corresponding to the peak output of hot, luminous stars 
-- are largely restricted to extragalactic cases.

R136 was included in the census of optically bright Magellanic Cloud stars
by Feast et al. (1960), and subsequently resolved into three components, 
a, b and c (Feitzinger et al. 1980), with component a subsequently shown to be
multiple (Weigelt \& Baier 1985).
R136a is the closest example of an extragalactic young massive star 
cluster (Hunter et al. 1995). It is the central 
cluster of the NGC~2070 complex within the 30 Doradus star forming 
region in the Large Magellanic Cloud (LMC, Massey \& Hunter 1998; Doran et al. 
2013), and is  relatively lightly reddened, facilitating study at 
ultraviolet wavelengths (de Marchi et al. 1993). However, the 
0.1\,pc core radius of R136a (Portegies Zwart et al. 
2010; Selman \& Melnick 2013) subtends
only 0.41$''$ at the 50\,kpc distance to the LMC (Pietrzynski et al. 2013), which led to 
the  cluster at one time being mistaken for a $\geq$1000 $M_{\odot}$ star 
(Cassinelli  et al. 1981, Feitzinger et al. 1983), although Walborn (1973) had earlier
predicted that R136a would have a structure analogous to the core of NGC~3603. 
Consequently, observations at high spatial resolution are 
required to resolve individual stars within the core of R136a, requiring 
either Hubble Space Telescope or large 
ground-based 
telescopes with adaptive optics (VLT/MAD: Campbell et al. 2010).

The star-formation history of NGC~2070 has recently been investigated by Cignoni et al. (2015)
using Hubble Tarantula Treasury Project (HTTP) imaging, revealing an upturn in star formation
$\sim$7 Myr ago, peaking 2--3 Myr ago, and a total stellar mass of 9$\times 10^{4} M_{\odot}$ assembled
within 20 pc of R136. The mass function for intermediate- and low-mass stars in R136 have
been obtained from HST imaging (Hunter et al. 1995; de Marchi et al. 
2011). However, the colour/reddening degeneracy of hot massive stars at 
ultraviolet and  optical wavelenths necessitates spectroscopy for robust 
physical  properties, and in turn, masses, measuring present-day mass 
functions (though see Ma\'{i}z Apell\'{a}niz et al. 2014).
To date, Koornneef \& Mathis (1981) and Vacca et al. (1995)
have presented UV scans across NGC~2070 with the International Ultraviolet Explorer (IUE),
while UV spectroscopy of individual stars of 
R136a has been limited to the brightest members with HST/GHRS (de Koter et al. 1997, 1998).

Meanwhile, Massey \& Hunter (1998) have obtained optical spectroscopy for dozens 
of stars within NGC~2070 using HST/FOS, which revealed large numbers of very 
early O stars, indicating a young cluster age ($<$1--2 Myr). More recently 
Schnurr et al. 
(2009) used VLT/SINFONI to obtain spatially resolved K-band spectroscopy 
of the brightest sources, namely R136a1, a2, a3 and c, which are 
WN-type emission-line stars. Analysis of the VLT and HST spectroscopy plus 
VLT/MAD imaging led Crowther et al. (2010) to conclude that such stars 
possess very high (initial) stellar masses of 160--320 $M_{\odot}$, results
largely supported by Bestenlehner et al. (2011) and Hainich et al. (2014).

\begin{figure*}
\begin{center}
 \includegraphics[height=1.0\columnwidth,angle=90]{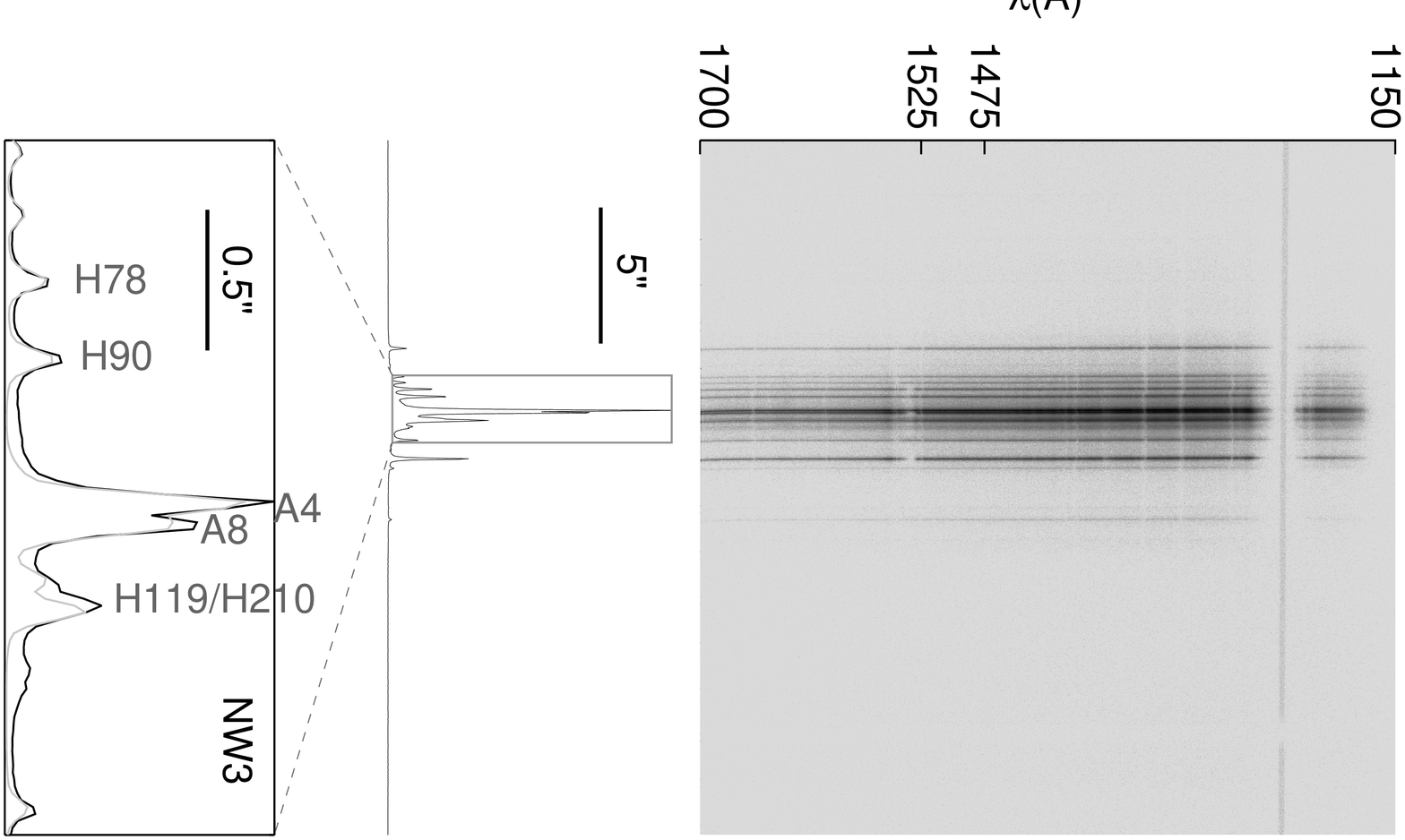} 
 \includegraphics[height=1.0\columnwidth,angle=90]{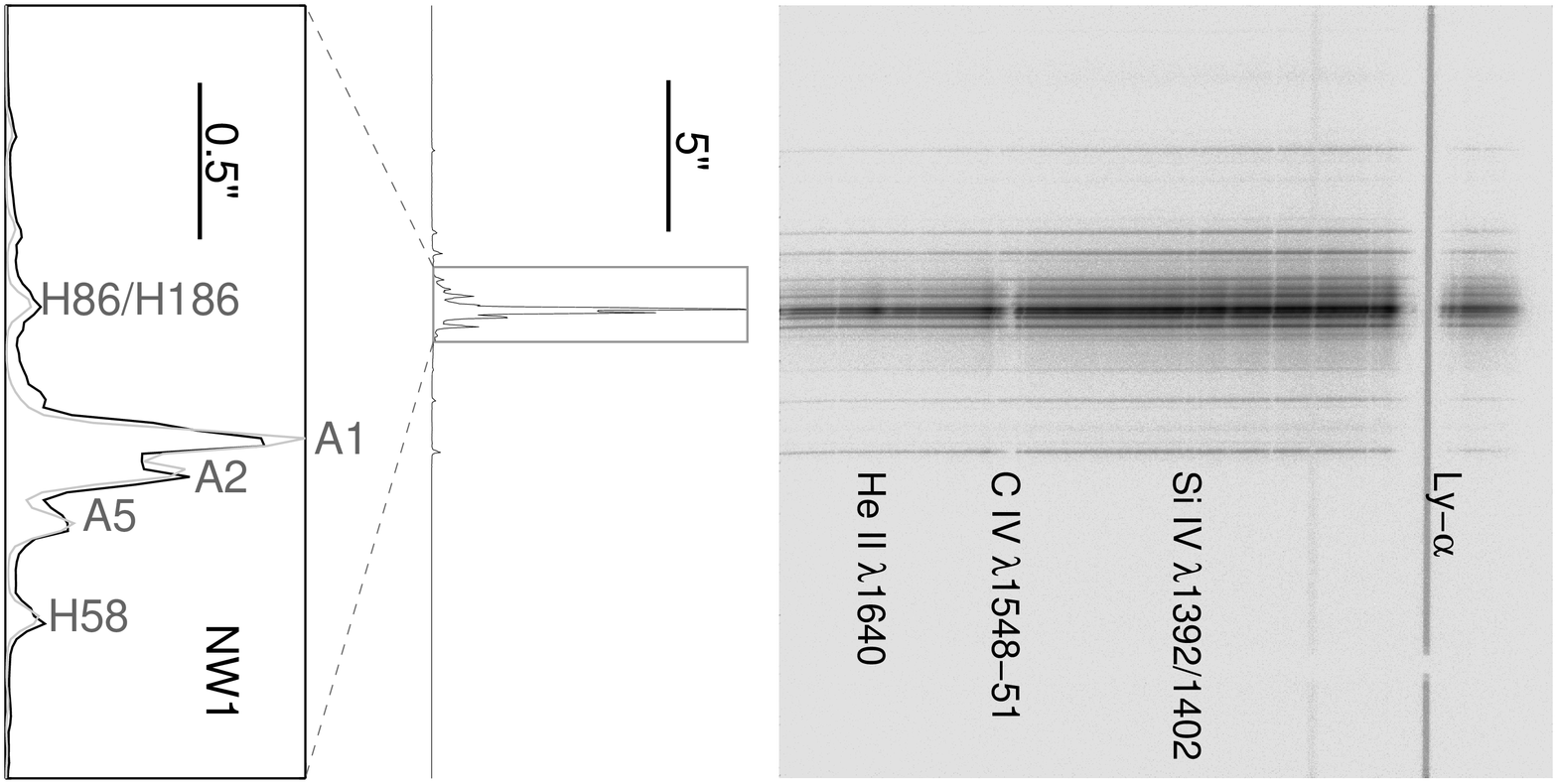} 
 \caption{
{\bf (top)}: Two dimensional spectral dataset for NW3 (left) and 
NW1 (right) slits, oriented such that the 
$\lambda\lambda$1150--1700 dispersion axis runs from top 
to bottom  (horizontal line near top is geocoronal  Ly$\alpha$  emission, occulting
bar lies to the right of each image) 
and the (25$''$ = 6.2 pc) spatial axis is oriented from left (SW) to right 
(NE). Prominent (stellar plus interstellar) 
absorption features include Ly$\alpha$ (top) and C\,{\sc iv} $\lambda$1548--51 (bottom); 
(centre): Collapsed spectral image for NW3 (left) 
and NW1  (right) between $\lambda\lambda$1475--1525; {\bf (bottom)} 
2.$''$5 (0.62 pc) zoom of collapsed spectral image for NW3 (left) and NW1 
(right) indicating the dense cores, resolving R136a1 and a2, and R136a4 
and a8, respectively, together with {\sc multispec} fits (faint gray line).
   }
\label{2D}
\end{center}
\end{figure*}

The existence of such high-mass stars in R136 challenges the previously 
accepted $\sim 150 M_{\odot}$ upper stellar mass limit (Figer 2005), unless they are  
products of stellar mergers (Banerjee et al. 2012; Schneider et al. 2014a).
If such stars are indeed merger remnants one would expect younger apparent ages for this
with respect to lower mass stars, and differences in the present day mass function with respect to
a Salpeter slope (Schneider et al. 2014a; Schneider et al. 2015). Massey \& Hunter (1998) suggested that
the mass function of R136 is normal, although their study only included a
subset of the brightest stars in the vicinity of R136a.

\begin{table*}
\begin{center}
\caption{Log of HST/STIS spectroscopic observations of R136 from Cycles 
19 (GO 12465) and 20 (GO 13052).}
\label{log}
\begin{tabular}{l
@{\hspace{3mm}}c
@{\hspace{3mm}}l
@{\hspace{3mm}}l
@{\hspace{3mm}}l
@{\hspace{3mm}}r
@{\hspace{3mm}}c
@{\hspace{3mm}}r
}
\hline
GO & Orbits & Target & Detector & Grating & PA & $T_{\rm exp}$ & Date\\
/Visit         &        &  && /Position & $\circ$ & sec & \\
\hline
12465/1 & 4 & SE9 .. NW1 & CCD  & G430M/3936 & 64 & 2$\times$(440 .. 
570) & 6 Apr 2012 \\
12465/2 & 4 & SE9 .. NW1 & CCD  & G430M/4451 & 64 & 2$\times$(440 .. 
570) & 4 Apr 2012 \\
12465/3 & 4 & SE9 .. NW1 & CCD  & G430M/4706 & 64 & 2$\times$(440 .. 
570) & 6 Apr 2012 \\
12465/4 & 4 & SE9 .. NW3 & MAMA  & G140L & 64 & 2$\times$(400 ..
470) & 7 Apr 2012 \\
12465/5 & 2 & NW4 .. NW8 & MAMA  & G140L & 64 & 2$\times$(470 .. 
520) & 8 Apr 2012 \\
12465/6 & 5 & SE9 .. NW8 & CCD  & G750M/6581 & 244 & 2$\times$(350 .. 
370) & 23 Oct 2012 \\
12465/7 & 3 & NW2 .. NW8 & CCD  & G430M/3936 & 244 & 2$\times$540 & 20 Oct 
2012 \\
12465/8 & 3 & NW2 .. NW8 & CCD  & G430M/4451 & 244 & 2$\times$540 &  20 
Oct 2012 \\
12465/9 & 3 & NW2 .. NW8 & CCD  & G430M/4706 & 244 & 2$\times$540 & 23 Oct 
2012 \\
13052/1 & 4 & SE9 .. NW1 & CCD & G430M/4194 & 244 & 2$\times$(440 ..
570) & 21 Oct 2012 \\
13052/2 & 3 & NW2 .. NW8 & CCD  & G430M/4194 & 64 & 2$\times$540 & 3 Apr 
2013 \\
\hline
\end{tabular}
\end{center}
\end{table*}

A census of the hot, luminous stars in R136a with STIS
is the focus of  the present series of papers. We discuss the suitability of ultraviolet spectroscopy
for spectral classification of O stars, based in part upon spectroscopic datasets in the usual blue visual
wavelength range. Wind velocities for O stars can uniquely be obtained from P Cygni profiles in the far-UV  (e.g. Groenewegen et al. 1989, Prinja et al. 1990).
Comparisons of empirical (reduced) wind momenta\footnote{Reduced wind momenta $\dot{M} v_{\infty} R^{1/2}$, where $\dot{M}$ is the mass-loss rate, $v_{\infty}$ is the
terminal wind velocity and $R$ is the stellar radius} with theoretical predictions
usually rely on approximate $v_{\infty}$ calibrations (e.g. Lamers et al. 1995). Indeed, high quality far-UV spectroscopy 
has been restricted to relatively few LMC O stars to date (Prinja \& Crowther 1998; Walborn et al. 1995; Massey et al. 2005). One byproduct of the present study is a
calibration of wind velocities for (mostly) early O dwarfs at the $\sim 0.5 Z_{\odot}$ metallicity of the LMC.

In the first study in this series, we introduce our new STIS observations of R136a in Sect.~\ref{obs}.
We present and analyse ultraviolet STIS/MAMA spectroscopy of individual stars in 
Sect.~\ref{mama}, including approximate spectral types and estimates of terminal wind velocities. An ultraviolet
spectral atlas of LMC O stars is compiled using literature spectra, in a few instances complemented with new STIS/CCD
spectroscopic datasets to provide templates of missing spectral types.
The current dataset also allows the contributions from individual stars to its integrated ultraviolet spectrum to be quantified. This is discussed in Sect.~\ref{cluster}, of relevance for other young massive star clusters in the local universe (e.g. Chandar et al. 2004; Leitherer et al. 2011) and the integrated rest-frame ultraviolet spectrum of high redshift star forming galaxies (e.g. Shapley et al. 2003). Finally, brief conclusions are drawn in Sect.~\ref{conclusions}.

\section{Observations}\label{obs}

Here we present our new long-slit HST Space Telescope Imaging Spectrograph  
(STIS) observations of R136, and use of archival WFC3/UVIS imaging discussed 
by de Marchi et al. (2011).

\subsection{Ultraviolet spectroscopy}

We have obtained ultraviolet STIS/G140L spectroscopy using the far-UV Multi-Anode 
Microchannel Array (MAMA) detector and a 52$\times$0.2 arcsec slit
at 17 contiguous pointings in R136, identified by their position within
the cluster, namely SE9, SE8, \ldots SE1, NW1, \ldots  NW7, NW8 (Fig~\ref{slits}).
The spectral coverage of the G140L grating is $\lambda\lambda$1150--1700\AA,
with a pixel scale of 0.6\AA/pixel, and notional point source
resolving power of 1250 at $\lambda$=1500\AA. Since 
the MAMA plate scale is 0.024 arcsec/pix, a slit length of $\sim$25 arcsec is sampled in all cases. 
All MAMA exposures were obtained in two visits, 4 and 5 from 
GO 12465, totalling 6 orbits, as set out in Table~\ref{log}. All  
observations 
were obtained  at a fixed position angle 64$^{\circ}$ (E of N),
to ensure that R136a1  and R136a2 lie within the same 
slit position (NW1) as shown in  Figure~\ref{slits}. For each visit the 
nearby isolated star Melnick 34 was acquired using the CCD and F28X50LP 
aperture, followed by a peakup with the CCD and 52$''\times$0\farcs1 
aperture, 
followed by an offset to  individual pointings, with two exposures 
acquired per pointing, the second spatially offset along the slit by 11 pix (0.27 arcsec),
with integration times shown in Table~\ref{log}.

\begin{figure}
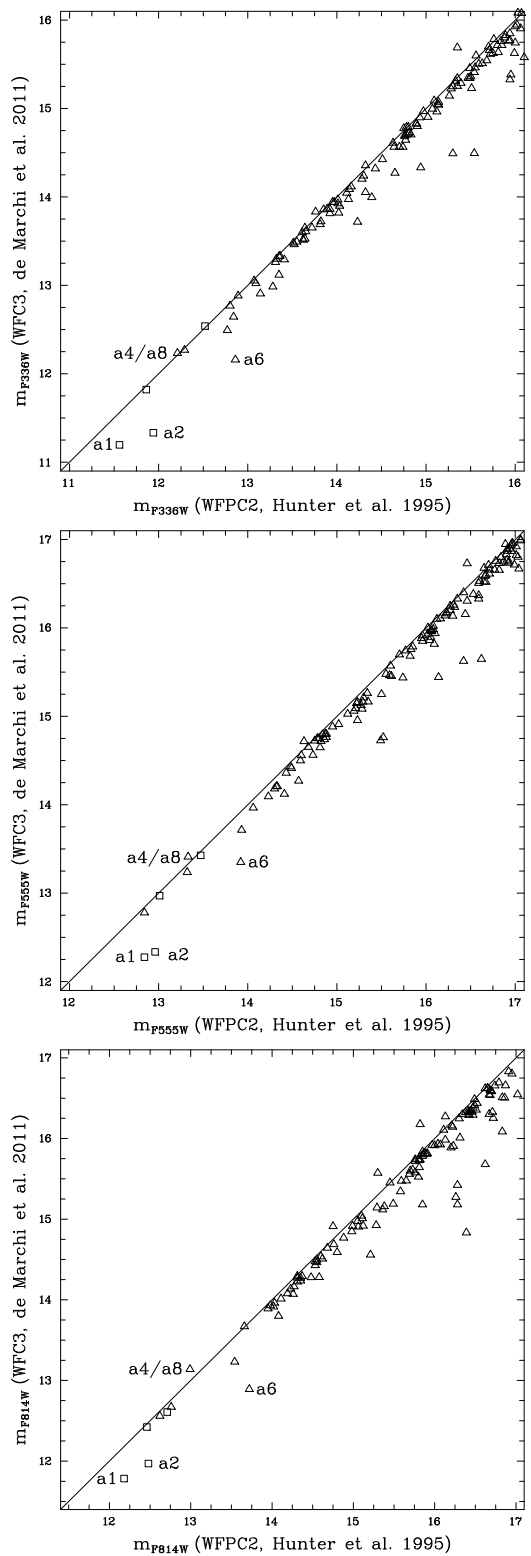

\begin{center}
 \includegraphics[bb=65 200 515 630,width=0.85\columnwidth]{wfpc2_wfc3_u.eps} 
 \includegraphics[bb=65 200 515 630,width=0.85\columnwidth]{wfpc2_wfc3_v.eps} 
 \includegraphics[bb=65 200 515 630,width=0.85\columnwidth]{wfpc2_wfc3_i.eps} 
 \caption{Comparison between WFPC2 (Hunter et al. 1995) and WFC3/UVIS 
photometry (de Marchi et al. 2011) for all sources in common within 10 arcsec of R136a1 (O stars: triangles, WN 
stars: squares) for F336W (top), F555W (middle) and F814W (bottom). Agreement is generally excellent aside from 
R136a1, a2 and a6 (see Sect.~\ref{census}).}
\label{wfpc2_wfc3}
\end{center}
\end{figure}

MAMA exposures were combined and extracted using the
specialised {\sc stistools} package  (Bostroem \& Proffitt 2011). No automated wavelength 
correction was made for sources not centred in individual slits while 
standard {\sc calstis} slit loss corrections were applied. Therefore, 
absolute wavelength scales may be incorrect by up to $\pm$2 pix 
(0.56\AA/pix) while absolute far-UV fluxes will be underestimated by up to 
a factor of two in extreme cases.

\begin{table*}
\begin{center}
  \caption{HST/WFC3 photometry (de Marchi et al. 2011) for sources ($m_{\rm F555W} \leq$  16.0 mag).
    within a  projected distance of 2.05 arcsec (0.5 parsec) from R136a1.
In case WFC3 photometry is unavailable, WFPC2 photometry from Hunter et al. 
(1995) is used, albeit offset in F555W by --0.17 mag (indicated in italic font, see 
Sect.~\ref{census}). Identifications are from Weigelt \& Baier  (1985, WB85) or Hunter et al. (1995, HSH95), while 
spectral types are obtained from optical spectroscopy. $A_{\rm F555W}$ is either obtained via the observed $m_{\rm F336W}$ -- $m_{\rm F555W}$ colour plus $R_{\rm 5495}$ = 4.1, or $A_{\rm F555W}$ = 1.72 mag is adopted (values for WN5 stars are from
Crowther et al. 2010).}
\label{targets}
\begin{tabular}{
c@{\hspace{1.5mm}}r@{\hspace{1.5mm}}c
@{\hspace{1.0mm}}r@{\hspace{1.0mm}}c
@{\hspace{1.5mm}}c@{\hspace{1.5mm}}c
@{\hspace{1.5mm}}c@{\hspace{1.5mm}}c
@{\hspace{1.5mm}}c@{\hspace{1.5mm}}l
@{\hspace{1.5mm}}c}
\hline
WB85 & HSH95 & Spectral& Ref & $r$ & $m_{\rm F555W}$ & $m_{\rm F336W - F438W}$ & $m_{\rm F438W - F555W}$ & 
$m_{\rm F555W - F814W}$ & $A_{\rm F555W}$ & $M_{\rm F555W}$ & Slit\\
     &       &Type & & arcsec & mag& mag& mag & mag & mag &  mag  &\\
\hline
a1&   3  &  WN5h & b & 0.00 & 12.28$\pm$0.01\phantom{$\ddag$} & -1.25$\pm$0.01 & 0.17$\pm$0.01 & 0.49$\pm$0.01 & 1.88 & --8.09 &  NW1\\
a2&   5  &  WN5h & b & 0.08 & 12.80$\pm$0.01$^{\ddag}$ & -1.23$\pm$0.01 & 0.23$\pm$0.01 & 0.37$\pm$0.01 & 
1.83 & --7.52$^{\ddag}$& NW1  \\
a5 & 20 & O2\,If* & e & 0.28& 13.71$\pm$0.01\phantom{$\ddag$} & --1.28$\pm$0.02 & 0.06$\pm$0.01 & 0.48$\pm$0.01 & 1.87& --6.65 & NW1 \\
a7 & 24 & O3\,III(f*) & a & 0.36& 13.97$\pm$0.01\phantom{$\ddag$} & --1.28$\pm$0.02 & --0.04$\pm$0.02 & 0.30$\pm$0.02 & 1.58& --6.10 & NW3 \\
a8 & 27  & \multirow{2}{*}{} &  \multirow{2}{*}{} & 0.39
& 14.42$\pm$0.01$^{\ddag}$ & \multirow{2}{*}{--1.30$\pm$0.01} & 
\multirow{2}{*}{0.12$\pm$0.01}
& \multirow{2}{*}{0.27$\pm$0.01} & \multirow{2}{*}{1.98} & --6.05$^{\ddag}$ & \multirow{2}{*}{NW3} \\ 
a4 & 21 &     &  &0.43 & 13.96$\pm$0.01$^{\ddag}$ & & & & & --6.51$^{\ddag}$ & \\
       & 186 &         &    & 0.43 & 15.63$\pm$0.02\phantom{$\ddag$} & --1.23$\pm$0.03 & 0.10$\pm$0.03 & 0.35$\pm$0.04 
& 1.75 & --4.62 & NW1\\
       & 86 &         &    & 0.45 & 14.73$\pm$0.01\phantom{$\ddag$} & --1.42$\pm$0.02 & 0.41$\pm$0.02 & & 2.15 & --5.91 & NW1 \\
a3 &  6  &  WN5h & b & 0.49 & 12.97$\pm$0.01\phantom{$\ddag$} &--1.31$\pm$0.01 & 0.16$\pm$0.01 & 0.55$\pm$0.01 & 1.87 & --7.39 & SE1\\
       & 66 & O3\,V      & a  & 0.49 & 15.06$\pm$0.02\phantom{$\ddag$} & --1.29$\pm$0.02 & 0.17$\pm$0.02 & 0.29$\pm$0.03 & 1.95 & --5.38 & SE2 \\
       & 119&          &    & 0.51 & {\it 15.75\phantom{$\pm$0.00$\ddag$} 
} 
&        &      &  & 1.72 & --4.46 & NW3\\
%
%
       & 58 & O3\,III(f*) & a & 0.59 & 14.80$\pm$0.01\phantom{$\ddag$} & --1.30$\pm$0.02 & 0.11$\pm$0.02 & 0.34$\pm$0.02 & 1.96 & --5.65 & NW1 \\
       & 30 & O7\,V  & a  & 0.62& 14.21$\pm$0.01\phantom{$\ddag$} & --1.31$\pm$0.01 & 0.13$\pm$0.01 & 0.29$\pm$0.01 & 1.61 & --5.89 & NW4 \\
       & 70 &  O5\,Vz   &  f  & 0.62 & 14.96$\pm$0.01\phantom{$\ddag$} & --1.27$\pm$0.02 & 0.13$\pm$0.02 & 0.37$\pm$0.03 & 1.79 & --5.32 & NW2 \\ 
       & 89 &          &    & 0.66 & 14.76$\pm$0.01\phantom{$\ddag$} & --1.11$\pm$0.02 & 0.34$\pm$0.02 & 0.20$\pm$0.02 & 2.52 &--6.25& SE2 \\
       & 62 &          &    & 0.66 & 14.91$\pm$0.01\phantom{$\ddag$} & --1.36$\pm$0.02 & 0.11$\pm$0.02 & 0.31$\pm$0.02 &1.85 & --5.43 & SE3 \\
a6 & 19  &      &  & 0.73 & 13.35$\pm$0.01\phantom{$\ddag$} & --1.26$\pm$0.01 & 0.07$\pm$0.01 & 0.46$\pm$0.01 
&1.82&--6.96 & SE1\\
       & 50 &     &   & 0.74 & 14.65$\pm$0.01\phantom{$\ddag$} & --1.25$\pm$0.01 & 0.11$\pm$0.01 & 0.37$\pm$0.02 & 2.05 & --5.89 & SE3 \\
       & 90 & O5\,V::    & a  & 0.77 & 15.48$\pm$0.02\phantom{$\ddag$} & --1.36$\pm$0.03 & 0.12$\pm$0.03 & 0.34$\pm$0.04 & 1.67 & --4.68 & NW3 \\
       & 149 &         &    & 0.78 & 15.44$\pm$0.02\phantom{$\ddag$} & --1.28$\pm$0.03 & 0.17$\pm$0.03 & 0.26$\pm$0.04 & 1.96 & --5.01 & SE2\\
       & 141 &         &    & 0.78 & 15.82$\pm$0.03\phantom{$\ddag$} & --1.35$\pm$0.03 & 0.09$\pm$0.03 & 0.29$\pm$0.05 & 1.55 & --4.22 & SE4\\
       & 80 & O8\,V    & f   & 0.87 & 15.17$\pm$0.02\phantom{$\ddag$} & --1.30$\pm$0.02 & 0.11$\pm$0.02 & 
0.25$\pm$0.03 & 1.53 & --4.85 & SE3 \\
       & 35 & O3\,V  & f  & 0.88& 14.32$\pm$0.01\phantom{$\ddag$} & --1.43$\pm$0.01& 0.11$\pm$0.01 &  & 1.64 & --5.81 &NW5 \\
       & 78 & O4:\,V    & f   & 0.97 & 15.26$\pm$0.02\phantom{$\ddag$} & --1.33$\pm$0.02 & 0.11$\pm$0.02 & 0.29$\pm$0.03 & 1.70 & --4.93 & NW3 \\
       & 73 & O9\,V      & a  & 1.01 & 15.13$\pm$0.02\phantom{$\ddag$} & --1.12$\pm$0.02 & 0.11$\pm$0.02 &  & 1.78 & --5.14 & SE4 \\
       & 92 & O3\,V      & a  & 1.05 & 15.46$\pm$0.02\phantom{$\ddag$} & --1.36$\pm$0.02 & 0.10$\pm$0.03 & 0.30$\pm$0.04 & 1.52 & --4.55 & NW6 \\
       & 143 &         &    & 1.05 & 15.94$\pm$0.03\phantom{$\ddag$} & --1.30$\pm$0.04 & 0.19$\pm$0.04 & 0.36$\pm$0.06 & 1.70 & --4.25 & SE2 \\
       &112 & O8.5\,III(f) & a & 1.06 & 15.74$\pm$0.03\phantom{$\ddag$} & --1.25$\pm$0.04 & 0.12$\pm$0.04 & 0.29$\pm$0.05 & 1.65 & --4.40 &NW2/3 \\
       & 135 &         &    & 1.08 & 15.90$\pm$0.03\phantom{$\ddag$} & --1.20$\pm$0.04 & 0.13$\pm$0.04 & & 1.67 & --4.26 & NW5\\
       & 69 & O3--6\,V   & a  & 1.09 & {\it 
15.05\phantom{$\pm$0.00$\ddag$} } 
& --1.32$\pm$0.04 &  &  & 1.72 & --5.16 & SE5 \\
       & 52 & O3\,V     & a & 1.13 & 14.72$\pm$0.01\phantom{$\ddag$} & --1.34$\pm$0.01 & 0.09$\pm$0.01 & 0.22$\pm$0.02 
& 1.71 & --5.48 & SE5 \\
       & 48 & O2--3\,III(f*) & f &  1.22 & 14.75$\pm$0.01\phantom{$\ddag$} & --1.26$\pm$0.01 & 0.10$\pm$0.01 & 0.46$\pm$0.02 & 2.01 & --5.75 & SE2 \\
       & 77 & O5.5\,V+O5.5\,V & d & 1.29 & 15.21$\pm$0.01\phantom{$\ddag$} & --1.31$\pm$0.02 & 0.04$\pm$0.02 & 0.30$\pm$0.02 & 2.87 & --6.15 & SE6 \\
       & 94 & O3\,V      & a   & 1.31 & 15.57$\pm$0.02\phantom{$\ddag$} & --1.31$\pm$0.03 & 0.10$\pm$0.03 & 0.28$\pm$0.04 & 1.69 & --4.61 &SE6 \\
      &115 &          &          &  1.41 & 15.76$\pm$0.02\phantom{$\ddag$} & --1.32$\pm$0.03 & 0.12$\pm$0.03 & 0.28$\pm$0.04 & 1.44 & --4.17 &NW8 \\
       & 132 &         &    & 1.48 & 15.86$\pm$0.03\phantom{$\ddag$} & --1.35$\pm$0.03 & 0.13$\pm$0.03 & 0.22$\pm$0.04 & 1.40 & --4.03 & NW8\\
       & 36 & O2\,If* & f & 1.51& 14.41$\pm$0.01\phantom{$\ddag$} & --1.26$\pm$0.01 & 0.11$\pm$0.01 & 0.48$\pm$0.01 & 1.89 & --5.97 & SE3 \\
       & 173  &        &    & 1.53 & 15.98$\pm$0.03\phantom{$\ddag$} & --1.32$\pm$0.04  & 0.19$\pm$0.04 & 0.25$\pm$0.05 & 1.56 & --4.07 & NW8\\
       & 75 &          &    & 1.56 & 15.08$\pm$0.01\phantom{$\ddag$} & --1.32$\pm$0.02 & 0.10$\pm$0.02 & 0.18$\pm$0.02 & 1.70 & --5.11 & NW8 \\
       &114 &          &    & 1.57 & 15.68$\pm$0.02\phantom{$\ddag$} & --1.35$\pm$0.03 & 0.10$\pm$0.03 & 0.21$\pm$0.04 & 1.44 & --4.25 &SE7 \\
       & 108 &         &    & 1.59 & 15.44$\pm$0.02\phantom{$\ddag$} & --1.35$\pm$0.02 & 0.18$\pm$0.02 & 0.25$\pm$0.03 & 1.60 & --4.65 & NW8\\
       & 31 &       &   & 1.64& 14.12$\pm$0.01\phantom{$\ddag$} & --1.31$\pm$0.01 & 0.10$\pm$0.01 & 0.32$\pm$0.01 & 1.92 & --6.29 & SE8 \\
       & 49 & O3\,V     & a  &  1.67 & 14.75$\pm$0.01\phantom{$\ddag$} & --1.41$\pm$0.01 & 0.15$\pm$0.01 & 0.27$\pm$0.02 & 1.60 & --5.34 & NW8 \\
       &207 &          &    & 1.72 & 15.65$\pm$0.02\phantom{$\ddag$} &        & 0.09$\pm$0.02 & 0.47$\pm$0.03 & {\it 1.72} & --4.56 &SE2 \\
       & 46 &     &   & 1.73 & 14.56$\pm$0.01\phantom{$\ddag$} & --1.25$\pm$0.01 & 0.21$\pm$0.01 & 0.49$\pm$0.01 & 2.14 & --6.07 & SE2 \\
       & 47 & O3\,III(f*) & a & 1.75 & 14.72$\pm$0.01\phantom{$\ddag$} & --1.22$\pm$0.01 & 0.15$\pm$0.01 & 0.45$\pm$0.01 & 2.10 & --5.87 & NW3 \\
       & 40 & O3\,V & a & 1.80& 14.56$\pm$0.01\phantom{$\ddag$} & --1.34$\pm$0.01& 0.08$\pm$0.01 & 0.29$\pm$0.01 & 1.83 & --5.76 & NW8 \\
       &118 &          &    & 1.81 &{\it 15.74\phantom{$\pm$0.01$\ddag$} } 
&        &  & & 1.72 & --4.49 &SE2 \\
        & 116 &         &    & 1.81 & 15.79$\pm$0.02\phantom{$\ddag$} & --1.32$\pm$0.03 & 0.10$\pm$0.03 &  & 1.51 & --4.21 & SE6 \\
        & 42 & O3\,V+O3\,V & d & 1.83 & 14.72$\pm$0.01\phantom{$\ddag$} & --1.36$\pm$0.01 & --0.03$\pm$0.01 & 0.44$\pm$0.01 & 1.45 & --5.22 &NW8 \\
     & 55 & O3\,V     & a  & 1.84 & 14.74$\pm$0.01\phantom{$\ddag$} & --1.31$\pm$0.01 & 0.09$\pm$0.01 & 0.31$\pm$0.01 & 1.91 & --5.66 & SE9 \\
       & 39 & O3\,V+O5.5\,V & d & 1.95 & 14.50$\pm$0.01\phantom{$\ddag$} & --1.32$\pm$0.01 & 0.08$\pm$0.01 & 0.21$\pm$0.01 &1.76 & --5.75 & \ldots \\
& 137   &      &    & 1.97 & 15.97$\pm$0.02 & --1.33$\pm$0.03 & 0.07$\pm$0.03 & 0.23$\pm$0.04 & 1.51 & --4.03 & \ldots \\
       & 71 & O3--6\,V   & a  & 2.03 & 15.16$\pm$0.01\phantom{$\ddag$} & --1.35$\pm$0.01 & 0.11$\pm$0.01 &  & 1.67 & --5.00 & SE9 \\
& 121 & O9.5\,V  & f   & 2.05 & 15.85$\pm$0.02\phantom{$\ddag$} & --1.24$\pm$0.03 & 0.12$\pm$0.03 &  & 1.58 & --4.22 & SE8\\
\hline
\hline
\multicolumn{12}{l}{
  \begin{minipage}{2\columnwidth}~\\
a: Massey \& Hunter (1998); b: Crowther \& Dessart (1998); c: Walborn et al. (2002b); d: Massey et al. 
(2002); e: Crowther \& Walborn (2011); f: Caballero-Nieves et al. (2016, in preparation) \\
$\ddag$: Absolute magnitudes have been adjusted for R136a2, a4 and a8 (see 
Sect.~\ref{census})
  \end{minipage}
}\\
\end{tabular}
\end{center}
\end{table*}


\begin{table*}
\begin{center}
\caption{HST/WFC3 photometry (de Marchi et al. 2011) for additional STIS/MAMA sources ($m_{\rm F555W} \leq$ = 16.0 
mag) beyond a projected distance of 2.05 arcsec (0.5 parsec) from R136a1.
In case WFC3 photometry is unavailable, WFPC2 photometry from Hunter et al. 
(1995) is used, albeit offset in F555W by --0.17 mag (indicated in italic font, see 
Sect.~\ref{census}). Identifications are from Weigelt \& Baier  (1985, WB85) or Hunter et al. (1995, HSH95), while 
spectral types are obtained from optical spectroscopy. $A_{\rm F555W}$ is either 
obtained via the observed $m_{\rm F336W}$ -- $m_{\rm F555W}$
colour plus $R_{\rm 5495}$ = 4.1, or $A_{\rm F555W}$ = 1.72 mag is adopted (values for WN5 stars are from
Crowther et al. 2010).}
\label{mama_targets}
\begin{tabular}{c@{\hspace{1.5mm}}r@{\hspace{1.5mm}}c@{\hspace{1.5mm}}r
@{\hspace{1.5mm}}c@{\hspace{1.5mm}}c
@{\hspace{2mm}}c@{\hspace{2mm}}c
@{\hspace{2mm}}c@{\hspace{2mm}}c
@{\hspace{2mm}}c@{\hspace{2mm}}l
@{\hspace{2mm}}c}
\hline
WB85 & HSH95 & Spectral& Ref & $r$ & $m_{\rm F555W}$ &$m_{\rm F336W - F438W}$ & $m_{\rm F438W - F555W}$ & 
$m_{\rm F555W - F814W}$ & $A_{\rm F555W}$ & $M_{\rm F555W}$ &  Slit\\
     &       &Type & & arcsec & mag& mag& mag & mag & mag &  mag  &\\
\hline
b  &  9  & O4\,If/WN8 & b & 2.12 & 13.24$\pm$0.01 &--1.14$\pm$0.01 & 0.18$\pm$0.01 & 0.57$\pm$0.01 & 2.10 &--7.31 & SE8 \\       
       & 65 &          &    & 2.20 & {\it 15.01\phantom{$\pm$0.01$\ddag$} } 
&  &      &      & 1.72 & --5.20 & SE7\\
        & 134   & O7\,Vz  &  b  & 2.28 & 15.97$\pm$0.01 & --1.36$\pm$0.02 & 0.10$\pm$0.02 & 0.23$\pm$0.03 & 1.57 & --3.97 & SE6\\
       & 64 & O7\,V((f)) & a & 2.44 & 15.03$\pm$0.01 & --1.19$\pm$0.01 & 0.14$\pm$0.01 & 0.38$\pm$0.01 & 1.96 & --5.42 & NW7 \\
       & 45 & O3\,V & a & 2.45 & 14.65$\pm$0.01 & --1.32$\pm$0.01 & 0.15$\pm$0.01 & 0.43$\pm$0.01 & 1.77 & --5.61 & SE7 \\
        & 123   & O6\,V     & b   & 2.47 & 15.91$\pm$0.01 & --1.32$\pm$0.02 & 0.10$\pm$0.02 & 0.09$\pm$0.03 & 1.58 & --4.16 & SE3\\ 
%
       & 68 &      &   & 4.61 & 15.15$\pm$0.01 &--1.24$\pm$0.01 & 0.18$\pm$0.01 & 0.46$\pm$0.01 & 1.88 & --5.22 & SE4 \\
       & 102&      &   & 4.83 & 15.70$\pm$0.01 &--1.18$\pm$0.02 & 0.26$\pm$0.01 & 0.66$\pm$0.01 & 2.44 & --5.24 & NW1 \\ 
\hline
\hline
\multicolumn{12}{l}{
  \begin{minipage}{2\columnwidth}~\\
a: Massey \& Hunter (1998); b: Caballero-Nieves et al. (2016, in preparation)
  \end{minipage}
}\\
\end{tabular}
\end{center}
\end{table*}

Extractions were also independently made using {\sc multispec} 
(Ma\'{i}z Apell\'{a}niz 2005; Knigge et al. 2008) which is particularly suited to the crowded 
region at the geometric centre of the cluster. Figure~\ref{2D} shows two 
dimensional spectral datasets for particularly crowded slits NW1 (a1, a2 
and a5) and NW3 (a4 and a8), including collapsed spectral images
between $\lambda\lambda$1475--1525. {\sc multispec} extractions relied 
upon empirical point spread functions for G140L/MAMA spectroscopy 
employing the position of individual sources in slits from short exposure 
HST WFC3/UVIS F336W imaging of R136 from Oct 2009 (for details see de 
Marchi et al. 2011). The latter provided wavelength corrections if sources 
were not centred in individual slits, but a standard slit loss correction 
is again adopted.

For relatively isolated sources, {\sc calstis} and {\sc multispec} 
extractions agree extremely well, while the latter approach usually 
performed  better in  situations where sources were closely spaced within 
individual  slits. For consistency, spectroscopic datasets  presented here 
were extracted with {\sc multispec}, except for R136 H70, 141, 149 for which 
{\sc multispec} extractions were problematic.

\subsection{Optical spectroscopy}

We have obtained optical STIS spectroscopy using the Charged Couple Device 
(CCD) detector and identical 52$''\times0.''$2 slit at identical 
positions and position angle (or PA + 180$^{\circ}$) to the UV 
spectroscopy, although the 
coarser plate scale of 0.05 arcsec/pix provided complete spatial coverage 
of the slit length. In total, two exposures at four grating positions with 
the G430M grating provided complete blue coverage 
$\lambda\lambda$3793-4849\AA, with a pixel scale of 0.28\AA/pixel, and 
point source resolving power of 7700 at $\lambda$=4400\AA. 
A single grating position with the 
G750M grating provided $\lambda\lambda$6482--7054\AA\ spectroscopy, 
with a pixel scale of 0.56\AA/pixel, and point source
resolving power of 6000 at H$\alpha$. 
For all CCD observations, the second exposure was offset spatially along the slit by 7 pix 
(0.35 arcec). Three of the four G430M observations were taken over the 
course of a few days, with the fourth (G430M/4194) obtained approximately 
6 months later to enable to search for low period binaries. 
All G750M exposures were taken at a single epoch. 

In total CCD exposures 
were split between 7 visits during Cycle 19 (GO 12465, 26 orbits), plus 2 
visits during Cycle 20 (GO 13052, 7 orbits), also set out in 
Table~\ref{log}. {\sc multispec} and {\sc calstis} extractions of CCD 
observations will be the focus of the next paper in this series (Caballero-Nieves et al. 2016, in prep). 

LMC stars that have been observed in the UV and for which reliable optical classifications are available are fairly scarce.
Our sample allows 10 stars with robust optical classifications spanning the full O2--O9.5 subtype sequence, which are included here
as LMC templates (see Sect.\ref{mama}).

\subsection{Stellar census of R136}\label{census}

Hunter et al. (1995) have used HST/WFPC2 to identify the visually brightest 
sources in R136, while deeper HST/WFC3 images have been analysed by de 
Marchi et al. (2011) to derive multicolour photometry. We compare WFPC2
and WFC3/UVIS photometry in F336W, F555W and F814W filters for all sources in common
within 10 arcsec (2.4 pc) of R136a1 in  Fig.~\ref{wfpc2_wfc3}.
Overall, the earlier WFPC2 results are confirmed by WFC3, albeit with modest
systematic offsets of 
$m_{\rm F336W}$(WFPC2) -- $m_{\rm F336W}$(WFC3) = +0.15 $\pm$ 0.20 mag,
$m_{\rm F555W}$(WFPC2) -- $m_{\rm F555W}$(WFC3) = +0.17 $\pm$ 0.21 mag,
$m_{\rm F814W}$(WFPC2) -- $m_{\rm F814W}$(WFC3) = +0.18 $\pm$ 0.22 mag.
In sources for which de Marchi et al. (2011) photometry is unavailable,
we have substituted Hunter et al. (1995) results to which these offsets have been applied. 
Overall, agreement between early WFPC2 and new WFC3 photometry is generally very good, with a few notable
exceptions -- WFC3 photometry from de Marchi et al. (2011) is significantly
brighter for R136a1, a2 and a6.

In Table~\ref{targets} we provide a catalogue of all 55 sources within 2.05 
arcsec (0.5 pc) of R136a1 brighter than $m_{\rm F555W} = 16.0 $ mag, while 
a further 8 bright sources observed with STIS/MAMA are presented in Table~\ref{mama_targets}. Additional 
sources brighter than $m_{\rm F555W} = 17.0 $ mag within 4.1 arcsec (1 pc) are 
listed in Appendix D (Table~\ref{faint_targets}). There is evidence for a non-standard extinction
law in 30 Doradus, and significant variation thereof (Ma\'{i}z Apell\'{a}niz et al. 2014). Specifically
for R136, Doran et al. (2013) favoured $R_{\rm V} \sim$ 4.2 mag for individual stars based on
a combination of optical (WFC3/UVIS) and near-IR (VLT/MAD, Campbell et al. 2010) photometry. 
We have applied {\sc chorizos} to 41 O stars with WFC3 photometry and 
approximate spectral types, and obtained good fits in 34 cases, from which average values of
$R_{\rm 5495}$ = 4.1$\pm$0.5 and $A_{\rm F555W}$ = 1.72$\pm$0.25 mag is obtained, in
good agreement with the Tatton et al. (2013) near-IR study, for which $A_{\rm V}$ = 1.75 
mag was inferred for R136. Here, we adopt tailored extinctions of WN5 stars from Crowther
et al. (2010), while for the O stars we adopt a reddening law of $R_{\rm 5495} = 4.1$ and individual
extinctions based upon temperature dependent F336W -- F555W intrinsic colours from multicolour
{\sc chorizos} (Ma\'{i}z Apell\'{a}niz 2004): 
\[ (m_{\rm F336W} - m_{\rm F555W})_{0} = - 1.475 - 0.155 (T_{\rm eff}/10000K) 
{\rm mag}, \]
plus $A_{\rm F555W}$ = $R_{\rm 5495} E_{\rm F336W-F555W}/2.28$ 
from the Seaton (1979)/Howarth (1983) LMC extinction law, plus an LMC  distance modulus 
of 18.49 mag (Pietrzynski et al. 2013). For stars without UV derived spectral types, we adopt
O6\,V ($T_{\rm eff}$ = 39,900 K) for stars brighter than $m_{\rm F555W} = 16.0 $ mag, and
O9.5\,V ($T_{\rm eff}$ = 32,900 K) for fainter stars. We adopt 
$A_{\rm F555W}$ = 1.72 mag for all sources without WFC3 photometry.

Severe crowding results in significant differences in photometry between Hunter et al.
(1995) and de Marchi et al. (2011) for some of the brightest sources in the core of R136a.
HST/WFC3 photometry implies F336W/F555W/F814W flux ratios of 0.88/0.95/0.84 for 
a2/a1, while K-band  VLT/SINFONI observations favour a significantly lower 
ratio of 0.76 (Crowther et al. 
2010). From our STIS/MAMA spectroscopy, the 1500\AA\ flux ratio inferred 
from the {\sc multispec} extraction is lower still (0.56), while Gaussian  
fits to  spatial profiles in our 2D ultraviolet NW1 dataset (lower left 
panel in Fig.~\ref{2D}) favour 0.62. We adopt this flux ratio, so $m_{\rm 
F555W}^{a1}  -  m_{\rm  F555W}^{a2}$ = 0.52 under the reasonable 
assumption that their interstellar extinctions 
are equivalent. From comparison between the UV fluxes of a1 and a2 and
relatively isolated sources in the NW1 slit, we support WFC3/F555W photometry of 
a1, resulting in a revised a2 apparent magnitude of $m_{\rm F555W}$ = 
12.80 mag in Table~\ref{targets}. This leads to a downward revision of 0.18 dex in
stellar luminosity, and corresponding 20\% lower mass estimate.

\begin{table*}
\begin{center}
\caption{STIS/MAMA-derived properties of R136 stars within a projected distance of 2.05 arcsec (0.5 pc), brighter than 
$m_{\rm F555W}$ = 16.0 mag), sorted by projected distance from R136a1. References to 
  optical (literature) spectral types are provided in Table~\ref{targets}.
  $F_{\rm 1500}$ fluxes are averages for the range 1500$\pm$25\AA~(estimated values are provided in parentheses). Aside from the WN5 stars (Crowther et al. 2010), 
temperatures and luminosities are from spectral type calibrations (Doran et al. 2013), while inferred {\it current}
masses $M$ and ages $\tau$ for O stars are from {\sc bonnsai} (Schneider et al. 2014b) using the Bonn evolutionary models (Brott
et al. 2011; K\"{o}hler et al. 2014). Velocities include $v_{\rm edge}$, $v_{\rm black}$ and $v_{\infty}$ (colons indicate less secure values).}
\label{targets2}
\begin{tabular}{l@{\hspace{1.5mm}}l@{\hspace{1.5mm}}l@{\hspace{1.5mm}}l@{\hspace{-1mm}}
r@{\hspace{2mm}}c@{\hspace{2mm}}c@{\hspace{2mm}}c@{\hspace{2mm}}c@{\hspace{2mm}}c@{\hspace{2mm}}
l@{\hspace{1mm}}l@{\hspace{1mm}}l@{\hspace{1mm}}r}
\hline
HSH95 & Optical& Ultraviolet & $r$ ($''$)& $10^{14} F_{\rm 1500}$ & $M_{\rm F555W}$ & $T_{\rm eff}$ & $\log L/L_{\odot}$ & $M_{\rm current}$ & $\tau$ & $v_{\rm edge}$ & $v_{\rm black}$ & $v_{\infty}$ & Fig\\
(WB85) &Sp Type &Sp Type && erg\,s$^{-1}$\,cm$^{-2}$\,\AA\ & mag & kK & & $M_{\odot}$ & Myr & km\,s$^{-1}$ & km\,s$^{-1}$ & km\,s$^{-1}$ & \\
\hline
3 (a1)  &  WN5h & WN5 & 0.00 & 22.9$\pm$1.4 & --8.09 & 53$\pm$3 & 6.94$\pm$0.09 & 315$^{+60}_{-50}$ & 0.0$^{+0.3}_{-0.0}$ & 3270: & 2600 & 2600 & \ref{r136-wn-uv} \\ 
5 (a2)  &  WN5h & WN5 & 0.08 & 13.1$\pm$1.2 & --7.52 & 53$\pm$3 & 6.63$\pm$0.09 & 195$^{+35}_{-30}$ & 0.3$^{+0.4}_{-0.3}$ & 3440 & 2425 & 2425 &  \ref{r136-wn-uv}\\ 
20 (a5) & O2\,If* & O2\,If/WN5 & 0.28 & 6.9$\pm$0.3& --6.65 & 50$^{+4}_{-5}$ & 6.32$^{+0.16}_{-0.15}$ & 101$^{+28}_{-26}$ & 0.8$^{+0.4}_{-0.7}$ & 3615 & 3045 & 3045 & \ref{r136-osuper-uv} 
\\ 
24 (a7) & O3\,III(f*) & O3--4\,V & 0.36& 4.4$\pm$0.4&--6.10 & 46$^{+4}_{-3}$ & 5.99$^{+0.10}_{-0.08}$ & 69$^{+12}_{-10}$ & 0.5$^{+0.1}_{-0.3}$ & 3385 & \ldots & 2710: & \ref{r136-o3-4uv} 
\\
27 (a8) &        & O2--3\,V  & 0.39 & 5.2$\pm$0.3& --6.05& 51$\pm$6 & 6.28$^{+0.13}_{-0.15}$ & 96$^{+27}_{-22}$ & 0.8$^{+0.5}_{-0.8}$ & 3550 &2980 & 2980 & \ref{r136-o2-3uv}\\ 
21 (a4)  &       & O2--3\,V & 0.43 & 7.8$\pm$0.3& --6.51 & 51$\pm$6 & 6.46$^{+0.13}_{-0.15}$ & 124$^{+37}_{-31}$ & 0.3$^{+0.2}_{-0.3}$ & 3275& 2475:& 2475: & \ref{r136-o2-3uv}\\ 
        186 &        & \ldots  & 0.43 &   (1.3)\phantom{$\pm$0.0}  & --4.62    &  \ldots   &  \ldots    & \ldots &\ldots & \ldots 
& \ldots & \ldots & --- \\
      86 &         &  O3--4\,V  & 0.45 & 2.1$\pm$0.2& --5.91 & 46$^{+4}_{-3}$ & 5.91$^{+0.10}_{-0.08}$ & 63$^{+10}_{-8}$ & $1.5^{+0.5}_{-0.9}$ & 3275:&2475 & 2475 &\ref{r136-o3-4uv} \\ 
6 (a3) &  WN5h & WN5 & 0.49 &16.3$\pm$2.1& --7.39 & 53$\pm$3 & 6.58$\pm$0.09 & 180$^{+30}_{-30}$ & 0.3$^{+0.4}_{-0.3}$ & 2950 & 2400 & 2400 & \ref{r136-wn-uv}\\ 
       66 & O3\,V      & O3--4\,V   & 0.49 & 2.2$\pm$0.2& --5.38 & 46$^{+4}_{-3}$ & 5.70$^{+0.10}_{-0.08}$ & 51$\pm 7$ & 1.4$^{+0.6}_{-1.1}$ & 3045 &2590& 2590  &\ref{r136-o3-4uv} \\ 
       119&          &  O4--5\,V  & 0.51 &  2.7$\pm$0.2 & --4.46 & 43$\pm$3 & 5.24$^{+0.08}_{-0.09}$ & 32$^{+4}_{-3}$ & 0.8$^{+1.4}_{-0.8}$ & 3075 &\ldots& 2460: & \ref{r136-o4-5uv}\\
      58 & O3\,III(f*) & O2--3\,V  & 0.59 & 2.7$\pm$0.3 & --5.65 & 51$\pm$6 & 5.93$^{+0.13}_{-0.15}$ & 63$^{+16}_{-12}$ & 0.6$^{+0.9}_{-0.6}$ & \ldots   & 2980& 2980& \ref{r136-o2-3uv}\\ 
       30 & O7\,V  & O6\,V   & 0.62& 5.7$\pm$0.3& --5.89 & 38$\pm$2 & 5.67$^{+0.05}_{-0.06}$ & 42$\pm$3 & 2.9$^{+0.3}_{-0.3}$ & 3110: &\ldots& 2490:  & \ref{r136-o5-6uv} \\ 
       70 & O5\,Vz & O5\,V & 0.62 & 2.6$\pm$0.2& --5.32 & 42$\pm$2 & 5.57$^{+0.07}_{-0.08}$ & 41$^{+5}_{-4}$ & 2.3$^{+0.5}_{-0.7}$ & 3340: &\ldots& 2670: &\ref{r136-o4-5uv}\\ 
       89 &          & O4\,V  & 0.66 & 1.3$\pm$0.1 & --6.25 & 44$\pm$2.5 & 5.99$^{+0.08}_{-0.07}$ & 65$^{+10}_{-9}$ & 1.7$\pm$0.4 &2425: &\ldots& 1940:& \ref{r136-o4-5uv}\\ 
       62 &          &  O2--3\,V  & 0.66 & 3.3$\pm$0.2 & --5.43 & 51$\pm$6 & 5.84$^{+0.13}_{-0.15}$ & 58$^{+13}_{-12}$ & 0.4$^{+1.2}_{-0.4}$ & 3340 &2770& 2770  & \ref{r136-o2-3uv}\\ 
19 (a6) &       & O2\,If & 0.73 & 11.6$\pm$0.4& --6.96 &  46$^{+4}_{-2}$ & 6.52$^{+0.10}_{-0.05}$ & 150$^{+36}_{-40}$ & 0.9$^{+0.5}_{-0.6}$ & 3440& 2650 & 2650 & \ref{r136-o2super-uv}\\ 
        50 &     & O2--3\,V  & 0.74 &3.4$\pm$0.2 &--5.89  & 51$\pm$6 & 6.02$^{+0.15}_{-0.13}$ & 70$^{+17}_{-14}$ & 0.8$^{+0.7}_{-0.8}$ &3390& 2620& 2620  &\ref{r136-o2-3uv}\\ 
        90 & O5\,V    &  O4\,V  & 0.77 & 1.7$\pm$0.1 & --4.68 & 44$\pm$2.5 & 5.36$^{+0.08}_{-0.07}$ & 36$\pm$4 & 1.0$^{+0.8}_{-1.0}$ & \ldots&2475:& 2475:   & \ref{r136-o4-5uv}\\ 
       149 &         &  O3--4\,V:: & 0.78 &  1.0$\pm$0.1& --5.01 & \ldots & \ldots & \ldots  &\ldots   &  2900& 2210& 2210 & \ref{r136-o3-4uv}\\ 
        141 &         & O5--6\,V & 0.78 &
1.0$\pm$0.1& --4.22 &  41$\pm$3 & 5.09$\pm$0.12 & 27$^{+4}_{-3}$ & 1.0$^{+1.5}_{-1.0}$  &  \ldots &\ldots & \ldots & \ref{r136-o5-6uv} \\ 
        80 & O8\,V  & O8\,V   & 0.87 & 2.3$\pm$0.1& --4.85 & 36$\pm$2 & 5.20$^{+0.08}_{-0.10}$ & 26$^{+3}_{-2}$ & 4.0$^{+0.8}_{-0.9}$ & 2070: &\ldots& 1655:& \ref{r136-o7-8uv}\\
        35 & O3\,V  & O3\,V  & 0.88& 5.5$\pm$0.2& --5.81 & 48$\pm$3 & 5.92$^{+0.08}_{-0.09}$ & 66$^{+10}_{-9}$ & 1.2$^{+0.5}_{-0.8}$ & 3440 &2770&  2770  &  \ref{r136-o2-3uv}\\ 
        78 & O4:\,V  & O4:\,V & 0.97 & 1.8$\pm$0.2 & --4.93 & 44$\pm$2.5 & 5.47$^{+0.08}_{-0.07}$ & 40$^{+5}_{-4}$ & 1.5$^{+0.7}_{-1.1}$ & 3000&2375:& 2375:  & 
\ref{r136-o4-5uv}\\ 
        73 & O9\,V      & O9+\,V   & 1.01 &2.0$\pm$0.1 & --5.14 & 33$\pm$2 & 5.21$^{+0.09}_{-0.10}$ & 24$\pm$2 & 4.8$\pm$0.7 & \ldots & \ldots & \ldots & 
\ref{r136-o9+uv}\\
       92 & O3\,V      & O6\,V   & 1.05 &  2.2$\pm$0.2 & --4.55 & 40$\pm$2 & 5.20$^{+0.07}_{-0.08}$ & 30$\pm$3 & 2.3$^{+0.9}_{-1.4}$  & \ldots &2080:& 2080:   & 
\ref{r136-o5-6uv}\\ 
        143 &         & O7--8\,V   & 1.05 & 0.8$\pm$0.1 & --4.25 & 39$\pm$3 & 4.99$^{+0.12}_{-0.14}$ & 24$\pm$3 & 1.6$^{+1.4}_{-1.6}$ & 1850:  & \ldots& 1480: & 
\ref{r136-o7-8uv}\\
       112 & O8.5\,III(f) & O7+\,V & 1.06 & 1.0$\pm$0.1 & --4.40 & 36$\pm$4 & 5.01$^{+0.10}_{-0.11}$ & 23$\pm$3
& 3.8$^{+1.6}_{-2.5}$ &\ldots &\ldots& \ldots & \ref{r136-o7-8uv}\\
        135 &         & O9+\,V   & 1.08 & 1.0$\pm$0.1 & --4.26 & 33$\pm$2 & 4.86$^{+0.09}_{-0.10}$ & 20$\pm$2
        & 5.3$^{+1.3}_{-1.7}$ & \ldots & \ldots & \ldots & 
\ref{r136-o9+uv} \\
        69 & O3--6\,V   & O5\,V  & 1.09 & 2.2$\pm$0.2 & --5.16 & 42$\pm$2 & 5.49$^{+0.07}_{-0.08}$ &38$\pm$4 & 2.2$^{+0.6}_{-0.9}$ & 3225 &\ldots & 2580  & 
\ref{r136-o5-6uv}\\ 
        52 & O3\,V     &  O3--4\,V & 1.13 & 4.0$\pm$0.3& --5.48 & 46$^{+4}_{-3}$ & 5.74$^{+0.10}_{-0.08}$ & 53$^{+8}_{-7}$ & 0.3$\pm$0.2 & 3260 &2820& 2820 & 
\ref{r136-o3-4uv}\\ 
        48 & O2--3\,III(f*) & O2--3\,III &  1.22 & 2.7$\pm$0.2 & --5.75
& 51$\pm$6 & 5.97$^{+0.13}_{-0.15}$ & 66$^{+16}_{-12}$ & 0.7$^{+0.8}_{-0.7}$ & 3615&3045& 3045  & 
\ref{r136-o2super-uv}\\ 
        77 & O5.5\,V+O5.5\,V & O6\,V & 1.29 &2.4$\pm$0.2&  --6.15 &\ldots  & \ldots & \ldots  & \ldots  & 1920:&1510:& 1510:  &  \ref{r136-o5-6uv}\\ 
        94 & O3\,V      &  O4--5\,V  & 1.31 & 1.5$\pm$0.1  & --4.61 & 43$\pm$3 & 5.31$^{+0.08}_{-0.09}$ & 34$\pm$4 & 1.3$^{+0.9}_{-1.2}$ & 3110: &\ldots& 2490:& 
\ref{r136-o4-5uv}\\
       115 &          & O9+\,V  &  1.41 & 1.4$\pm$0.1 & --4.17 & 33$\pm$2 & 4.82$^{+0.09}_{-0.10}$ & 19$^{+2}_{-1}$ & 5.3$^{+1.4}_{-1.9}$ & \ldots & \ldots& \ldots & 
\ref{r136-o9+uv}\\
        132 &         & O9+\,V   & 1.48 &1.1$\pm$0.1 & --4.03 & 33$\pm$2 & 4.76$^{+0.09}_{-0.10}$ & 18$^{+2}_{-1}$ & 6.2$^{+1.6}_{-0.5}$ & \ldots &\ldots& \ldots & 
\ref{r136-o9+uv} \\
        36 & O2\,If* & O2\,If* & 1.51 &3.6$\pm$0.2 & --5.97 & 46$^{+4}_{-2}$ & 5.94$^{+0.10}_{-0.05}$ & 66$^{+10}_{-9}$ & 1.5$^{+0.4}_{-0.9}$ & 4625:& 3500 & 3500 & \ref{r136-o2super-uv}\\ 
      173 &         & O9+\,V   & 1.53 & 0.7$\pm$0.1 & --4.07 & 33$\pm$2 & 4.78$^{+0.09}_{-0.10}$ & 19$\pm$1 & 6.6$^{+1.4}_{-0.8}$ & \ldots &\ldots &  \ldots &\ref{r136-o9+uv} \\
        75 &          & O4\,V   & 1.56 & 2.2$\pm$0.2 & --5.11& 44$\pm$2.5 & 5.54$^{+0.08}_{-0.07}$ &  42$\pm$5 & 1.7$^{+0.7}_{-1.1}$ & 3000 &2550& 2550  &\ref{r136-o4-5uv}\\ 
       114 &          & O7--8\,V   & 1.57 & 1.6$\pm$0.1 & --4.25 & 37$\pm$3 & 4.99$^{+0.12}_{-0.14}$ & 23$\pm$3 & 3.1$^{+1.4}_{-2.2}$ &2215: & \ldots & 1770: &\ref{r136-o7-8uv}\\
       108 &         &  O7--8\,V  & 1.59 &  1.5$\pm$0.1 & --4.65 & 37$\pm$3 & 5.15$^{+0.12}_{-0.14}$ & 25$^{+4}_{-3}$ & 3.6$^{+1.2}_{-1.9}$ & 1300: &\ldots & 1040: &\ref{r136-o7-8uv}\\
        31 &       & O2--3\,V & 1.64 &6.0$\pm$0.3 & --6.29 & 51$\pm$6 & 6.19$^{+0.13}_{-0.15}$ &87$^{+21}_{-20}$ & 0.8$^{+0.5}_{-0.8}$ & 3500& 2815:& 2815:  &\ref{r136-o2-3uv}\\ 
        49 & O3\,V     &  O4--5\,V  &  1.67 & 0.9$\pm$0.1& --5.34 & 43$\pm$3 & 5.60$^{+0.08}_{-0.09}$ &43$^{+6}_{-5}$ & 2.0$^{+0.7}_{-1.0}$ & 3780&2980:& 2980:&\ref{r136-o4-5uv} \\
       207 &          & \ldots & 1.72 & 0.3$\pm$0.1 & --4.56 &\ldots & \ldots & \ldots & \ldots & 4560: &\ldots & 3650: &---\\
       46 &     & O2\,III-If  & 1.73 & 2.3$\pm$0.2& 
--6.07  & 48$\pm4$ & 6.02$^{+0.12}_{-0.09}$ & 74$^{+14}_{-11}$ & 1.2$^{+0.4}_{-0.9}$ &4450:&3440& 3440  & \ref{r136-o2super-uv}\\ 
       47 & O3\,III(f*) & O2\,III-If & 1.75 & 2.6$\pm$0.3& --5.87 & 48$\pm$4 & 5.95$^{+0.14}_{-0.13}$ & 65$^{+14}_{-11}$ & 1.1$^{+0.5}_{-0.9}$  & 3615&3045& 3045 & \ref{r136-o2super-uv}\\ 
       40 & O3\,V & O2--3\,V & 1.80 & 4.5$\pm$0.2&  
--5.76 & 51$\pm$6 & 5.97$^{+0.13}_{-0.15}$ & 66$^{+16}_{-12}$ &0.7$^{+0.8}_{-0.7}$  & 3435 &2750& 2750  & \ref{r136-o2-3uv}\\ 
       118 &          & O7--8\,V   & 1.81 & 1.1$\pm$0.2 & --4.49 & 39$\pm$3 & 5.07$^{+0.12}_{-0.14}$ & 25$^{+4}_{-3}$ & 2.2$^{+1.2}_{-1.8}$  &1460: &\ldots& 1170: &\ref{r136-o7-8uv} \\
       116 &         &  O7--8\,V  & 1.81 & 1.3$\pm$0.1 & --4.21 & 37$\pm$3 & 4.97$^{+0.12}_{-0.14}$ & 23$\pm$3 &3.0$^{+1.4}_{-2.2}$ &1200:  &\ldots& 960: &\ref{r136-o7-8uv} \\
       42 & O3\,V+O3\,V & O3--4\,V & 1.83 & 3.5$\pm$0.2& --5.22 &46$^{+4}_{-3}$ & 5.64$^{+0.10}_{-0.08}$ & 48$^{+7}_{-6}$ & 0.3$^{+0.2}_{-0.3}$  &3045 &2245& 2245 & \ref{r136-o3-4uv}\\ 
       55 & O3\,V     & O2--3\,V  & 1.84 & 3.6$\pm$0.2 & --5.66 & 51$\pm$6 & 5.93$^{+0.13}_{-0.15}$ & 63$^{+15}_{-12}$ & 0.6$^{+0.9}_{-0.6}$  &3390 &2880& 2880 & \ref{r136-o2-3uv}\\ 
       39 & O3V+O5.5V & \ldots & 1.95&(3.7)\phantom{$\pm$0.0}& --5.75 &\ldots  & \ldots & \ldots & \ldots & \ldots &\ldots 
&\ldots & ---\\
       137 &          &\ldots & 1.97 &  (1.0)\phantom{$\pm$0.0} & --4.03 & \ldots & \ldots & \ldots &  \ldots &  \ldots &  \ldots 
&   \ldots & --- \\
       71 & O3--6\,V   & O4\,V   & 2.03 & 2.4$\pm$0.2& --5.00 & 44$\pm$2.5 & 5.49$^{+0.08}_{-0.07}$ & 40$^{+5}_{-4}$ & 1.6$^{+0.7}_{-1.1}$   &3030 &2475& 2475  &\ref{r136-o4-5uv}\\
       121 & O9.5\,V  & O9.5\,V   & 2.05 & 1.2$\pm$0.1 & --4.22  &  33$\pm$1.5 & 4.84$^\pm$0.07 & 19$^{+2}_{-1}$ & 5.3$^{+1.1}_{-1.3}$ & \ldots & \ldots & \ldots &\ref{r136-o9+uv}\\
       \hline
\hline
\end{tabular}
\end{center}
\end{table*}


\begin{table*}
\begin{center}
  \caption{STIS/MAMA-derived properties of R136 stars beyond a projected distance of 2.05 arcsec (0.5 pc) from R136a1, brighter
    than $m_{\rm F555W}$ = 16.0 mag). References to 
  optical (literature) spectral types are provided in Table~\ref{mama_targets}.
  $F_{\rm 1500}$ fluxes are averages for the range 1500$\pm$25\AA. Aside from the WN5 stars (Crowther et al. 2010), 
temperatures and luminosities are from spectral type calibrations (Doran et al. 2013), while inferred {\it current}
masses $M$ and ages $\tau$ for O stars are from {\sc bonnsai} (Schneider et al. 2014b) using the Bonn evolutionary models (Brott
et al. 2011; K\"{o}hler et al. 2014). Velocities include $v_{\rm edge}$, $v_{\rm black}$ and $v_{\infty}$ (colons indicate less secure values).}
\label{mama_targets2}
\begin{tabular}{l@{\hspace{1.5mm}}l@{\hspace{1.5mm}}l@{\hspace{1.5mm}}l@{\hspace{-1mm}}
r@{\hspace{2mm}}c@{\hspace{2mm}}c@{\hspace{2mm}}c@{\hspace{2mm}}c@{\hspace{2mm}}c@{\hspace{2mm}}
l@{\hspace{1mm}}l@{\hspace{1mm}}l@{\hspace{1mm}}r}
\hline
HSH95 & Optical& Ultraviolet & $r ('')$ & $10^{14} F_{\rm 1500}$ & $M_{\rm F555W}$ & $T_{\rm eff}$ & $\log L/L_{\odot}$ & $M$ & $\tau$ & $v_{\rm edge}$ & $v_{\rm black}$ & $v_{\infty}$ & Fig\\
 (WB85)&Sp Type &Sp Type & & erg\,s$^{-1}$\,cm$^{-2}$\,\AA\ & mag & kK & $L_{\odot}$ & $M_{\odot}$ & Myr & km\,s$^{-1}$ & km\,s$^{-1}$ & km\,s$^{-1}$ &\\
\hline
9 (b)  & O4\,If/WN8 & O4\,If/WN8 & 2.12 & 7.5$\pm$0.6&--7.31& 41$\pm$3 & 6.30$^{+0.11}_{-0.12}$ & 93$\pm$19 & 1.5$\pm$0.3  & 2080&1400&1400 &\ref{r136-osuper-uv} \\ 
       65 &          &  O4\,V  & 2.20 &1.9$\pm$0.2& --5.20 & 44$\pm$2.5 & 5.56$^{+0.08}_{-0.07}$ & 43$\pm$5 & 1.7$^{+0.6}_{-1.0}$  &3110 &2540&2540  &\ref{r136-o4-5uv}\\ 
        134& O7\,Vz & O7\,V & 2.28 &1.2$\pm$0.1 & --3.97 & 38$\pm$2 & 4.91$^{+0.05}_{-0.06}$ &  23$\pm$2 & 1.8$^{+1.3}_{-1.6}$  &1465    &\ldots& 1170:& \ref{r136-o7-8uv} \\
        64 & O7\,V((f)) &  O5\,V  & 2.44 &1.7$\pm$0.1 & --5.42  &  42$\pm$2 & 5.60$^{+0.07}_{-0.08}$  & 42$^{+5}_{-4}$ &  2.3$^{+0.5}_{-0.6}$  & 2215:& \ldots &1770:  &\ref{r136-o5-6uv}\\ 
        45 & O3\,V & O4--5\,V & 2.45 & 
3.0$\pm$0.2& --5.61 & 43$\pm$3 & 5.76$^{+0.08}_{-0.09}$ & 50$^{+7}_{-6}$ & 2.0$^{+0.6}_{-0.7}$  & 3275: & \ldots & 2620:&\ref{r136-o4-5uv}\\
       123 & O6\,V & O6\,V & 2.47 & 
1.1$\pm$0.1 & --4.16 & 40$\pm$2 & 5.04$^{+0.07}_{-0.08}$ &  27$\pm$2 & 1.2$^{+1.1}_{-1.2}$ & 2020  &\ldots & 1615: &\ref{r136-o5-6uv}\\
       68  &           & O6\,V & 4.61 &  1.6$\pm$0.1& --5.22 & \ldots & \ldots  &\ldots    & \ldots  & 2390   & \ldots  & 1910:  & \ref{r136-o5-6uv} \\
       102 &          & O2--3\,III & 4.83 &0.6$\pm$0.1 & --5.24 & \ldots  & \ldots  &\ldots   &   \ldots & 2865: &  2540 & 2540 & \ref{r136-o2super-uv} \\
\hline
\hline
\end{tabular}
\end{center}
\end{table*}

In addition, WFC3 photometry of a6 from de Marchi et al. (2011) is significantly brighter than
Hunter et al. (1995). From our MAMA  SE1 dataset, we obtain a 1500\AA\ flux ratio of 0.70 for a6/a3,
in perfect agreement with $m_{\rm F555W}^{a3} - m_{\rm F555W}^{a6}$ = --0.38 from WFC3, so adopt
photometry from de Marchi et al. Finally, de 
Marchi et al. (2011) provide a single photometric measurement for a4+a8, which agrees satisfactorily with 
the combination of these sources in Hunter et al. (1995). Again, we utilise UV datasets (lower right panel in 
Fig.~\ref{2D}) to verify their relative fluxes. From our  {\sc multispec} extraction we obtain a  
1500\AA\ flux ratio of 0.66 for  a8/a4, in good agreement with 0.65 from 
Gaussian fits to their spatial profiles in the NW3 slit. We obtain 
$m_{\rm F555W}^{a4} - m_{\rm F555W}^{a8}$ = 0.46, again assuming  
identical interstellar extinctions for these stars. From the combined WFC3 photometry of
$m_{\rm F555W}^{a4+a8}$ = 13.41$\pm$0.01, we obtain revised a4 and a8 apparent 
magnitudes  of $m_{\rm F555W}$ = 13.91 and 14.42 mag (Table~\ref{targets}).

\section{Spatially resolved far-ultraviolet spectroscopy}\label{mama}

We have classified individual R136 stars, based upon a comparison 
between  their ultraviolet morphology and LMC template stars with reliable 
optical classifications. As noted above, we have incorporated 10 reference
O2--O9.5 stars based on optical classifications provided in 
Caballero-Nieves et al. (2016, in prep) which are indicated in 
Table~\ref{targets}--\ref{mama_targets}. Our approach therefore complements the 
UV clasification system of LMC OB stars by Smith Neubig \& 
Bruhweiler (1999) using IUE SWP/LORES spectroscopy. We adopt existing
(optical) subtypes for the three WN5 stars in  R136a (Crowther \& Dessart 1998).

\begin{figure}
\begin{center}
 \includegraphics[bb=20 90 475 
680,height=1.0\columnwidth,angle=90]{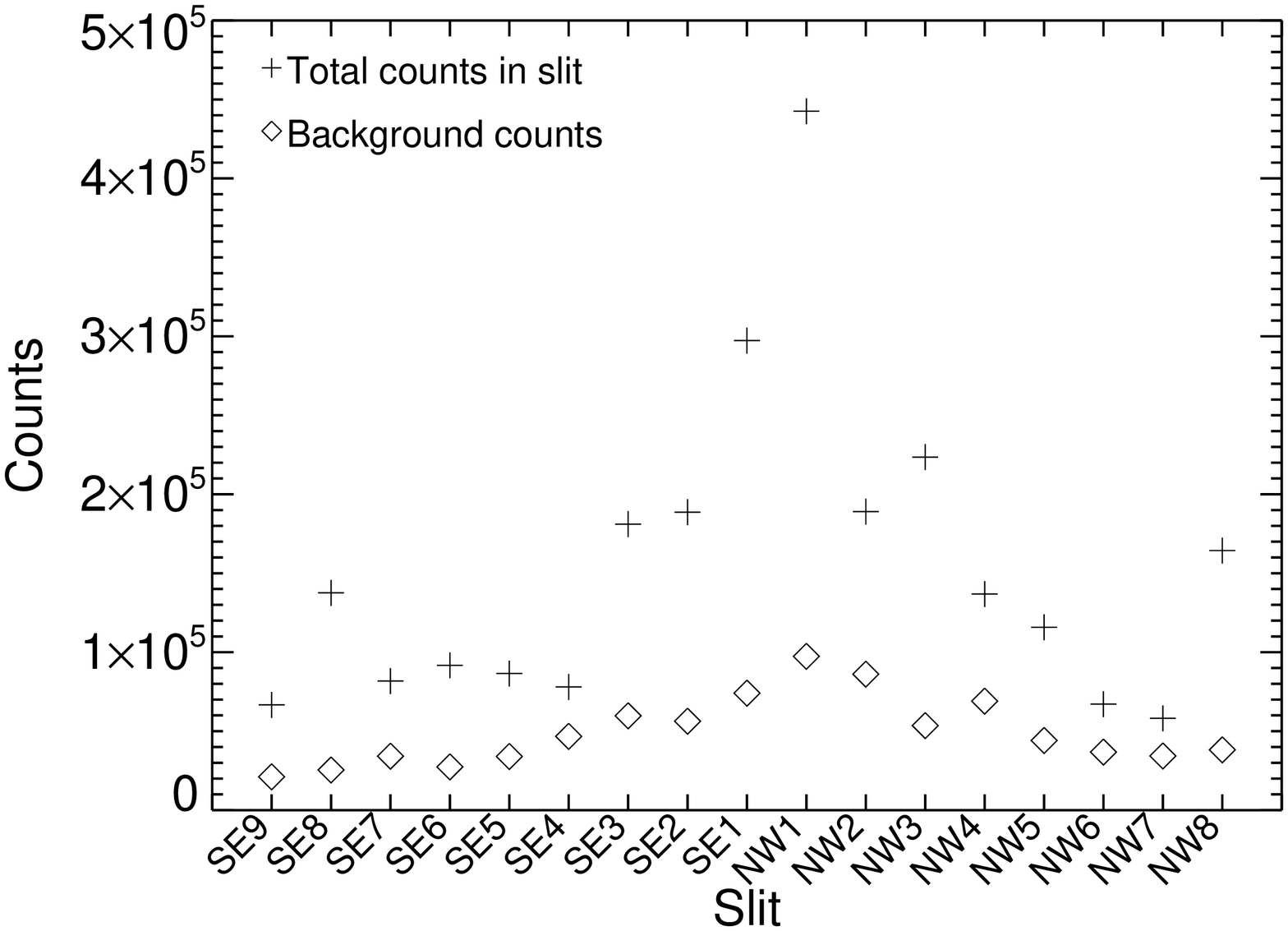} 
 \caption{Comparison of the background (diamonds) to total (pluses) far-ultraviolet 
counts in the central 4$''$ ($\sim$1 pc) of each slit. This background represents a combination of unresolved
cluster members (20\%) plus extended emission from bright stars in adjacent slits (15\%).}
\label{diffuse}
\end{center}
\end{figure}

\begin{figure}
\begin{center}
 \includegraphics[bb=15 70 485 775,height=1.0\columnwidth,angle=-90]{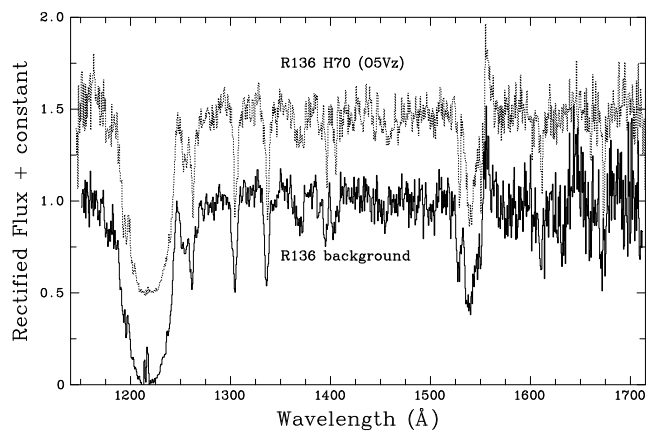} 
 \caption{Intra-cluster far-UV spectra of selected inner regions from SE3 and SE4 slits, in close
proximity (0.7--1$''$, solid) to R136a1 together with R136 H70 (O5\,Vz, dotted) for reference.
Extended He\,{\sc ii} $\lambda$1640 emission from R136a1 is apparent in the extracted background spectra.}
\label{background}
\end{center}
\end{figure}

In addition to the point sources, a significant fraction of the
far-UV spectrum of the central parsec of R136 arises from the intra-cluster background, on the basis of
the {\sc multispec} fits to the 100 brightest cluster members. 
This is illustrated in Fig.~\ref{diffuse}, and represents
a combination of unresolved cluster members plus extended emission from bright cluster stars originating in adjacent
slits. In Fig.~\ref{background}, we present the
far-UV spectra from the sum of selected background regions of SE3 (1$''$ E of R136a1)
and SE4 (0.7$''$ SE of R136a1), reminiscent of R136 H70 (O5\,Vz), except for extended
He\,{\sc ii} emission originating from either R136a1 or a2. 
$\sim$20\% of the integrated far-UV continuum arises from the diffuse
background, with bright stars in adjacent slits contributing the remaining 15\% of the total.


\subsection{UV morphology of O stars in the LMC}

Walborn et al. (1985) have previously presented an atlas of O 
stars based on IUE/HIRES spectroscopy. The overwhelming majority of these
stars are located in the Milky Way. More recently, Walborn et al. (1995, 
2002a, 2004) have presented HST and FUSE ultraviolet spectroscopy of Magellanic Cloud O stars, 
although subtypes and luminosity classes are poorly sampled. Since then, a 
number of HST programmes have obtained ultraviolet spectroscopy for a 
range of LMC O stars (e.g. Massey et al. 2004, 2005, 2009). In 
Appendix A we present a  montage of ultraviolet spectroscopic sequences 
for dwarfs, giants and  supergiants 
(Figs.~\ref{oearlydwarf-uv}-\ref{osuper-uv}), with luminosity sequences 
shown in Figs.~\ref{o2uv}--\ref{o9+uv}. A log of 
observations of these datasets is presented in Table~\ref{lmc_uv_atlas}.

\begin{figure}
\begin{center}
 \includegraphics[bb=10 77 544 761,width=1.0\columnwidth]{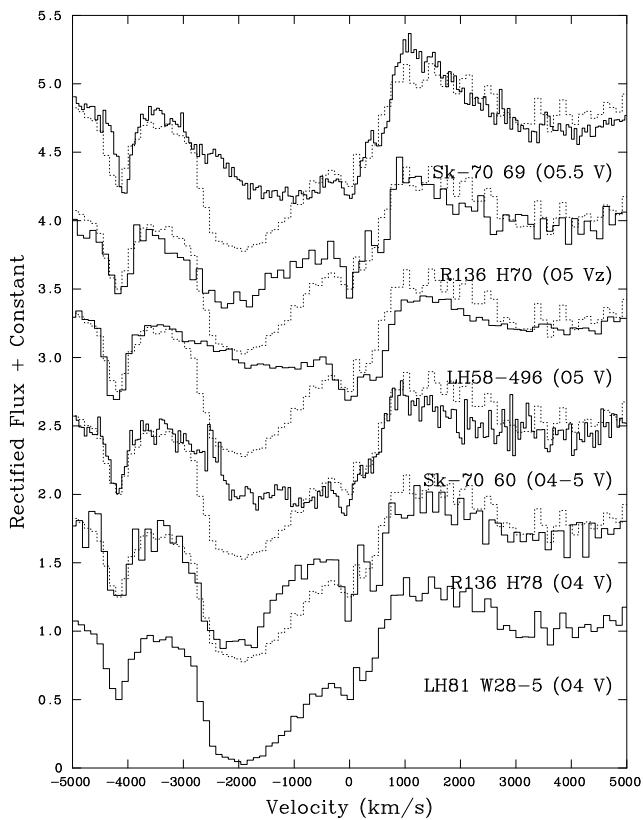} 
 \caption{Variation in 
   spectral morphology of C\,{\sc iv} 1548--51 amongst O4--5.5 dwarf stars in the LMC. The O4 dwarf
   LH81/W28-5 (Massey et al. 2004) is shown for reference in each case (dotted lines).}
\label{o4-5-morph}
\end{center}
\end{figure}

There is a clear variation in UV morphology amongst O stars from early to late subtypes
and from dwarf to supergiant, although we wish to highlight wind line profile variation at 
an individual subtype/luminosity class. 
By way of example, we present C\,{\sc iv} $\lambda\lambda$1548--51
P Cygni line profiles amongst six O4--5.5 dwarfs in the LMC  in Fig.~\ref{o4-5-morph}. Both
LH81/W28-5 and R136 H78 possess saturated P Cygni (black) absorption troughs, whereas
other stars possess shallow extended 
absorption troughs which may be flat (e.g. Sk--70$^{\circ}$ 60, LH 58--496, Sk--70$^{\circ}$ 69) or 
deeper at high velocities (R136 H70). Within this sample, P Cygni emission components are 
relatively weak, with the exception of Sk--70$^{\circ}$ 69 for which C\,{\sc iv} emission is very prominent.

This variation indicates (at least) a third parameter affecting the wind profiles beyond ionization and luminosity class. 
While further investigation is required, it may well be age, i.e. evolutionary distance from the ZAMS, since some of 
these stars are in very young regions. While LH\,81/W28--5 does have a strong He\,{\sc ii} $\lambda$4686 absorption line 
(Massey et al. 2004, their Fig 17), it is substantially weaker than He\,{\sc ii} $\lambda$4541, indicating a higher 
luminosity and/or stronger wind than for typical class V. It is well known that mass-loss rates of O stars at a given 
temperature and luminosity span a wide range of values (Hillier et al. 2003). Binaries and/or binary evolution may also 
be relevant contributors to this diversity.

\subsection{UV morphology of stars in R136}

In Appendix B we present a series of montages of ultraviolet spectroscopy of
stars in R136 (Figs.~\ref{r136-wn-uv}-\ref{r136-o9+uv}), together with UV templates where
appropriate.

In Figure~\ref{r136-wn-uv} we present HST/STIS spectroscopy of the three 
luminous WN stars R136a1, a2 and a3 (Crowther et al. 2010). These stars 
have previously been  observed in the ultraviolet with HST/GHRS (de Koter 
et al. 1997), although a1 and a2 were blended in the GHRS SSA 
0.22$''$ entrance aperture. Our long-slit STIS spectroscopy was obtained at a 
position angle of  64$^{\circ}$ and extracted with {\sc multispec} to 
disentangle a1 from a2.  Nevertheless, the ultraviolet morphologies of all 
WN stars are extremely  similar to one another, as is the case optically 
(HST/FOS, de Koter et al.  1997) and in the near-IR (VLT/SINFONI, Schnurr 
et al. 2008). We include an IUE/LORES spectrum of R146 (Brey 88, WN5) for 
comparison.

In Figure~\ref{r136-osuper-uv} we present HST/STIS spectroscopy of two
Of/WN stars, Melnick~37 and Sk--67$^{\circ}$ 22, together with a5 and b. 
R136a5 has 
previously been observed in the ultraviolet by HST/GHRS and 
classified as O3f/WN (de Koter et al. 1994). From optical HST/FOS 
spectroscopy R136a5 has been assigned O3\,If/WN (Massey \& Hunter 1998) or 
O2\,If* (Crowther \&  Walborn 2011). The closest LMC analogue 
to R136a5 in the ultraviolet is Sk--67$^{\circ}$ 22 (O2\,If/WN5). R136b 
has also 
previously been  observed with HST/GHRS (GO 6018/Heap) while Massey \& 
Hunter (1998)  assigned O4\,If$^{+}$ from HST/FOS spectroscopy (see also 
Crowther \&  Walborn 2011). From STIS/CCD spectroscopy the ratio of 
N\,{\sc iv} $\lambda$4058/N\,{\sc iii} $\lambda\lambda$4634-41 is 
characteristic of O4 and WN8 subtypes (rather than O3.5 and WN7) so
O4\,If/WN8 is newly inferred here. He\,{\sc 
ii} $\lambda$1640 and  Si\,{\sc iv} $\lambda\lambda$1394-1402 P Cygni 
profiles are much stronger in  R136b than for conventional LMC O4 supergiants 
(e.g. Sk$-$65$^{\circ}$ 47, Fig.~\ref{o4uv}).

HST/STIS spectroscopy of O2--3 (super)giants in R136 is presented in 
Fig.~\ref{r136-o2super-uv}, including the LMC O2--3\,III and O2\,If reference stars R136 H48
and H36, from our new STIS/CCD dataset. Massey \& Hunter (1998) previously assigned O3\,III for H48 while
Walborn et al. (2002b) assigned O2--3\,If for H36. The O2 supergiant 
classification of R136a6 is new here, with R136 H46 and H47 assigned O2\,III-If, 
owing to their morphological similarity to the candidate runaway star VFTS 16 (Evans et al. 2010).
Another example of a O2--3\,III star is
observed beyond the central parsec, namely R136 H102 located 4.8 arcsec to the NE.

Figure~\ref{r136-o2-3uv} presents STIS/MAMA spectroscopy 
of 8 O2--3\,V stars in R136
together with the templates BI 237 (O2\,V) and R136 H35 (O3\,V) based on our new STIS/CCD datasets. 
All stars exhibit prominent O\,{\sc v} $\lambda$1371 
absorption, P Cygni C\,{\sc iv} $\lambda\lambda$1548--51, N\,{\sc v} 
$\lambda\lambda$1238--42 and negligible/weak 
absorption at He\,{\sc ii} $\lambda$1640. Of these, six stars have previously been 
spectroscopically observed with HST/FOS by Massey \& Hunter (1998), who assigned 
O3\,V or O3\,III classifications. R136a4, a8, H31, H50 and H62 are newly classified as O2--3\,V 
in this study. 


HST/STIS spectroscopy of 7 UV-classified O3--4 dwarfs in R136 is presented 
in  Fig.~\ref{r136-o3-4uv} together with the reference stars R136 H35 (O3\,V) and H78 (O4\,V)
from our STIS/CCD spectroscopy. All O3--4 dwarfs possess weak O\,{\sc v} $\lambda$1371 absorption,
plus prominent, albeit unsaturated C\,{\sc iv} $\lambda\lambda$1548--51 P 
Cygni profiles (Fig.~\ref{o3uv}). Of these, four have previously been spectroscopically 
observed by Massey  \& Hunter (1998), who assigned O3\,V, except for R136a7 for which O3\,III was proposed 
from HST/FOS spectroscopy.  R136 H86 and H149 are 
newly  classified here as O3--4\,V from STIS/MAMA datasets, although the latter, centrally 
located in the SE2 slit,  suffers from significant He\,{\sc ii} $\lambda$1640 contamination by R136a3 (SE1).

Figure~\ref{r136-o4-5uv} presents STIS/MAMA spectroscopy for 
dwarf O4--5 stars in R136 together with new templates R136 H78 (O4\,V) and H70 (O5\,V)
from new STIS/CCD observations. We favour O4\,V for R136 H65, H71, H75, H89 and H90 from their morphological
similarity to H78, while O4--5\,V is preferred for H45, H49, H94 and H119. 
Classifications are in good agreement with  Massey \& Hunter (1998) who proposed O3\,V for H45, H49 and H94,
O5\,V for H90 and O3--6\,V for H71. Figure~\ref{r136-o5-6uv} extends this sequence to O5\,V and O6\,V subtypes, 
exploiting a STIS/CCD subtype of H123 (O6\,V). We propose O5\,V subtypes for H64 and H69, O6\,V subtypes for 
H30, H77 and H92, and O5--6\,V for H141, again in reasonable agreement with Massey \& Hunter (1998) who adopted
O7\,V for H30 and H64, O3--6\,V for H69, O3\,V for H92. Massey et al. (2002) obtained O5.5\,V+O5.5\,V for H77.
The UV  morphology of R136 H68, located 4.6$''$ from R136a1, is also consistent with O6\,V.

Figure~\ref{r136-o7-8uv} presents STIS spectroscopy for O7--8 dwarfs in R136, exploiting new optical subtypes of O7\,V
and O8\,V for R136 H134 and H80, respectively. All stars are  newly classified except for 
H112, for which Massey \& Hunter (1998) proposed O8.5\,III. A very weak wind signature of C\,{\sc iv} 
$\lambda\lambda$1548--51 is observed, via either very shallow P Cygni absorption or a clear 
blueshifted absorption. We assign O9+\,V subtypes for the final group of stars, presented in 
Figure~\ref{r136-o9+uv}, for which solely photospheric 
C\,{\sc iv} $\lambda\lambda$1548--51  absorption is observed, in common with 
H121 for which O9.5\,V is obtained from our STIS/CCD spectroscopy. 
Low S/N becomes increasingly problematic for these 
stars, whose far-UV fluxes are typically $F_{\rm 1500} \sim 10^{-14}$  
erg\,s$^{-1}$\,cm$^{-2}$\,\AA$^{-1}$ or lower.

Of the remaining stars within 2.05 arcsec of R136a1 that are listed in 
Table~\ref{targets2} two lie beyond our HST/STIS slit spectroscopy, namely 
R136 H39 and H137, of which H39 has been 
observed previously by  Massey et al. (2002) and classified O3\,V+O5.5\,V. 
Finally, H186 lies close to a much brighter source (H86), within its slit, so we are unable to isolate its UV spectrum.

\begin{table}
\begin{center}
\caption{Terminal wind velocities (km\,s$^{-1}$) and standard deviations ($\sigma$) for O stars in R136 (this study) and other O stars in the LMC (Walborn et al. (1995); Massa et al. (2003); 
Massey et al. (2004, 2005, 2009), Evans et al. (2010); Bestenlehner et al. (2014). N is the sample size for each category.}
\label{wind_velocities}
\begin{tabular}{
l@{\hspace{2.5mm}}
r@{\hspace{1.5mm}}r@{\hspace{1.5mm}}r@{\hspace{3.5mm}}
r@{\hspace{1.5mm}}r@{\hspace{1.5mm}}r@{\hspace{3.5mm}}
r@{\hspace{1.5mm}}r@{\hspace{1.5mm}}r@{\hspace{3.5mm}}
}
\hline
Subtype & \multicolumn{3}{c}{dwarf} & \multicolumn{3}{c}{giants} & \multicolumn{3}{c}{supergiants}\\
        & $v_{\infty}$ & $\sigma$ & N & $v_{\infty}$ & $\sigma$ & N & $v_{\infty}$ & $\sigma$ & N \\
\hline
\multicolumn{10}{c}{--- R136 (HST/STIS) --- } \\
O2--3      &  2780 & 160 & 9 & 3020 & 370 & 4 & 3065 & 425 & 3 \\
O3--4      &  2525 & 235 &  8 &   \ldots  &  \ldots  & \ldots & 1400 & \ldots & 1 \\
O4--5      &  2475 & 245 & 11 & \ldots      &  \ldots   & \ldots &  \ldots  & \ldots &  \ldots\\
O5--6      &  2095 & 415 &   8 & \ldots     & \ldots    & \ldots & \ldots & \ldots  & \ldots \\
O7--8      &  1320 & 315 &   7 & \ldots     & \ldots   & \ldots  & \ldots & \ldots & \ldots \\
\multicolumn{10}{c}{ --- LMC field (HST FOS/STIS/COS; FUSE ) --- } \\
O2           & 3290 & 155 & 2  & 3310 & 180 & 7 & 2600 & 345 & 4 \\ 
%
O3--3.5    & 2950  & 375 & 4 & 2880 & 450 & 2 & 2000     & \ldots   & 1 \\
%
O4--6       & 2510 & 205 & 5 & 2360       & 245    & 3 & 1900 & 130 & 5 \\ 
O6.5--8   & 1950 & \ldots & 1 & 2675     & 1025 &  4  & 1950 & 70 & 2 \\
O8.5--9.7 & \ldots & \ldots & \ldots & \ldots & \ldots & \ldots & 1375 & 375 & 6 \\
\hline
\hline
\end{tabular}
\end{center}
\end{table}

\subsection{Wind velocities for R136 stars}

The presence of strong, saturated P Cygni profiles amongst the O stars of R136a allows
us to measure their wind velocities. To date, wind velocities for large numbers of 
Galactic OB stars have been obtained from IUE SWP/HIRES spectroscopy (Prinja et al. 
1990; Howarth et al. 1997), whereas wind velocities for O stars in the Magellanic Clouds have been 
relatively scarce (e.g. Prinja \& Crowther 1998; Massey et al. 2004, 2005). 

Here we measure the maximum blueward extend of saturated P Cygni absorption lines (usually C\,{\sc iv} 
$\lambda\lambda$1548--51), $v_{\rm black}$, and use it as a probe of terminal wind velocity, i.e. $v_{\infty} = v_{\rm 
black}$ (Prinja et al. 1990). The velocity at which the violet absorption meets the continuum, the edge velocity $v_{\rm 
edge}$, is usually larger than $v_{\rm black}$ as a result of turbulent motions in the outflow. Such stochastic motions 
cause some of the material to move at speeds beyond the local flow speed, hence will cause absorption beyond the terminal 
speed if the line is fully saturated. If P Cygni profiles are not fully saturated $v_{\infty}$ can be approximated as a 
function of $v_{\rm edge}$ (see below). In such cases the terminal velocity could be underestimated because the ion may 
no longer be present in the outer wind.

We measure $v_{\rm black}$ and/or $v_{\rm edge}$ with respect to (LMC) 
interstellar absorption lines C\,{\sc iv} $\lambda\lambda$1548--51 and/or 
Si\,{\sc ii} $\lambda$1527 (e.g. Howarth \& Phillips 
1986). The systemic radial velocity of R136 is $\sim$268 km\,s$^{-1}$ (H\'{e}nault-Brunet et al. 2012). 
Individual measurements should be reliable to $\pm$100 km\,s$^{-1}$, so $v_{\infty}$ should be accurate
to $\pm$200 km\,s$^{-1}$.

\begin{figure}
\begin{center}
 \includegraphics[bb=60 60 520 780,width=1.\columnwidth]{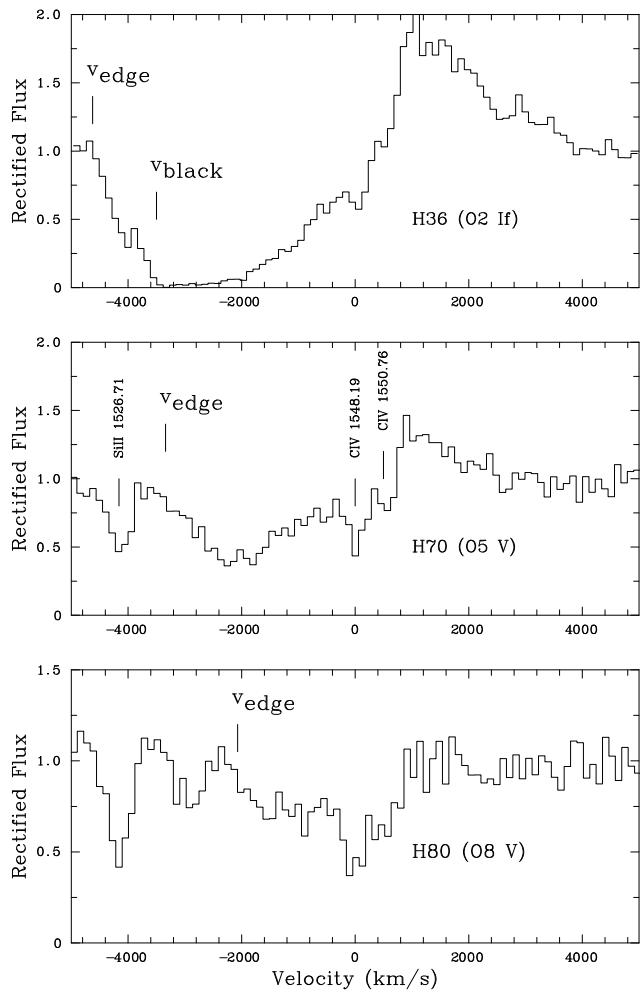} 
 \caption{Representative HST/STIS C\,{\sc iv} $\lambda\lambda$1548--51 
profiles for stars with saturated P Cygni profiles
(top), unsaturated P Cygni profiles (centre) and non-P Cygni profiles (bottom). Profiles are shown in
velocity space in the LMC rest frame of interstellar C\,{\sc iv} 1548.2\AA\ 
in all cases.}
\label{wind_velocity}
\end{center}
\end{figure}

For the present sample, C\,{\sc iv} $\lambda\lambda$1548--51 exhibits one of three spectral morphologies. A significant 
subset, typically O2--4 stars and supergiants, exhibit saturated P Cygni profiles, from which $v_{\rm black}$ (and 
$v_{\rm edge}$) can be measured, as shown for R136 H36 in the top panel of Figure~\ref{wind_velocity}. A further subset 
possess strong P Cygni profiles, albeit without saturated absorption features, from which $v_{\rm edge}$ can be measured, 
typical of O5--6 dwarfs, as shown for R136 H70 in the central panel of Fig.~\ref{wind_velocity}. O7--8 dwarfs tend not to 
exhibit prominent P Cygni profiles, but some show blue absorption that are sufficiently distinct from interstellar 
C\,{\sc iv}, as shown in the lower panel of Fig.~\ref{wind_velocity} for H80.

\begin{figure}
\begin{center}
 \includegraphics[bb=65 200 515 630,width=1.0\columnwidth]{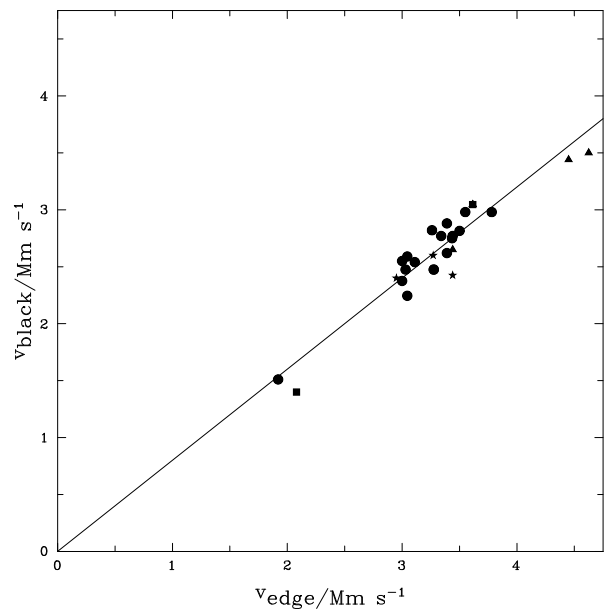} 
 \caption{Comparison between edge and black velocities (in Mm\,s$^{-1}$) for R136 
WN (stars), Of/WN (squares),
(super)giants (triangles) and dwarfs (circles), together with the best fitting
$v_{\rm black} = (0.8 \pm 0.05) v_{\rm edge}$ relationship for those stars without saturated black absorption troughs (solid line).
}
\label{vblack_vedge}
\end{center}
\end{figure}

Individual results are presented in Table~\ref{targets2}. The highest terminal wind velocities are $v_{\infty} \sim$ 3500 
km\,s$^{-1}$ for the O2 (super)giants H36 and H46, while late O dwarfs often possess $v_{\rm edge} \leq 1500$ 
km\,s$^{-1}$, or no wind measurements are possible.  Figure~\ref{vblack_vedge} shows edge and black velocities for 28 
R136 stars, from which we obtain $v_{\rm black}/v_{\rm edge} = 0.8 \pm 0.05$, which is adopted to estimate wind 
velocities for stars in which solely edge velocities are measured. For reference, Prinja \& Crowther (1998) obtained 
$v_{\rm black}/v_{\rm edge} \sim$ 0.7 for LMC stars and $v_{\rm black}/v_{\rm edge} \sim 0.9$ for 6 R136 stars (in common 
with the present sample).

In Table~\ref{wind_velocities} average values of wind velocities in R136 are provided for O2--3, O3--4, O4--5, O5--6, 
O7--8 dwarfs, plus early O giants and supergiants, including cases for which wind velocities are based upon $v_{\infty} = 
(0.8\pm 0.05) v_{\rm edge}$. Amongst dwarfs, the trend towards lower wind velocities for later subtypes is apparent. 
The standard deviation at O5--6 is unusually large since individual estimates span a wide range from 1510 km\,s$^{-1}$ 
(H77, O6\,V from UV morphology) to 2580 km\,s$^{-1}$ (H69, O5\,V from UV morphology). Statistics of wind velocities for 
'field' O stars elsewhere in the LMC are included, drawn from Walborn et al. (1995), de Koter et al. (1998), Prinja \& 
Crowther (1998), Massa et al. (2003), Massey et al. (2004, 2005, 2009), Evans et al. (2010) and Bestenlehner et al. 
(2014), with individual measurements from the literature and comparisons with present results provided in 
Table~\ref{winds} in Appendix C.

The wind velocity of 1400 km\,s$^{-1}$ for R136b (O4\,If/WN8) is extremely
low, even with respect to O2--3.5\,If/WN stars in the LMC (Table~\ref{winds}). In contrast, O2--3 supergiants within R136 possess higher wind
velocities than dwarfs, which is the reverse of O2--3 stars elsewhere in the LMC, albeit based on low number statistics.
For completeness, we obtain an average wind terminal velocity of 2475 $\pm$ 110 km\,s$^{-1}$ for the three WN5  stars 
within R136a. Crowther et al. (2010) obtained $v_{\infty} = 2200 - 2600$ km\,s$^{-1}$ for these stars from 
their UV to near-IR spectroscopic analysis. 

Our new dataset effectively double the sample of LMC early O stars with measured wind velocities. 
It is important to note that a range of wind velocities have been obtained for individual stars. In part this 
arises from the methodology, either direct measurement as applied here or line fitting based on the
Sobolev with Exact Integration method (SEI, Groenewegen \& Lamers 1989). Still, different authors 
applying the same approach to the same dataset have obtained significantly different wind velocities. By way of 
example, de Koter et al. (1998) obtained $v_{\infty}$ = 3000 km\,s$^{-1}$
from HST GHRS spectroscopy of R136a5, while Massey et al. (2004) obtained 
3400 km\,s$^{-1}$ from the same dataset; we find 3045 $\pm$ 200 km\,s$^{-1}$ from STIS spectroscopy. 
Mindful of concerns about systematic differences between alternative approaches, an SEI analysis of H36
was carried out (R.K.~Prinja, priv. comm.) revealing $v_{\infty}$ = 3500 km\,s$^{-1}$, $v_{\rm turb}/v_{\infty}$ = 
0.15 for a $\beta$=1 velocity law, in agreement with the $v_{\rm black}$ result. This does not imply that
all such measurements will be entirely consistent, although it does suggest that there is no inherent systematic 
offset.

Despite the low number statistics, wind velocities of O2--4 dwarfs in R136 are somewhat lower than elsewhere in the LMC. 
By way of example, the average  wind velocity of 8 R136 O3--4 dwarfs is 2525 km\,s$^{-1}$, 
versus 2900 km\,s$^{-1}$ for 5 counterparts elsewhere in the LMC (and 3040 km\,s$^{-1}$ for Galactic O3--4 dwarfs).
The reason for this is not clear, if not simply due to small sample sizes.
If physical, it is likely not the result of a lower metallicity in 30 Doradus
($v_{\infty} \propto Z^{0.13}$: Leitherer et al. 1992) relative to the rest of the LMC, although youth
may be a factor, recalling the discussion linked to Figure~\ref{o4-5-morph}. 

\begin{figure}
\begin{center}
 \includegraphics[bb=65 200 515 630,width=1.0\columnwidth]{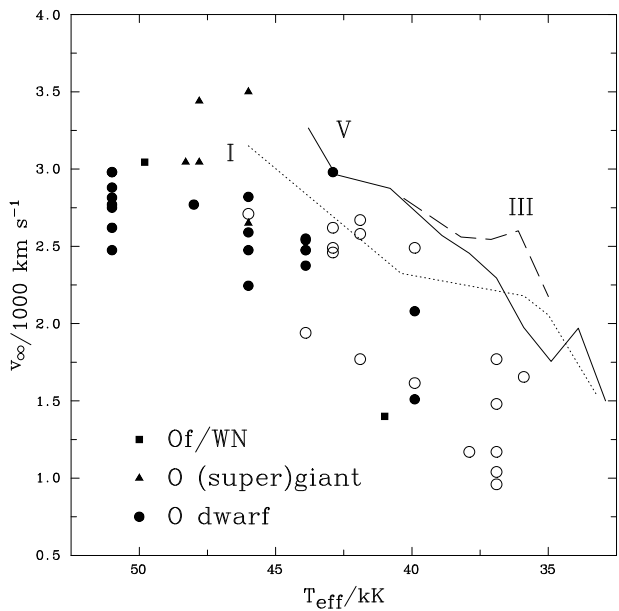} 
 \caption{Wind velocities (in km\,s$^{-1}$) versus estimated stellar temperature (Of/WN stars: squares, (super)giants: 
triangles, dwarfs: circles) based on $v_{\rm black}$  (filled) or $v_{\rm edge}$ (open) 
assuming the temperature scale of 
Doran et al. (2013). Average velocities versus stellar temperature for Galactic O stars from Prinja et al. (1990)
are indicated as dotted lines (supergiants), dashed lined (giants) and solid lines (dwarfs), incorporating 
the updated classification scheme of Walborn et al. (2002b) and temperature scale of 
Martins et al. (2005) plus $T_{\rm eff}$ (O2\,I) = 46\,kK.}
\label{r136_vely}
\end{center}
\end{figure}

Wind velocities for individual R136 stars are presented as a function of stellar temperature in 
Fig.~\ref{r136_vely}, based on UV-derived subtypes and the temperature calibration of LMC O stars of Doran 
et al. (2013), in which results from $v_{\rm black}$ (filled symbols) and $v_{\rm edge}$ (open symbols) are 
distinguished. We have included average UV-derived wind velocities for Galactic O stars from Prinja et al. 
(1990) -- based on the same methodology as used here -- updated for O2--4 stars
following Walborn et al. (2002b), and adopting the Martins et al. (2005) Galactic observational
temperature calibration adapted for LMC O stars by Doran et al. (2013). 

To date, comparisons of empirical (reduced) wind momenta of O stars with theoretical predictions either adopt
wind velocities or are based on calibrations.
Pre-empting our stellar mass estimates from Section~\ref{parameters}, we are able
to compare terminal wind velocities with escape velocities.
Lamers et al. (1995) obtained $v_{\infty}/v_{\rm esc}$
= 2.65 $\pm$ 0.2 for 16 Galactic O stars earlier than O7. From 35 O2--6 stars in R136, we find
$v_{\infty}/v_{\rm esc}$ = 2.5 $\pm$ 0.4, in
agreement with the Galactic sample. Therefore, although
R136 early O stars possess $\sim$15\% lower wind velocities than Galactic counterparts, the ratio
$v_{\infty}/v_{\rm esc}$ is unchanged. Still,
considerable variation {\it is} observed for other O stars in metal-poor galaxies (Garcia et al. 2014).

\subsection{Physical parameters for R136 stars}\label{parameters}

De Koter et al. (1998) provided UV-derived parameters of the brightest 
dozen or so members in R136, while 
Massey \& Hunter (1998) provided a more complete Hertzsprung-Russell diagram 
from HST/FOS spectroscopy. Since then, 
Doran et al. (2013)  have provided updated physical parameters for stars within
R136 using contemporary calibrations for O stars in the LMC, based on the Galactic scale from Martins et al. (2005).

We defer more robust R136 O-type classifications until future papers in 
this series, but already our UV morphological classification allows us to 
present a preliminary, near complete Hertzsprung-Russell diagram for 
luminous stars  within a radius of 2.5 arcsec (0.6 parsec) of R136a1. Temperatures and bolometric 
corrections of O  stars are adopted from Tables 5--7 of Doran et al. 
(2013), while  properties of Of/WN stars are adopted from their Appendix 
A. HST/WFC3 photometry (de Marchi  et al. 2011), except in particularly crowded regions (recall a1/a2, a4/a8),
and interstellar extinctions provide  absolute magnitudes as set out in Sect.~\ref{census}.

\begin{table*}
\begin{center}
\caption{Comparison of mass and age estimates for the WN5h stars in R136a from Geneva (Crowther et al. 2010; Yusof et al. 
2013) and Bonn (K\"{o}hler et al. 2015) based on physical and chemical abundances obtained by Crowther et al. (2010), 
incorporating the adjustment to the absolute magnitude for R136a2, as discussed in the present study.}
\label{wn5}
\begin{tabular}{
l@{\hspace{2.5mm}}
c@{\hspace{1.5mm}}c@{\hspace{1.5mm}}c@{\hspace{3.5mm}}
r@{\hspace{1.5mm}}l@{\hspace{1.5mm}}l@{\hspace{3.5mm}}
l@{\hspace{1.5mm}}l}
\hline
Star        & $T_{\ast}$ & $\log L$   & $X_{\rm H}$ & Code     & $M_{\rm init}$ & $v_{\rm init}$ &  $M_{\rm current}$ & $\tau$ \\
            & kK        & $L{\odot}$ &  \%        &          & $M_{\odot}$    & km\,s$^{-1}$  &  $M_{\odot}$    & Myr   \\
\hline
R136a1      & 53$\pm$3   & 6.94$\pm$0.09 & \ldots& Bonn& 325$^{+55}_{-45}$  & 100$^{+180}_{-60}$ & 315$^{+60}_{-50}$ & 0.0$^{+0.3}_{-0.0}$\\
            & 53$\pm$3   & 6.94$\pm$0.09 & 40$\pm$10 & Bonn   & 315$^{+50}_{-20}$ & 440$^{+20}_{-85}$ & 280$^{+35}_{-30}$ & 0.8$\pm$0.2 \\
            & 53$\pm$3   & 6.94$\pm$0.09 & 40$\pm$5 & Geneva & 320$^{+100}_{-40}$ & 400   & 265$^{+80}_{-35}$  & 1.4$^{+0.2}_{-0.1}$ \\
R136a2      & 53$\pm$3   & 6.63$\pm$0.09 & \ldots & Bonn & 195$^{+35}_{-30}$  & 100$^{+325}_{-55}$ & 190$^{+35}_{-35}$  & 0.3$^{+0.4}_{-0.3}$ \\
            & 53$\pm$3   & 6.63$\pm$0.09 & 35$\pm$5 & Bonn   & 160$^{+25}_{-20}$ & 380$^{+85}_{-20}$ & 130$^{+20}_{-20}$ & 1.6$\pm$0.2 \\
            & 53$\pm$3   & 6.63$\pm$0.09 & 35$\pm$5 & Geneva & 180$^{+35}_{-30}$  & 400   & 150$^{+30}_{-25}$  & 1.7$\pm$0.1 \\
R136a3      & 53$\pm$3   & 6.58$\pm$0.09 & \ldots & Bonn  & 180$\pm$30        & 100$^{+330}_{-55}$ & 175$^{+35}_{-35}$  & 0.3$^{+0.4}_{-0.3}$ \\
            & 53$\pm$3   & 6.58$\pm$0.09 & 40$\pm$5 & Bonn   & 155$^{+25}_{-20}$ &  370$^{+80}_{-30}$ & 130$^{+25}_{-15}$  & 1.5$\pm$0.2 \\
            & 53$\pm$3   & 6.58$\pm$0.09 & 40$\pm$5 & Geneva & 165$\pm$30        & 400   & 135$^{+25}_{-20}$  & 1.7$\pm$0.1 \\
\hline
\hline
\end{tabular}
\end{center}
\end{table*}

In Figure~\ref{r136_hrd} we present 
the first spectroscopically-derived Hertzsprung-Russell diagram 
for stars brighter  than $m_{\rm F555W}$ = 16.0 mag within 0.6 pc from 
R136a1. A dotted line corresponding to $m_{\rm F555W}$ = 16.0 mag for an adopted $A_{\rm F555W}$ = 1.72 mag
is indicated, highlighting the region where our sample is complete.
A total of 25 stars (26 including R136c), mostly early O dwarfs, are more massive than  $\sim$50\,$M_{\odot}$.
The remaining stars are mid to late O dwarfs with  initial masses above $\sim$20\,$M_{\odot}$.

\subsection{Ages and masses of individual stars in R136}

LMC isochrones for rotating stars with $v_{\rm eq}$ = 100 km\,s$^{-1}$ from Brott et al. (2011) and K\"{o}hler et al. 
(2015) are included, together with tracks for non-rotating 25, 40, 60, 100 and 200 $M_{\odot}$. These tracks were 
selected as respresentative since they closely match the average rotation rate of O stars in 30 Doradus 
(Ram\'irez-Agudelo et al. 2013), though do not differ significantly from non-rotating models. It is apparent that, R136b 
aside, all stars with initial masses above 100 $M_{\odot}$ lie between the 0--1 Myr isochrones, whereas stars with masses 
between 40--100 $M_{\odot}$ span 0--2.5 Myr, and those below 40 $M_{\odot}$ typically span 0--4 Myr, although 
uncertainties in physical parameters do not exclude ages below $\sim$2 Myr, with the exception of R136 H73 (O9+\,V) for 
which an age of $\sim$5 Myr is favoured.

\begin{figure}
\begin{center}
 \includegraphics[bb=65 200 515 630,width=1.0\columnwidth]{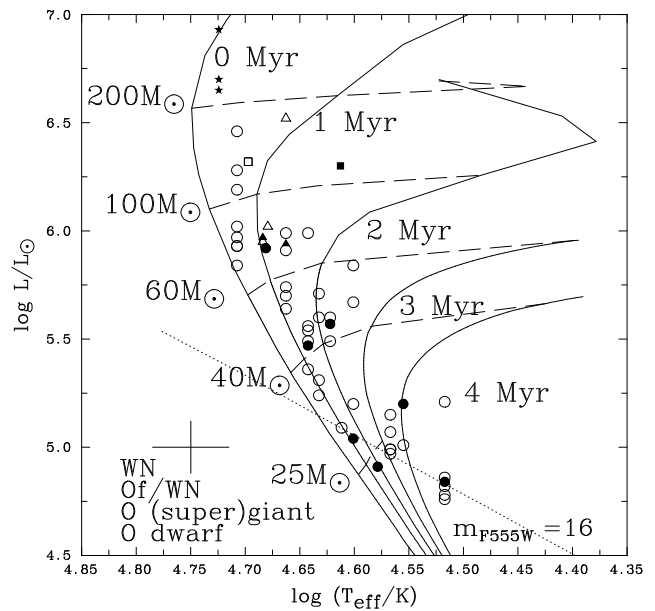} 
 \caption{Hertzsprung-Russell diagram (WN: stars; Of/WN stars: squares;
(super)giants: triangles; dwarfs: circles) brighter  than  $m_{\rm 
     F555W}$ = 16.0 mag within 0.6 pc  from R136a1, based on either optically-derived (filled) or
   UV-derived (open) spectral types using the temperature calibrations from Doran et al. (2013).   LMC 
isochrones for slowly rotating stars ($v_{\rm eq}$ = 100 km\,s$^{-1}$) at ages of 0, 1, 2, 3 and 4 Myr are overlaid (solid),
together with individual tracks for 25, 40, 60, 100 and 200 $M_{\odot}$ 
stars (dashed), from Brott et al. (2011) and K\"{o}hler et al. (2015).
Representative uncertainties are indicated in the lower left. A dotted line marking $m_{\rm F555W}$ = 16.0 mag
for an adopted $A_{\rm F555W}$ = 1.72 mag is indicated, emphasising that our census is sensitive to all luminous stars in R136a.
}
\label{r136_hrd}
\end{center}
\end{figure}

For quantitative mass and age determinations we have exploited {\sc bonnsai}\footnote{{\sc bonnsai} is available at 
www.astro.uni-bonn.de/stars/bonnsai} (Schneider et al. 2014b) which implements a Bayesian method of quantifying ages and 
masses from comparison between UV-estimated temperatures and luminosities and evolutionary models, involving two key 
priors: a Salpeter mass function and an adopted rotation distribution of O stars from VFTS (Ram\'irez-Agudelo et al. 
2013). Individual age and mass estimates are presented in Table~\ref{targets2}, and include results for the WN5h stars 
solely based upon their positions in the HR diagram, i.e. for consistency initially neglecting their He-abundances.

As suggested by Fig.~\ref{r136_hrd}, application of the Bonn evolutionary models (via {\sc bonnsai}) to the three WN5h 
stars implies very young ages of $\leq$0.3 Myr in all cases. Low initial rotation rates -- favoured by observations of O 
stars in 30 Dor (Ram\'irez-Agudelo et al. 2013) -- will lead to a redward evolution on the main sequence in evolutionary 
models, so young ages are required to explain the blue location of the WN5h stars. The next most luminous stars are 
extreme O supergiants, R136a5 (O2\,If/WN5), a6 (O2\,If) and b (O4\,If/WN8), each with current masses of 
100--150\,$M_{\odot}$, plus O2--3\,V stars R136a4 and a8. Once again, young ages of 0.6$\pm$0.3 Myr are inferred for 
these stars, with the exception of R136b whose lower temperature suggests an age of $\sim$1.5 Myr. Schnurr et al. (2009) 
have previously argued that R136b, offset by a projected distance of 0.5 pc is more evolved than other core members of 
R136 (though see Appendix of Hainich et al. 2014).

Consequently, the HR diagram positions of the WN5h stars and the most luminous O stars in R136a favours very young ages. 
However, it should be emphasised that envelope inflation is implemented in the Bonn models, which is most pronounced for 
the most massive models ($\geq 50 M_{\odot}$), since these lie closest to the Eddington limit (Sanyal et al. 2015). 
Inflation moves isochrones to lower temperatures, such that younger ages are inferred for stars with models accounting 
for inflation, compared with those that do not.

Turning to the lower mass stars, all O2--6 stars lie between the 0--3 Myr isochrones in Fig.~\ref{r136_hrd}, although all 
stars do not lie along a single isochrone. This is unsurprising given the approximate physical parameters of individual 
stars, initial rotation rate distributions, and uncertain binarity. For an extreme mass ratio system, the primary would 
completely dominate the spectrum of the system, so its location in the HR diagram would not be significantly displaced. 
In contrast, for a mass ratio of unity, individual stars would lie 0.3 dex lower in the HR diagram than the composite 
system, so the age would be overestimated. The effect would be modest for the highest mass systems at very young ages, 
but would become more significant for lower mass systems. Indeed, later O stars possess higher age estimates (3--6 Myr), 
although the inferred properties for such stars are least secure of the entire sample.

Our STIS/CCD datasets will help identify close binaries in R136 (Caballero-Nieves et al. 2016, in prep), while new HST 
Fine Guidance Sensor (FGS) observations -- currently being analysed (GO 13477, PI, S.E. de Mink) -- seek to reveal 
binaries with separations down to 0.02$''$ (1000\,AU), following on from Cycle 3 FGS observations of R136 (Lattanzi et 
al. 1994).

Individual age estimates follow solely from the position of stars in the HR diagram. We are in the process of 
analysing STIS/CCD datasets of R136a members, from which rotational velocities and helium abundances can be 
obtained. For the moment, He abundances for the WN5h stars are available (Crowther et al. 2010). From their 
luminosities and He-rich atmospheres, we can infer minimum ages since the surface helium mass fraction sets a lower 
limit to the core helium mass fraction. The minimum ages of R136a1--3 are 1--1.3 Myr, in reasonable agreement with 
results for the O stars. In order to simultaneously reproduce the high surface temperatures of the WN5hs stars, 
Bonn evolutionary models favour rapid initial rotation ($\sim$400 km\,s$^{-1}$), and ages of 1.1--1.6 Myr, as shown 
in Table~\ref{wn5}. These results compare favourably with Crowther et al. (2010) who applied Geneva evolutionary 
models for very massive stars (Yusof et al. 2013), also requiring rapid rotation ($v_{\rm rot}/v_{\rm crit}$ = 0.4 
or $v_{\rm rot} \sim$ 400 km\,s$^{-1}$) and ages of $\sim$1.5 Myr (see Table~\ref{wn5}). Details of the Bonn and 
Geneva codes differ significantly regarding the consideration of mixing, core overshooting and the treatment of 
envelope inflation. Both employ the metallicity-dependent Vink et al. (2001) prescription for mass-loss on 
the main-sequence, so estimates of $M_{\rm init}$ from current luminosities ought to be consistent, even if there
are uncertainties in their validity for very massive stars (see Crowther et al. 2010; Gr\"{a}fener \& Vink 2013).

\begin{figure}
\begin{center}
 \includegraphics[bb=0 0 353 248,width=1.0\columnwidth]{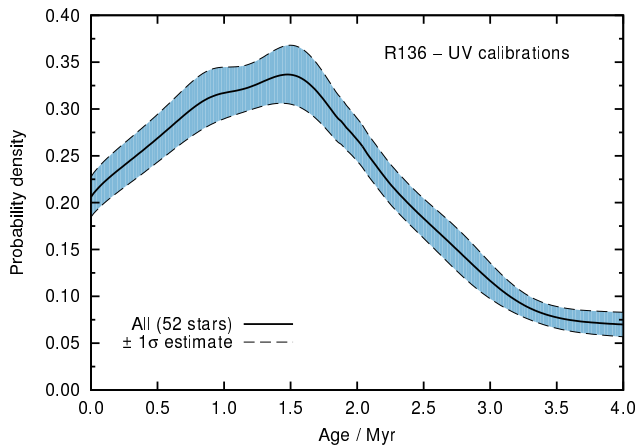} 
 \caption{Probability density distribution of the apparent ages of UV-derived spectral types of individual
O stars (excluding R136a1, a2, a3) based on {\sc bonnsai} (Schneider et al. 2014b) and main-sequence
evolutionary models (Brott et al. 2011; K\"{o}hler et al. 2015), favouring an age of 1.5$^{+0.3}_{-0.7}$ Myr. The shaded
area is a bootstrapped $\pm1\sigma$ estimate of the age distribution.}
\label{bonnsai}
\end{center}
\end{figure}


\subsection{R136 cluster age and mass}

Early estimates of the age of R136 were 3--4 Myr on the basis that the WN stars were classical He-burning stars. 
Subsequently, a lower age of $<$1--2 Myr was obtained once it was recognised that the WN stars are core H-burning stars, 
supported by a rich early O-type population (de Koter et al. 1998, Crowther \& Dessart 1998). The most extensive study of 
massive stars to date has been by Massey \& Hunter (1998) who inferred a very young age of $\leq$1 Myr using the Vacca et 
al. (1996) O star temperature scale, or $\leq$2 Myr from the Chlebowski \& Garmany (1991) scale, although the majority of 
their sample was drawn from the R136 region, rather than R136a itself.

Here, we have combined the age estimates of 52 O stars in R136a (excluding WN5h stars) from {\sc bonnsai} to produce a 
probability density distribution, presented in Fig.~\ref{bonnsai}, from which an age of 1.5$^{+0.3}_{-0.7}$ Myr is 
obtained. The $\pm1\sigma$ estimate represents the standard deviation in each bin of the age distribution computed with a 
bootstrapping method. As discussed above, lower ages are obtained for the highest luminosity stars, although the 
treatment of envelope inflation for the most massive stars ($\gg 50 M_{\odot}$) remains uncertain.

Discrepant age estimates for massive stars in other young massive star clusters (e.g. Arches and Quintuplet in Galactic 
Centre) have been obtained from their Wolf-Rayet and O star contents, in the sense that the luminous WR stars possess 
apparent ages which are younger than O stars by 1--2 Myr (Martins et al. 2008; Liermann et al. 2012). Schneider et al. 
(2014a) have interpreted this as rejuvenation of the most massive stars through stellar mergers. Banerjee et al. (2012) 
have suggested the very luminous WN5h stars in R136 are post-merger systems. If this were the case, one might anticipate 
younger (apparent) age estimates for these rejuvenated stars with respect to the massive O star population.

R136 differs from the Quintuplet and Arches clusters in the sense that the luminous O stars also appear young, yet the 
He-enrichment of WN5h stars favours ages of $\sim$1.5 Myr and unusually rapid rotation, the latter predicted by stellar 
mergers (de Mink et al. 2013). According to Schneider et al. (2014a), the probability that the most massive star in a 
5$\times 10^{4} M_{\odot}$ star cluster is a binary product rises from 50\% at 1 Myr to 100\% at 3 Myr. However, only the 
most massive system is anticipated to have merged after 1.5 Myr (4 are anticipated after 3 Myr) for a Salpeter-type mass 
function in the core. Therefore, the youth of R136a does not support a scenario in which all the very massive stars are 
merger products. Indeed, as we discuss below, the mass within the central parsec is significantly lower than 5$\times 
10^{4} M_{\odot}$ further strengthening this conclusion. A more definitive answer to this question is deferred until the 
analysis of STIS/CCD datasets, but the apparent youth of the most luminous stars likely arises from the treatment of 
inflation in single star models than the physics of binary mergers.

A variety of mass estimates of R136 have been obtained. Hunter et al. (1995) infer a mass of $2.2\times 10^{4} M_{\odot}$ 
for stars with $>2.8 M_{\odot}$ out to a radius of 5 pc from HST/WFPC2, suggesting a total mass of $\sim 5 \times 10^{4} 
M_{\odot}$, while Andersen et al. (2009) derived a mass of $5 \times 10^{4} M_{\odot}$ down to 2.1 $M_{\odot}$ out to a 
radius of 7 pc from HST/NICMOS. The actual mass of the R136 cluster may be somewhat lower, as it may suffer contamination 
from the surrounding halo of NGC~2070, for which Cignoni et al. (2015) infer a global mass of 9$\times 10^{4} M_{\odot}$. 
Indeed, the two components contribute equally to the surface brightness at a radius of $\sim$5 pc from R136a1 according 
to Mackey \& Gilmore (2003).

\begin{figure*}
\begin{center}
 \includegraphics[bb=15 70 485 775,height=1.5\columnwidth,angle=-90]{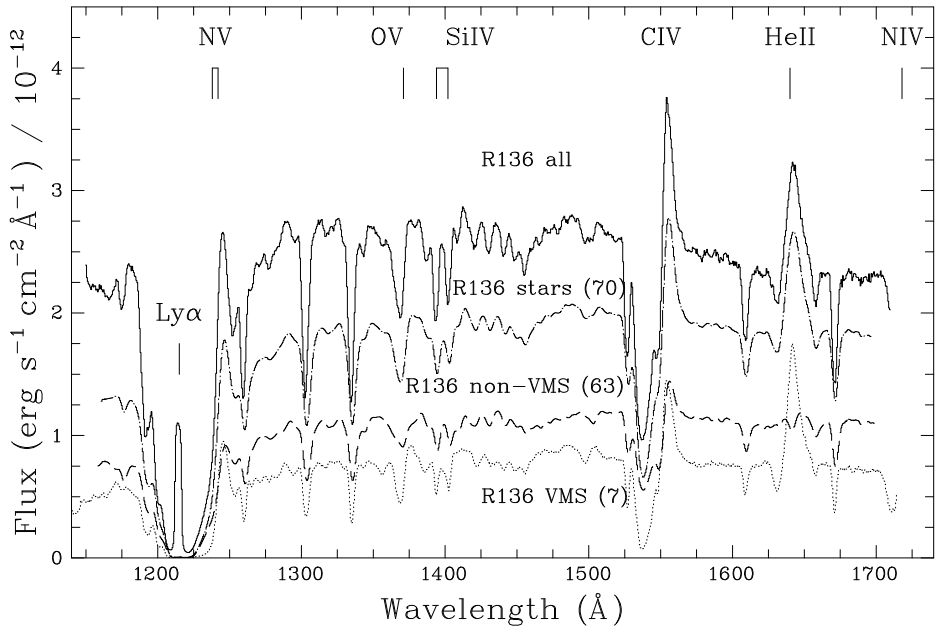} 
 \caption{Integrated HST/STIS spectrum of all sources within 0.5 parsec of R136a1 (solid, all stars), the
composite spectrum of all 70 bright stars with    $F_{\rm 1500} \geq 5 \times 10^{-15}$ erg\,s$^{-1}$\,cm$^{-2}$\,\AA$^{-1}$ (dot-dash), comprising 
7 very massive stars (VMS, dotted),
   and the remaining 63 far-UV bright stars (dashed). He\,{\sc ii} 
$\lambda$1640 emission in R136 is totally dominated by VMS. The difference between the `stars' and `all' arises
from the contribution from UV faint O late-type stars and unresolved early B stars.}
\label{r136_cluster}
\end{center}
\end{figure*}

\begin{table}
\begin{center}
\caption{Census of the most luminous ($\log (L/L_{\odot}) \geq$ 6.2) objects in 30 Doradus, 
sorted by projected distance from R136a1, drawn from the present study, Crowther et al. (2010), 
Doran et al. (2013) or Bestenlehner et al. (2014). Initial mass estimates from {\sc bonnsai} (based 
upon estimated temperatures and luminosities) are 
limited to apparently main-sequence stars.}
\label{vms}
\begin{tabular}{l@{\hspace{-1mm}}r@{\hspace{1.5mm}}r@{\hspace{1.5mm}}
c@{\hspace{1.5mm}}r@{\hspace{1.5mm}}r}
\hline
Star, Alias & r (pc) & Subtype & $\log  L/L_{\odot}$ & $M/M_{\odot}$ & Ref \\
\hline
R136a1, H3  & 0.00   & WN5h    & 6.94$\pm$0.09 & 325$^{+55}_{-45}$ & a \\ 
R136a2, H5  & 0.02   & WN5h    & 6.63$\pm$0.09 & 195$^{+35}_{-30}$ & a, e \\ 
R136a5, H20 & 0.07   & O2\,If/WN5 & 6.32$^{+0.15}_{-0.16}$ & 106$^{+30}_{-24}$   & e \\ 
R136a8, H27 & 0.10   & O2--3\,V & 6.28$^{+0.13}_{-0.15}$  &  102$^{+26}_{-22}$   & e \\
R136a4, H21 & 0.11   & O2--3\,V & 6.46$^{+0.13}_{-0.15}$  &   132$^{+37}_{-29}$  & e \\
R136a3, H6  & 0.12   & WN5h    & 6.58$\pm$0.09 & 180$\pm$30 & a, e \\ 
R136a6, H19 & 0.18   & O2\,If  & 6.52$^{+0.10}_{-0.05}$ &   158$^{+35}_{-28}$   & e\\
R136b, H9 & 0.52    & O4\,If/WN8 & 6.30$^{+0.11}_{-0.12}$ & 100$^{+22}_{-17}$ & e \\
R136c, H10 & 0.83   & WN5h    & 6.75$\pm$0.11 & 230$^{+50}_{-45}$ & a \\ 
\\
Mk 42, H2 &  1.96   & O2\,If   & 6.56$\pm$0.1  &  170$^{+40}_{-30}$    & d \\ 
Mk 34, H8 &  2.61   & WN5h    & 6.85$\pm$0.1 &   275$^{+55}_{-45}$   &c  \\ 
R134, BAT99-100& 2.86   & WN6(h)  & 6.2\phantom{$\pm$0.1}
  &   \ldots   & d  \\ 
Mk 37Wa, H11&2.86   & O4\,If & 6.34$\pm$0.1  &    110$\pm$20   &  d \\ 
Mk 37Wb, H44&2.94   & O2\,If/WN5 & 6.21$\pm$0.1 &   100$\pm$15  & d \\ 
Mk 39, VFTS 482&  3.01   & O2.5\,If/WN6   & 6.4$\pm$0.1    &  125$^{+25}_{-20}$  & d \\ 
Mk 37, H14& 3.02   & O3.5\,If/WN7 & 6.48$\pm$0.1  &    140$^{+40}_{-25}$   & d   \\ 
Mk 35, VFTS 545& 3.07   & O2\,If/WN5 & 6.3$\pm$0.1 &   110$\pm$20  & d \\ 
Mk 30, VFTS 542 & 4.15   & O2\,If/WN5 & 6.16$\pm$0.1  &  90$\pm$15  & d   \\ 
\\
Mk 25, VFTS 506  & 10.9  & ON2\,V((f*))  & 6.24$\pm$0.1  & 100$^{+20}_{-15}$  & d  \\ 
R140b, BAT99-103 & 12.0  &WN5(h)+O & 6.4\phantom{$\pm$0.1}   &   \ldots   & d  \\ 
Mk 51, VFTS 457 &12.6   & O3.5\,If/WN7 &6.2$\pm$0.1 &  90$\pm$15  &d  \\
\multirow{2}{*}{R139, VFTS 527}  &   \multirow{2}{*}{16.2}      & 
O6.5\,Iafc & \multirow{2}{*}{6.3\phantom{$\pm$0.1}} & $>$78 & b\\ 
         &           &  O6\,Iaf &   &  $>$66 & b \\
R145, VFTS 695 & 19.6 & WN6h+    & 6.5\phantom{$\pm$0.1}   &   \ldots  & d 
\\ 
R135, VFTS 402& 21.6   & WN7h+OB  & 6.2\phantom{$\pm$0.1}  &  \ldots   & 
c\\ 
VFTS 682 & 29.8& WN5h  &  6.51$\pm$0.1  &    155$^{+30}_{-25}$   & d  \\ 
\multirow{2}{*}{R144, BAT99-118} & \multirow{2}{*}{62.3}   &  
WN5--6  &  \multirow{2}{*}{6.4\phantom{$\pm$0.1}} &  
\ldots    & c\\ 
    &      &    WN6--7 &     & \ldots & c \\
R147, VFTS 758 & 71.7  &  WN5h    &  6.36$\pm$0.1 & 125$^{+25}_{-20}$     & d  \\ 
R146, VFTS 617 & 84.7 &  WN5ha   & 6.29$\pm$0.1  &   115$\pm$20   & d \\ 
VFTS 16 &      120: & O2\,III-If & 6.23$\pm$0.1 &   105$^{+20}_{-15}$   & d  \\ 
\hline
\hline
\multicolumn{6}{l}{
  \begin{minipage}{\columnwidth}~\\
a: Crowther et al. (2010); b: Taylor et al. (2011); c: Doran et al. (2013); d: Bestenlehner et al. (2014); e: This work; 
  \end{minipage}
}\\
\end{tabular}
\end{center}
\end{table}

For a total cluster mass of $5 \times 10^{4} M_{\odot}$, an Initial Mass Function following Kroupa (2001) with an upper 
mass limit of 300 $M_{\odot}$, we would anticipate $ 10^{4} / 1250 = 40$ stars with initial masses in excess of 50 
$M_{\odot}$. Since only 25 stars with such high masses are observed within the central 0.5 pc radius, the actual R136a 
cluster mass may be somewhat lower. Of course, some massive stars may have been ejected from the cluster (e.g. VFTS 16, 
Evans et al. 2010). Fujii \& Portegies Zwart (2011) suggest a few percent of stars with $\sim 50 M_{\odot}$ would be 
dynamically ejected from dense clusters of this mass. If we were to consider a radius of 5 pc, analogous to the region 
surveyed by Hunter et al. (1995), there are another $\sim$30 stars with masses in excess of 50 $M_{\odot}$ (from the 
census of Doran et al. 2013), supporting their high enclosed mass for the R136 `region'.

Sabbi et al. (2012) have proposed that R136 forms part of an ongoing merging binary cluster system within NGC~2070, on 
the basis of a clump of older stars located $\sim$5 pc to its NE (see also Cignoni et al. 2015). If this clump were to 
reflect stars in the process of dynamically accreting, one might expect older stars to have already merged with R136 
itself (see Gieles 2013). Indeed, some of the apparently older O stars in R136 (e.g. R136 H73) may arise from this clump, 
although the determination of physical parameters for these late-type O stars awaits detailed spectroscopic analysis of 
our optical STIS/CCD datasets. The evolutionary status of some stars in the NE clump (e.g. Mk~33Sb, WC5) implies an 
apparent age of 3--4 Myr, providing these are coeval (Melnick 1985; Walborn \& Blades 1997). The age of this region may 
be higher, should the highest luminosity stars be rejuvenated binary products.


\subsection{Census of very luminous stars in 30 Doradus}\label{30dor_census}

Doran et al.  (2013) have undertaken a study of the massive-star content of 30 Doradus, together with the contribution of 
R136 to the ionizing and mechanical feedback of the entire star forming region, albeit reliant upon an incomplete census 
of R136 itself. We are now able to reassess the total number of very luminous stars in 30 Doradus, updated to include 
tailored parameters of massive stars from Bestenlehner et al. (2014) based on VFTS observations (Evans et al. 2011). 
Table~\ref{vms} lists 29 stars within 30 Doradus which are more luminous than $\log (L/L_{\odot}) \approx 6.2$, 
corresponding to initial masses of $\sim 100 M_{\odot}$ or more for zero age main-sequence stars. Only 9 (31\%) of the 
most luminous stars in 30 Doradus lie within R136 itself!

A similar number (10) of very luminous stars are located in the halo around R136, at projected distances of 1.5--5 
pc\footnote{We have excluded VFTS 512 (O2\,V) from the Doran et al. (2013) statistics since the luminosity obtained by 
Bestenlehner et al. (2014) is $\log (L/L_{\odot}$) = 6.0. Source 1222 from Parker (1993) has also been excluded owing to 
its O9\,V spectral type from Walborn et al. (2002c), as has VFTS 1014 (Parker 863) despite Bestenhehner et al. obtaining 
$\log (L/L_{\odot}$) = 6.2 since this ARGUS spectrum is likely composite (O3\,V + mid/late O). Similarly, we favour the 
lower luminosities ($\log L/L_{\odot}$) = 6.0 -- 6.1) inferred by Doran et al. (2013) for R140a2 (WN5+O), Mk~49 (WN6) and 
BAT99-96 (WN8) over those by Hainich et al. (2014), who obtained significantly higher values by assuming the WR dominated 
the continuum flux (R140a2) or on the basis of high interstellar extinctions (Mk~49, BAT99-96).} The other (11) high 
luminosity stars in 30 Doradus lie $\geq$10 pc away from R136a1, approximately half within NGC~2070 and the remainder 
elsewhere. The majority of these are luminous WN-type stars (see also Hainich et al. 2014; Bestenlehner et al. 2014).

The overwhelming majority of the very massive stars in 30 Doradus are early O supergiants, transition Of/WN stars or 
luminous WN stars. Crowther et al. (2010) proposed that very massive stars may possess O-supergiant spectral appearances 
on the main sequence, owing to their proximity to the (classical) Eddington limit. However, R136 a4, a8 and Mk~25 
indicate that very luminous O2--3 dwarf stars {\it do} exist despite having classical Eddington parameters of $ 
\Gamma_{e} \sim$0.5, based upon {\sc bonnsai}-derived masses.



\section{Integrated far-ultraviolet spectroscopy}\label{cluster}

Ultraviolet spectroscopy of the central region of NGC\,2070 was first obtained with IUE (e.g. Koornneef \& Mathis 1981; 
Vacca et al. 1995), while the advent of HST/GHRS permitted high resolution spectroscopy (Walborn et al. 1992). The 
integrated R136 spectrum revealed prominent He\,{\sc ii} $\lambda$1640 emission, together with strong N\,{\sc v} 
$\lambda\lambda$1238--42 and C\,{\sc iv} $\lambda\lambda$1548--51 P Cygni features, with stellar Si\,{\sc iv} 
$\lambda\lambda$1393--1402 absent.


Figure~\ref{r136_cluster} presents the integrated far-UV spectrum of R136 considering all sources up to 0.5 pc from 
R136a1. Its appearance closely reflects earlier descriptions, He\,{\sc ii} $\lambda$1640 emission is relatively strong 
($W_{\lambda} \sim$ 4.5\AA) and broad (FWHM$\sim$10\AA\ or 1800 
km\,s$^{-1}$), so the integrated spectral morphology is closer to an Of/WN 
supergiant than a genuine WN star - recall Figure 4 of Walborn et al. (1992) which compares R136a to Melnick 42 (O2\,If). A comparison between the 2$\times2$ arcsec aperture GHRS/G140L observation of R136a from April 1991 (GO 1188, PI 
D.C.  Ebbetts) and an equivalent region of our STIS dataset reveals excellent agreement, and is morphologically identical to the 
integrated spectrum shown in Fig.~\ref{r136_cluster} aside from a 30\% lower continuum flux.

Fig.~\ref{r136_cluster} includes the co-added spectrum of all 70 stars for which $F_{\rm 1500} \geq 5 \times 
10^{-15}$ erg\,s$^{-1}$\,cm$^{2}$\,\AA$^{-1}$, which collectively contribute $\sim$77\% of the integrated continuum 
flux, with the difference resulting from UV faint late O-type stars, plus the unresolved early B population.
The cumulative spectrum of the  7 most luminous members of R136a -- each with inferred initial masses in excess of 
100 $M_{\odot}$ -- comprising three WN stars (a1--a3), one Of/WN star (a5), an O2 supergiant (a6) plus two O2--3 dwarfs 
(a4, a8) -- is also included in Fig.~\ref{r136_cluster}. 
Collectively, these most closely resemble a weak lined mid-WN integrated
star. It is apparent that the He\,{\sc ii} $\lambda$1640 emission 
line flux, and one third of the continuum flux originates from these few 
stars, the former dominated by R136a1--a3. The remaining 63 far-UV 
bright stars cumulatively resemble an early 
non-supergiant O star, with prominent O\,{\sc v} $\lambda$1371, and P Cygni N\,{\sc v} $\lambda\lambda$1238--42, C\,{\sc 
iv} $\lambda\lambda$1548--51.

\begin{figure}
\begin{center}
 \includegraphics[bb=65 200 515 
630,width=1.0\columnwidth]{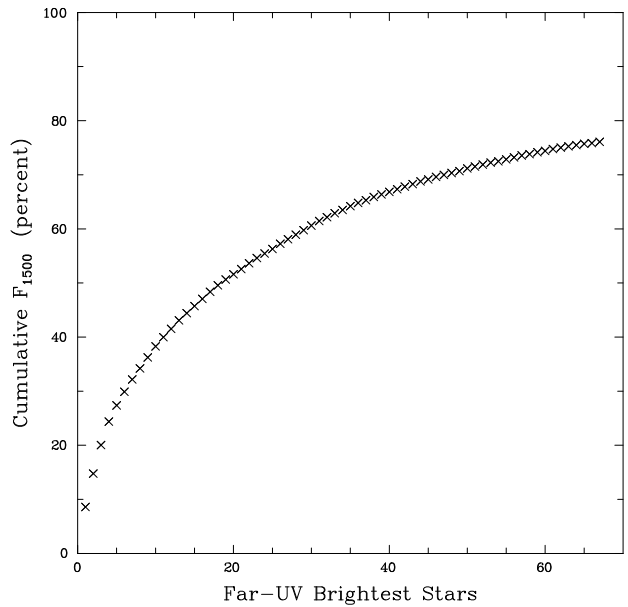} 
 \caption{Cumulative $F_{\rm 1500}$ flux for the brightest 70 far-UV sources 
within  2.05 arcsec (0.5 pc) of R136a1, including estimates for stars 
beyond our slit spectroscopy. The three WN stars R136a1--a3 contribute
20\% of the total, while collectively the 18 brightest far-UV sources provide 50\% of 
the integrated continuum flux.} 
\label{r136_cumulative}
\end{center}
\end{figure}

It is apparent that prominent He\,{\sc ii} $\lambda$1640 emission in R136 arises from the presence of VMS (initial masses 
$\geq 100 M_{\odot}$). Usually, the upper mass limit assumed in population synthesis models is 100 $M_{\odot}$ so a 
synthetic model for R136 would fail to reproduce its He\,{\sc ii} $\lambda$1640 emission.  Indeed, if this feature is 
observed in the integrated light of sufficiently young (unresolved) high mass star clusters, one might legitimately 
conclude that the stellar mass function extends well beyond 100 $M_{\odot}$.

At very young ages, early O stars will dominate the UV appearance of massive clusters, so strong O\,{\sc v} $\lambda$1371 
absorption will be present. The absence of mid to late O supergiants will ensure that P Cygni Si\,{\sc iv} 
$\lambda\lambda$1393--1402 is absent. At later ages (3--5 Myr) O\,{\sc v} will weaken while Si\,{\sc iv} will strengthen, 
plus He\,{\sc ii} $\lambda$1640 from classic WR stars will appear. Therefore, the simultaneous presence of O\,{\sc v} 
absorption, He\,{\sc ii} emission together with the absence of P Cygni Si\,{\sc iv} would suggest both a very young age 
and the presence of very massive stars. To date, the most promising example of such a star cluster is A1 within NGC~3125 
(Chandar et al. 2004; Hadfield \& Crowther 2006; Wofford et al. 2014).

In addition, since our datasets enable spatially resolved spectroscopy of the constituents of R136, we are able to 
quantify the contribution of individual stars to the integrated UV continuum. 5 stars brighter than $m_{\rm F555W}$ 
= 17.0 mag located within 2.05 arcsec of R136a1 lie beyond our STIS slit spectroscopy, so we have estimated their 
fluxes by adopting an average $F_{\rm 1500}/F_{\rm F555W}$ ratio from cluster members; they are presented in parentheses 
in Tables~\ref{targets2} and B1. We estimate $\sim$2\% of the integrated UV flux originates from these stars, primarily 
R136 H39 (O3\,V+O5.5\,V, Massey et al. 2002).

Figure~\ref{r136_cumulative} presents the cumulative 1500\AA\ flux of R136a to a radius of 0.5 pc, versus a ranked list of 
the 70 far-UV  brightest stars in this region. The three brightest sources alone (R136a1-a3) contribute 20\% of the far-UV 
continuum flux, while the 18  brightest far-UV sources provide 50\% of the integrated far-UV flux. Walborn et al. (1992) cite an 
higher (pre-COSTAR)  estimate of 40\% for R136a1-a3 from GHRS acquisition images, according to D.~Ebbets and E.~Malumuth.

\section{Summary}\label{conclusions}

We have introduced a census of the visually brightest members of the R136 star cluster based on new long-slit HST/STIS 
spectroscopy. We have obtained ultraviolet, blue visual and H$\alpha$ spectroscopic datasets using 17 contiguous 
52$\times$0.2$''$ slits sampling the core of R136. In the present study we reveal the following:
\begin{enumerate}
\item We classify  51 of the 57 stars brighter than $m_{\rm 
F555W} = 16.0$ mag within a radius of 0.5 pc of R136a1 based on 
their ultraviolet spectroscopic morphology. We confirm previous results of 
Massey \& Hunter (1998) that the bulk of the visually brightest members of 
R136 are O2--3 stars. For the first time we classify all Weigelt \& Baier 
(1985) stars within R136a, which comprise three WN5 stars (a1-a3), two O 
supergiants (a5-a6) and three early O dwarfs (a4, a7--a8). Eight 
additional sources are spectrally classified within the central $\sim$pc, 
including R136b (O4\,If/WN8).
\item We obtain wind terminal velocities from measurements of $v_{\rm 
black}$ (or $v_{\rm edge}$) from P Cygni C\,{\sc iv} 
$\lambda\lambda$1548--51 profiles, from which we obtain average O2--3 wind 
velocities of 2750 and 3120 km\,s$^{-1}$ for dwarfs and (super)giants, 
respectively. Comparisons with early O dwarfs elsewhere in the LMC and 
Milky Way indicate somewhat lower wind velocities in R136.
\item We estimate physical parameters for the visually brightest members 
of R136 based on the temperture calibration of LMC O-type stars from Doran 
et al. (2013) and exploit {\sc bonnsai} (Schneider et al. 2014b) to 
estimate masses and ages based on rotating evolutionary models appropriate 
for the LMC (Brott et al. 2011, K\"{o}hler et al. 2015). Three WN stars 
(R136a1-a3), two Of/WN stars (R136a5 and R136b), newly identified O2\,If 
R136a6 and O2--3\,V stars R136a4 and a8 possess initial masses $\geq$100 
$M_{\odot}$.
\item We obtain a cluster age of 1.5$^{+0.3}_{-0.7}$ Myr from the O star 
content, which is consistent with that estimated for the luminous WN5h 
stars by Crowther et al. (2010), although apparent ages of these and other 
luminous O stars is 0.6$\pm$0.3 Myr, likely to arise from the treatment of 
envelope inflation in evolutionary models (Sanyal et al. 2015), rather 
than rejuvenation following binary mergers. The mass and youth of R136a 
indicate that no more than the most massive system is potentially a binary 
product (Schneider et al. 2014a). More definitive conclusions await 
analysis of STIS/CCD spectroscopy, permitting rotation rates and He 
abundances of O stars to be obtained.  We identify 25 stars more massive 
than 50 $M_{\odot}$ within 0.5 pc of R136a1 indicating a lower enclosed 
mass than previous estimates, although reasonable consistency is achieved 
with respect to Hunter et al. (1995) and Andersen et al. (2009) in the 
R136 `region' from the total number of high mass stars within a 5 pc 
radius.
\item Incorporating results from our earlier VFTS-based census (Doran 
et al. 2013; Bestenlehner et al. 2014) we estimate a total of 29 stars 
more luminous than $\log (L/L_{\odot}$) = 6.2 in the entire 30 Doradus nebula,
of which 26 possess  O2--4\,If, Of/WN or WN subtypes. Only 9 stars are 
located within R136 itself, with an additional 9 stars within the extended 
halo, 1.5--5 pc away from R136a1 (Melnick 1985) and
another 11 stars at much greater distance (10--120 pc).
\item We have also considered the integrated ultraviolet spectrum of R136a 
in the context of its constituent members. The prominent He\,{\sc ii} 
$\lambda$1640 emission feature arises primarily from the most massive 
stars. Specifically, close to 100\% of the He\,{\sc ii} $\lambda$1640 flux 
and 32\% of the far-UV continuum originates in the 7 stars for which we 
estimate (initial) masses above 100 $M_{\odot}$, the former dominated by 
from R136a1--a3. Usually, such stars are neglected in conventional 
population synthesis models. Prominent He\,{\sc ii} $\lambda$1640 emission 
in the integrated spectrum of suitably young star clusters -- prominent 
O\,{\sc v} $\lambda$1371 absorption without strong Si\,{\sc iv} 
$\lambda\lambda$1393-1402 P Cygni emission -- would favour a mass function 
that extends well above this value. Indeed, extremely strong $\lambda$1640 
emission in cluster A1 within NGC~3125 favours extremely massive stars 
therein (Chandar et al. 2004; Hadfield \& Crowther 2006; Wofford et al. 
2014).
 \end{enumerate}

In future papers in this series we will present the corresponding blue, visual and H$\alpha$ 
STIS/CCD spectroscopic datasets. These will be used to refine spectral classifications, obtain more robust
physical 
and wind properties, and search for close binary systems. These will enable a refined 
Hertzsprung-Russell diagram for R136a.
The multi-epoch nature of the CCD datasets will enable an attempt at identifying close binaries (see
Cottaar \& H\'{e}nault-Brunet 2014), together new HST FGS observations of R136a (S.E. de Mink, 
P.I.).


\section*{Acknowledgements}

Observations were taken with the NASA/ESA Hubble Space Telescope, obtained 
from the data archive at the Space Telescope Institute. STScI is operated 
by the association of Universities for Research in Astronomy, Inc. under 
the NASA contract NAS 5-26555. We wish to thank Guido De Marchi
for providing WFC3/UVIS photometry of bright stars in NGC~2070, 
Elena Sabbi for permitting inspection of datasets from the Hubble Tarantula Treasury Project ahead of publication, and Raman Prinja for calculating an SEI model for R136 H36. 
Financial support was provided to SCN by the Science and Technology Facilities Council. JMA acknowledges support from
(a) the Spanish Government Ministerio de Econom{\'\i}a y Competitivad (MINECO) through grants AYA2010-15\,081, AYA2010-17\,631, and AYA2013-40\,611-P, and
(b) the Consejer{\'\i}a de Educaci{\'o}n of the Junta de Andaluc{\'\i}a through grant P08-TOC-4075, while AH acknowledges support from grants
AYA2012-39364-C02-01 and SEV2011-0187-01.

\appendix

\section{Ultraviolet atlas of LMC O stars}

\begin{table*}
\begin{center}
\caption{Log of ultraviolet spectroscopic observations used in 
the LMC O star montages}
\begin{small}
\begin{tabular}{l@{\hspace{3mm}}l@{\hspace{2.5mm}}l@{\hspace{2.5mm}}l
@{\hspace{2.5mm}}l@{\hspace{2.5mm}}l@{\hspace{2.5mm}}l@{\hspace{2.5mm}}
l@{\hspace{2.5mm}}l@{\hspace{2.5mm}} 
}
\hline 
Star & Alias & Sp Type & Ref & Tel & Inst & HST GO/PI & Ref & Fig\\
\hline
BI 237 &      & O2\,V & W02b & HST &STIS/G140L & 9412/Massey & M04 & 
\ref{oearlydwarf-uv}, \ref{o2uv}  \\
R136 H35 &      & O3\,V & CN16 & HST&STIS/G140L & 12465/Crowther & This Study 
& \ref{oearlydwarf-uv}, \ref{o3uv} \\
LH 81/W28-23 &    & O3.5\,V((f+)) & M05 & HST&STIS/G140L & 9412/Massey & 
M05  & \ref{oearlydwarf-uv}, \ref{o3p5uv} \\
LH 81/W28--5 &   & O4\,V((f+)) & M04 & HST & STIS/G140L & 8633/Massey & M04 & \ref{o4-5-morph} \\
R136 H78  &    & O4:\,V & CN16 & HST&STIS/G140L & 12465/Crowther & This Study
& \ref{oearlydwarf-uv}, \ref{o4uv} \\ 
Sk-70$^{\circ}$ 60 &  & O4--5\,V((f)) & M09 & HST & FOS/G130H & 5444/Robert & PC98 & \ref{o4-5-morph} \\
LH 58--496 &  LH58--10a & O5\,V(f) & M05 & HST & STIS/G140L & 9412/Massey & M05 & \ref{o4-5-morph} \\
R136 H70   &  & O5\,Vz & CN16 & HST&STIS/G140L & 12465/Crowther & 
This Study & \ref{oearlydwarf-uv}, \ref{olatedwarf-uv} \\
Sk--70$^{\circ}$ 69 &  & O5.5\,V((f)) & M09 & HST & FOS/G130H & 2233/Kudritzki & W95 & \ref{o4-5-morph} \\
R136 H123      &       & O6\,V & CN16 & HST&STIS/G140L & 12465/Crowther & 
This Study & \ref{olatedwarf-uv}, \ref{o6uv} \\
R136 H134      &       & O7\,Vz & CN16 & HST & STIS/G140L & 12465/Crowther & 
This Study  & \ref{olatedwarf-uv}, \ref{o7-8uv} \\
R136 H80       &       & O8\,V  & CN16 & HST & STIS/G140L & 12465/Crowther &
Ths Study   & \ref{olatedwarf-uv}, \ref{o7-8uv} \\
R136 H121      &       & O9.5\,V & CN16 & HST & STIS/G140L & 12465/Crowther &
This Study & \ref{olatedwarf-uv}, \ref{o9+uv} \\
\\
HDE 269810 & Sk--67$^{\circ}$ 211 & O2\,III(f) & W02b & HST&FOS/G130H 
                                             & 4110/Kudritzki & W95 & 
\ref{ogiant-uv}, \ref{o2uv} \\
R136 H48   &                    & O2--3\,III & CN16 & HST&STIS/G140L & 12465/Crowther 
& This Work & \ref{ogiant-uv}, \ref{o3uv} \\ 
LH90 ST2-22 &          & O3.5\,III(f+) & M05 & HST&STIS/G140L & 
9412/Massey & M05 & \ref{ogiant-uv}, \ref{o3p5uv} \\
HDE 269676 & Sk--71$^{\circ}$ 45  & O4--5\,III   &  W77   & IUE&SWP/HIRES  
&  Nandy & G80 & \ref{ogiant-uv}, \ref{o4uv} \\
Sk--66$^{\circ}$ 100 &             & O6\,II(f)    & W95 & HST&FOS/G130H & 
2233/Kudritzki & W95 & \ref{ogiant-uv}, \ref{o6uv} \\
Sk--67$^{\circ}$ 101 &             & O8\,II(f)    & R03  & HST&STIS/E140M 
& 7299/Bomans & M03  & \ref{ogiant-uv}, \ref{o7-8uv} \\
\\
R136 H36   & &O2\,If & CN16 & HST&STIS/G140L & 12465/Crowther & 
This Study & \ref{osuper-uv}, \ref{o2uv} \\
Sk--65$^{\circ}$ 47      & LH 43-18 & O4\,I(n)f+p & W10 & HST&STIS/G140L & 
9412/Massey & M05 & \ref{osuper-uv}, \ref{o4uv} \\
Sk--67$^{\circ}$ 111     &          & O6\,Ia(n)fpv& W02a & HST&STIS/E140M 
& 12218/Massa & & \ref{osuper-uv}, \ref{o6uv} \\
HDE 270952    & Sk--65$^{\circ}$ 22 & O6\,Iaf+    & W77  & IUE&SWP/HIRES &    
& C02 & \ref{o6uv} \\
Sk--69$^{\circ}$ 50      &          & O7(n)(f)p   & W10 & HST&STIS/E140M & 
12218/Massa  & & \ref{o7-8uv} \\
HDE 269702 & Sk--67$^{\circ}$ 168 & O8\,I(f)p    &W10      & 
HST&STIS/E140M & 
12218/Massa & & \ref{osuper-uv}, \ref{o7-8uv} \\
Sk--69$^{\circ}$ 124  &  & O9\,Ib       & C86   & HST&FOS/G130H & 
5444/Robert  & PC98 & \ref{o9+uv} \\
Sk--66$^{\circ}$ 169 &   & O9.7\,Ia+    & F88 & IUE&SWP/HIRES &     Patriarchi
&C02 & \ref{osuper-uv}, \ref{o9+uv} \\
Sk--68$^{\circ}$ 41    &              & B0.5\,Ia     & F88  & 
HST&FOS/G130H 
&4110/Kudritzki &W95 & \ref{o9+uv} \\
\\
Sk--67$^{\circ}$ 22       & &O2\,If/WN5 & CW11 & HST&STIS/G140L & 
9412/Massey & M05 & \ref{o2uv} \\
Mk 39      & R136 H7  & O2.5\,If/WN6 & CW11 & HST&STIS/G140L & 9412/Massey 
& M05 & \ref{o3uv} \\
Mk 37       & R136 H14 & O3.5\,If/WN7 & CW11 & HST&STIS/G140L & 
9412/Massey & M05 & \ref{o3p5uv} \\
\hline
\hline
\multicolumn{8}{l}{
  \begin{minipage}{1.6\columnwidth}~\\
C86: Conti et al. 1986; C02: Crowther et al. 2002;
CW11: Crowther \& Walborn 2011;
CN16: Caballero-Nieves et al. 2016 (in prep);
G80: Gondhalekar et al. 1980;
M03: Massa et al. 2003;
M04: Massey et al. 2004;
M05: Massey et al. 2005;
M09: Massey et al. 2009;
PC98: Prinja \& Crowther 1998;
R03: Robert et al. 2003;
W77: Walborn 1977;
W95: Walborn et al. 1995;
W02a: Walborn et al. 2002a; 
W02b: Walborn et al. 2002b; 
W10: Walborn et al. 2010
  \end{minipage}
    }\\
\end{tabular} 
\label{lmc_uv_atlas}
\end{small}
\end{center}

\end{table*}

\begin{figure*}
\begin{center}
 \includegraphics[bb=15 45 500 785, 
height=1.75\columnwidth,angle=-90]{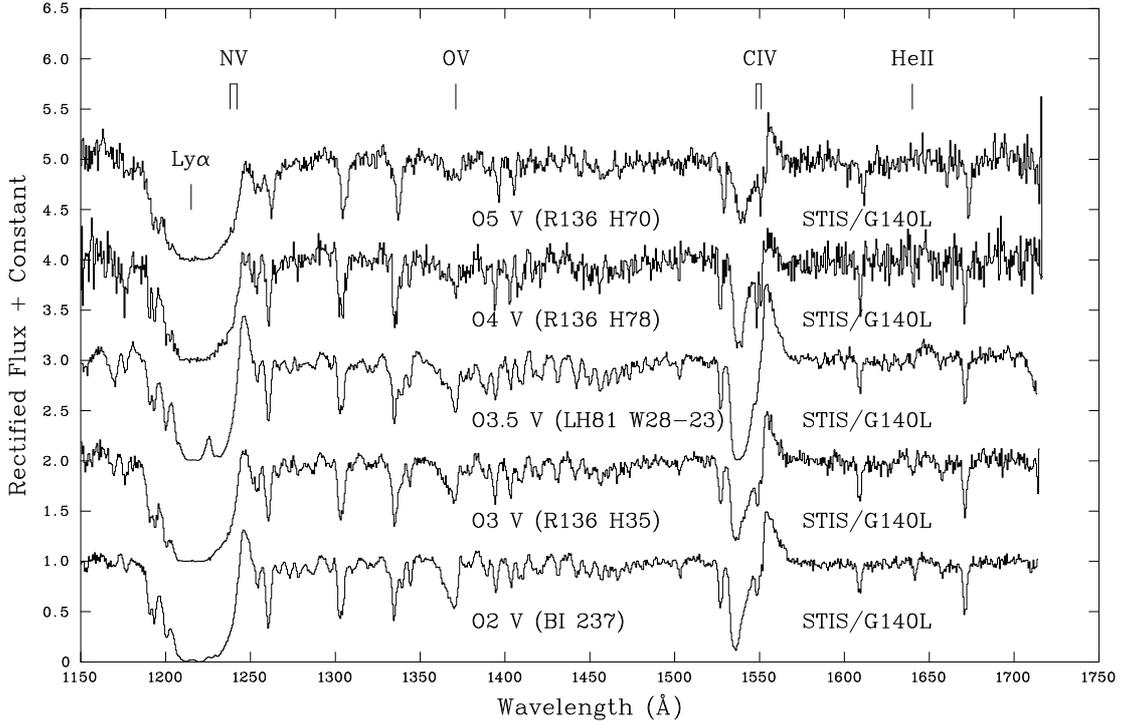} 
 \caption{Ultraviolet morphological progression of O2--5 dwarfs, showing
strong, broad O\,{\sc v} $\lambda$1371 absorption at O2--3, together with 
strong P Cygni N\,{\sc v} $\lambda\lambda$1238--42 and C\,{\sc iv} 
$\lambda\lambda$1548-51. O\,{\sc v} weakens at 
O3.5 and disappears at O4, while C\,{\sc iv} P Cygni weakens at O5.}
\label{oearlydwarf-uv}
\end{center}
\end{figure*}

\begin{figure*}
\begin{center}
 \includegraphics[bb=15 45 500 785, 
height=1.75\columnwidth,angle=-90]{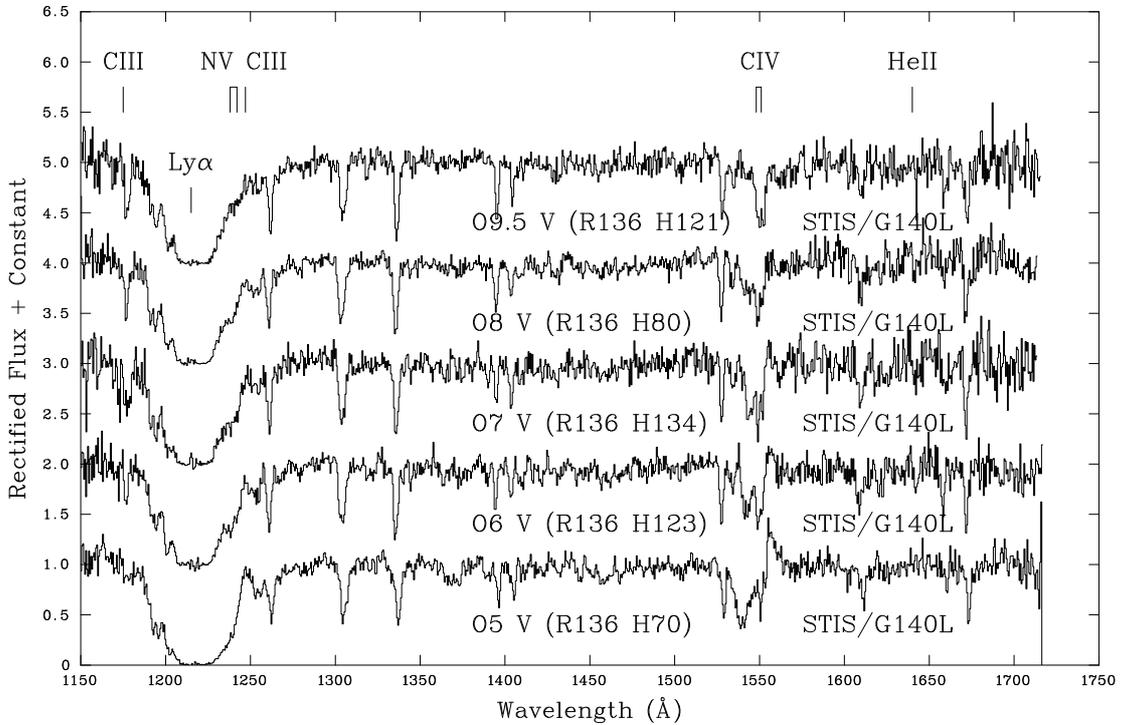} 
 \caption{Ultraviolet morphological progression of O5--9.5 dwarfs, showing
prominent broad, unsaturated P Cygni 
C\,{\sc iv} $\lambda\lambda$1548-51 at O5\,V, which
weakens in both emission strength and absorption width at O6--7\,V and becomes
photospheric at O9.5V. N\,{\sc v} $\lambda\lambda$1238--42 is apparent at O5\,V
but also weakens at O6--8\,V and is undetected at O9.5\,V.}
\label{olatedwarf-uv}
\end{center}
\end{figure*}

\begin{figure*}
\begin{center}
 \includegraphics[bb=15 45 500 785, 
height=1.75\columnwidth,angle=-90]{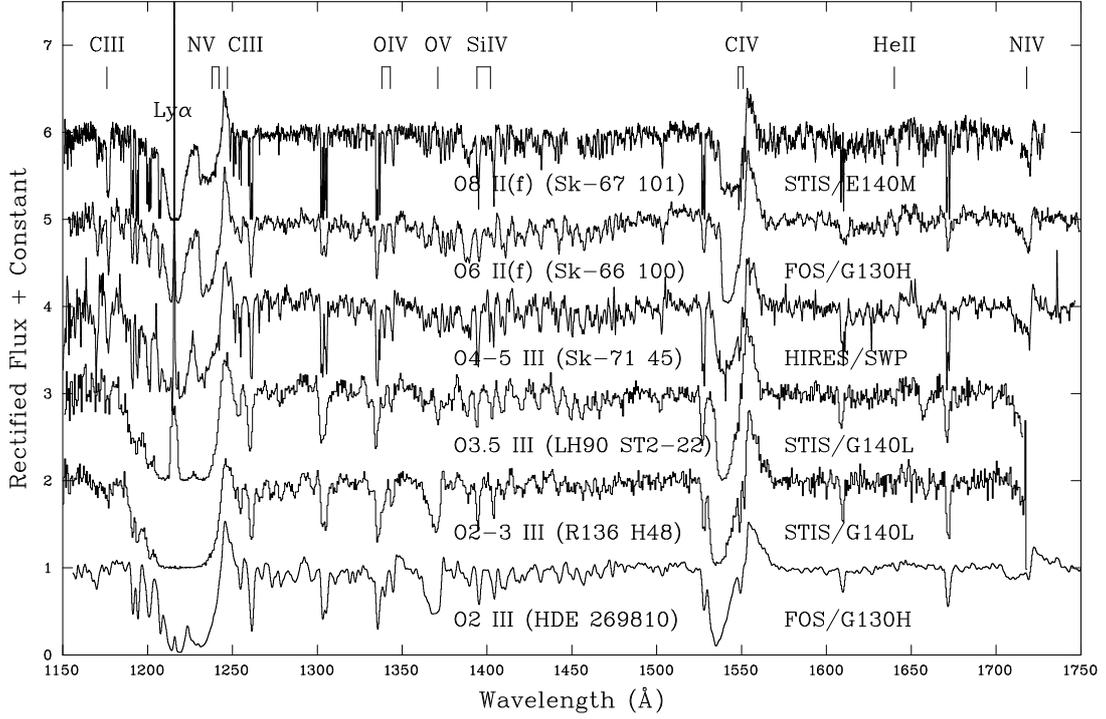} 
 \caption{Ultraviolet morphological progression of O2--8 giants, showing
strong, broad O\,{\sc v} $\lambda$1371 absorption at O2--3, with 
prominent N\,{\sc v} $\lambda\lambda$1238--42 and C\,{\sc iv} 
$\lambda\lambda$1548--51 P Cygni profiles 
until O6 giants, and unsaturated thereafter.}
\label{ogiant-uv}
\end{center}
\end{figure*}

\begin{figure*}
\begin{center}
 \includegraphics[bb=15 45 500 785, 
height=1.75\columnwidth,angle=-90]{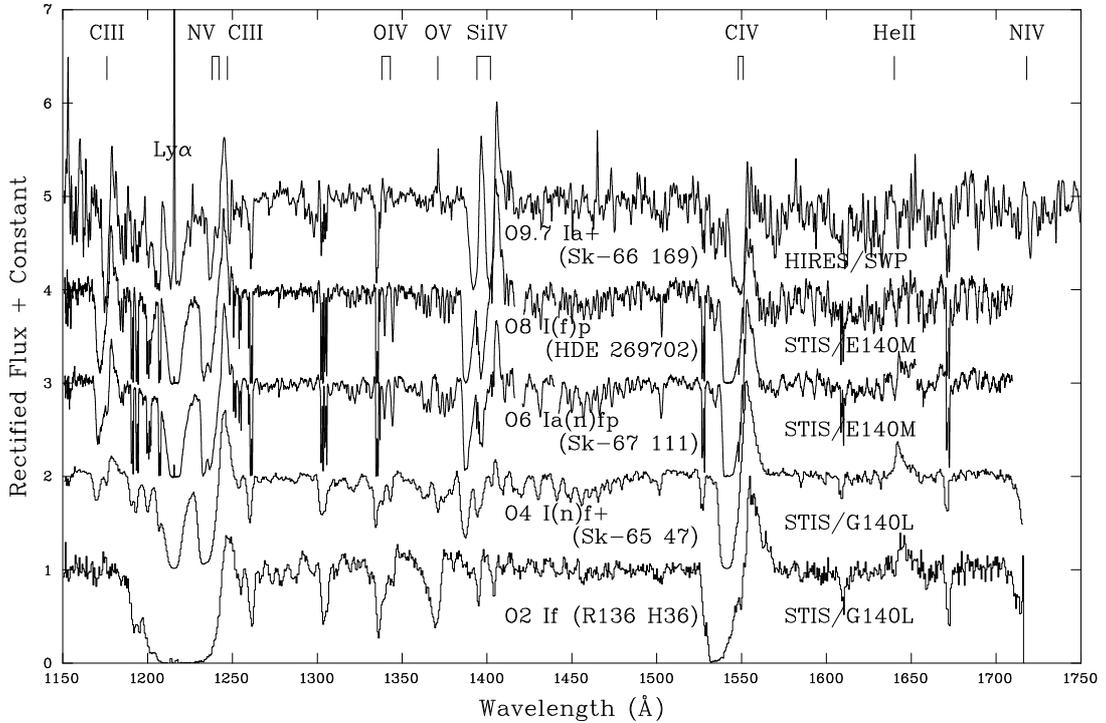} 
 \caption{Ultraviolet morphological progression of O2--9.7 supergiants, 
showing prominent O\,{\sc v} $\lambda$1371 absorption at O2, strong, saturated 
C\,{\sc iv} $\lambda\lambda$1548--51 P 
Cygni profiles until O8 supergiants, Si\,{\sc iv} $\lambda\lambda$1393--1402 P Cygni 
appearing weakly at O4 
and saturating for O6 and later subtypes. He\,{\sc ii} $\lambda$1640 emission is 
prominent between O2--6, but thereafter blended with iron forest features.} \label{osuper-uv} \end{center}
\end{figure*}

\begin{figure*}
\begin{center}
 \includegraphics[bb=15 45 500 
785,height=1.75\columnwidth,angle=-90]{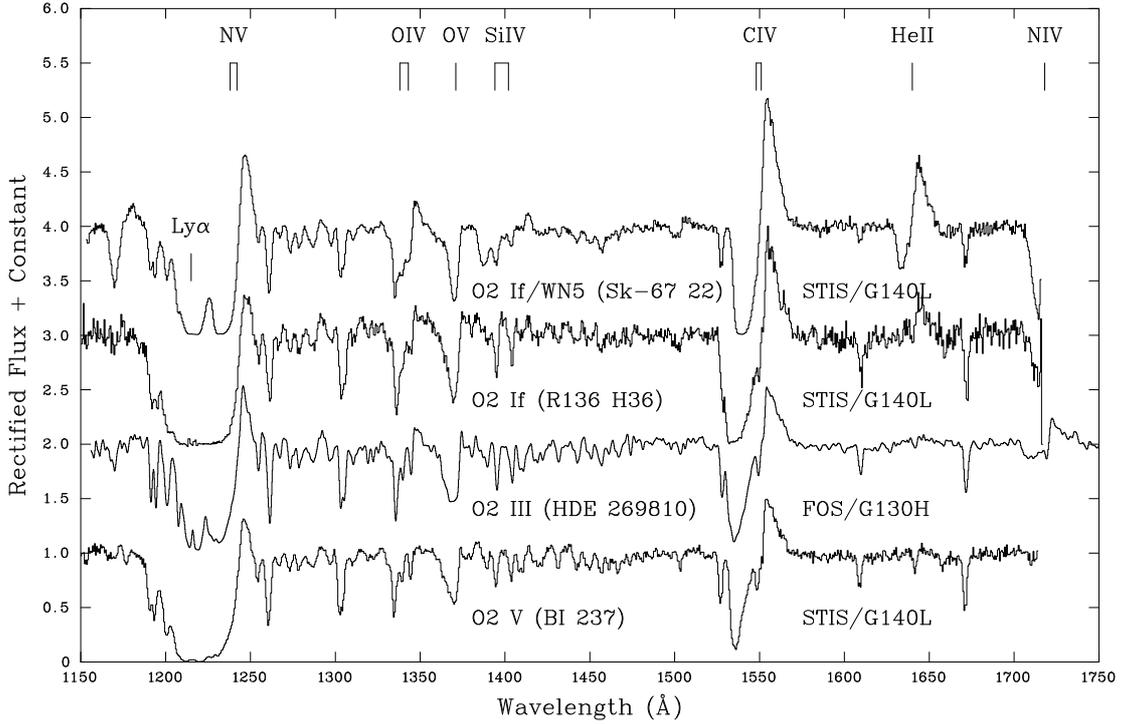} 
 \caption{Montage of O2 luminosity sequence. Strong/broad O\,{\sc v} 
$\lambda$1371 absorption is common to all O2 
subtypes. C\,{\sc iv} $\lambda\lambda$1548--51 
and N\,{\sc v} $\lambda\lambda$1238--42 are prominent P Cygni profiles at all luminosity classes,
although the former is unsaturated in dwarfs/giants. Note the reversal in the He\,{\sc ii} $\lambda$1640 
line from absorption to strong P Cygni emission from dwarf to Of/WN.}
\label{o2uv}
\end{center}
\end{figure*}

\begin{figure*}
\begin{center}
 \includegraphics[bb=15 45 500 
785,height=1.75\columnwidth,angle=-90]{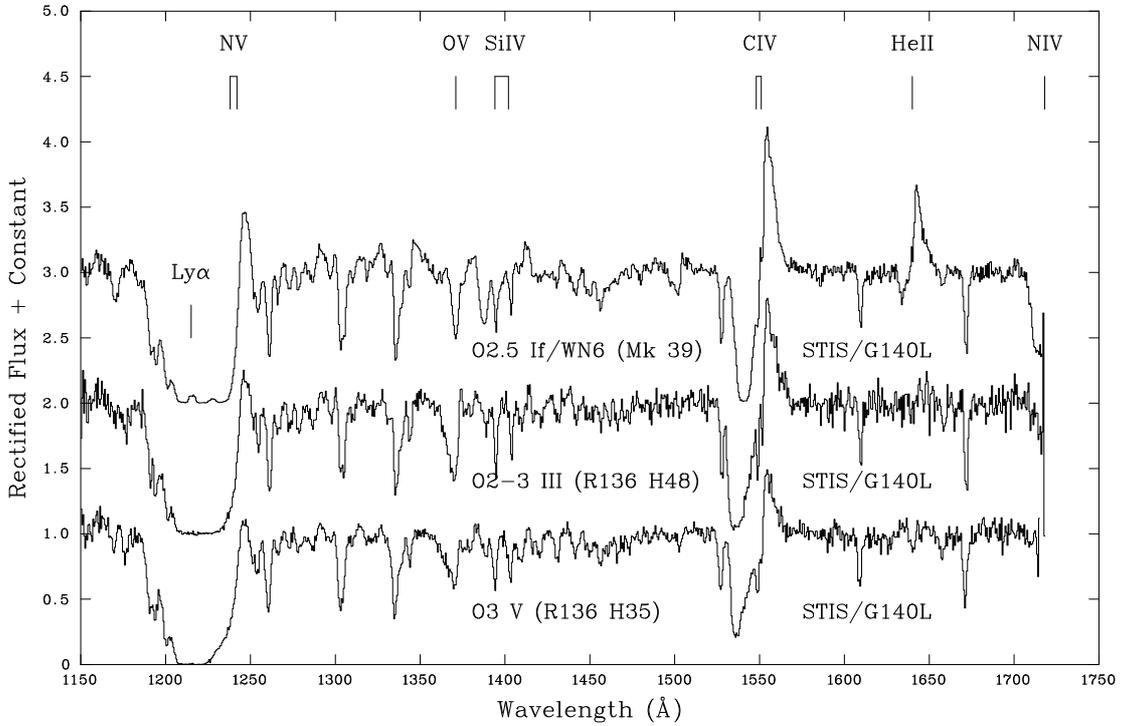} 
 \caption{Montage of O2--3 luminosity sequence.
O\,{\sc v} $\lambda$1371 absorption is common to all luminosity classes. Note the 
reversal in the He\,{\sc ii} $\lambda$1640 line from absorption to strong P Cygni emission
from dwarf to Of/WN.}
\label{o3uv}
\end{center}
\end{figure*}

\begin{figure*}
\begin{center}
 \includegraphics[bb=15 45 500 
785,height=1.75\columnwidth,angle=-90]{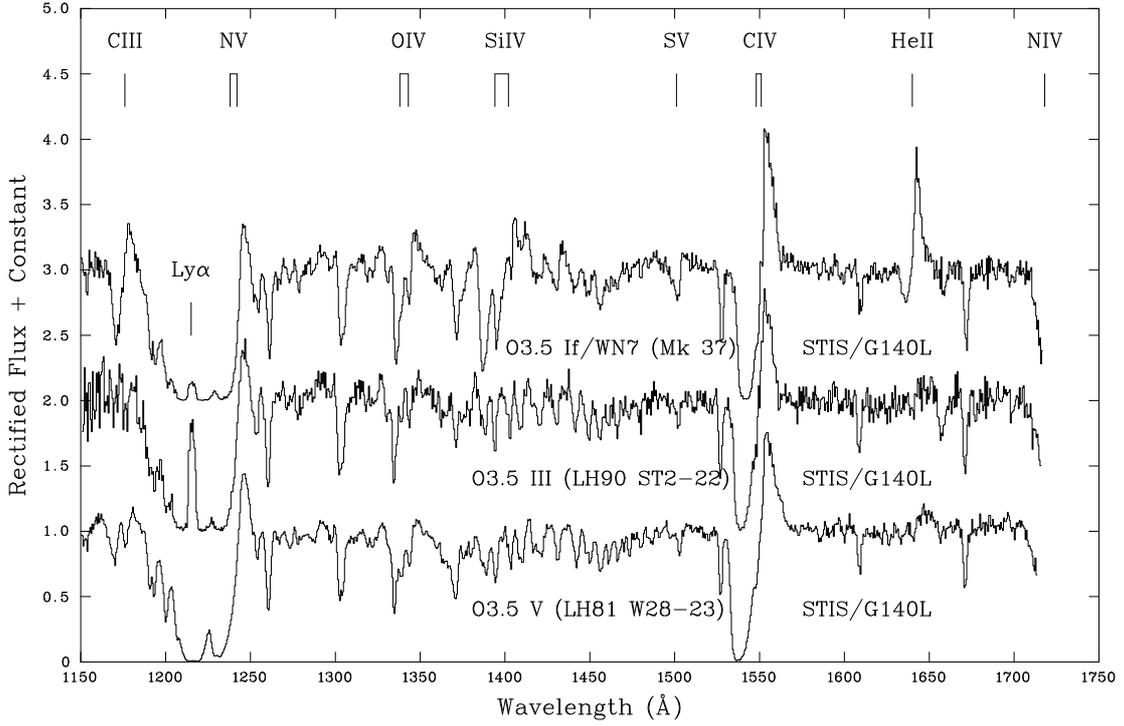} 
 \caption{Montage of O3.5 luminosity sequence.
O\,{\sc v} $\lambda$1371 absorption weakly is present, with Si\,{\sc iv} 
$\lambda\lambda$1393--1402 P Cygni appearing in Of/WN stars together with prominent
He\,{\sc ii} $\lambda$1640 P Cygni emission.}
\label{o3p5uv}
\end{center}
\end{figure*}

\begin{figure*}
\begin{center}
 \includegraphics[bb=15 45 500 
785,height=1.75\columnwidth,angle=-90]{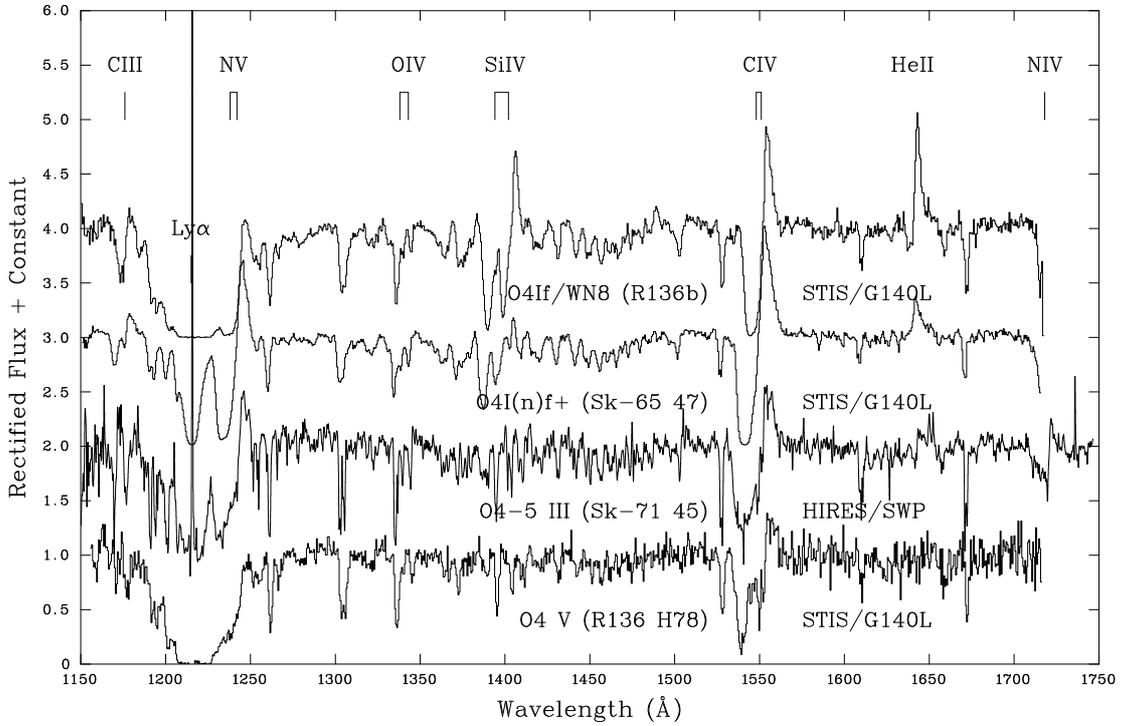} 
 \caption{Montage of O4--5 luminosity class.
O\,{\sc v} $\lambda$1371 absorption is absent, although P Cygni
N\,{\sc v} 
$\lambda\lambda$1238--42 
and C\,{\sc iv} $\lambda\lambda$1548--51 remain prominent for all classes, albeit
unsaturated in dwarfs/giants. P Cygni Si\,{\sc 
iv} $\lambda\lambda$1393--1402 and He\,{\sc ii} $\lambda$1640 emission are seen in O4 supergiants.}
\label{o4uv}
\end{center}
\end{figure*}

\begin{figure*}
\begin{center}
 \includegraphics[bb=15 45 500 
785,height=1.75\columnwidth,angle=-90]{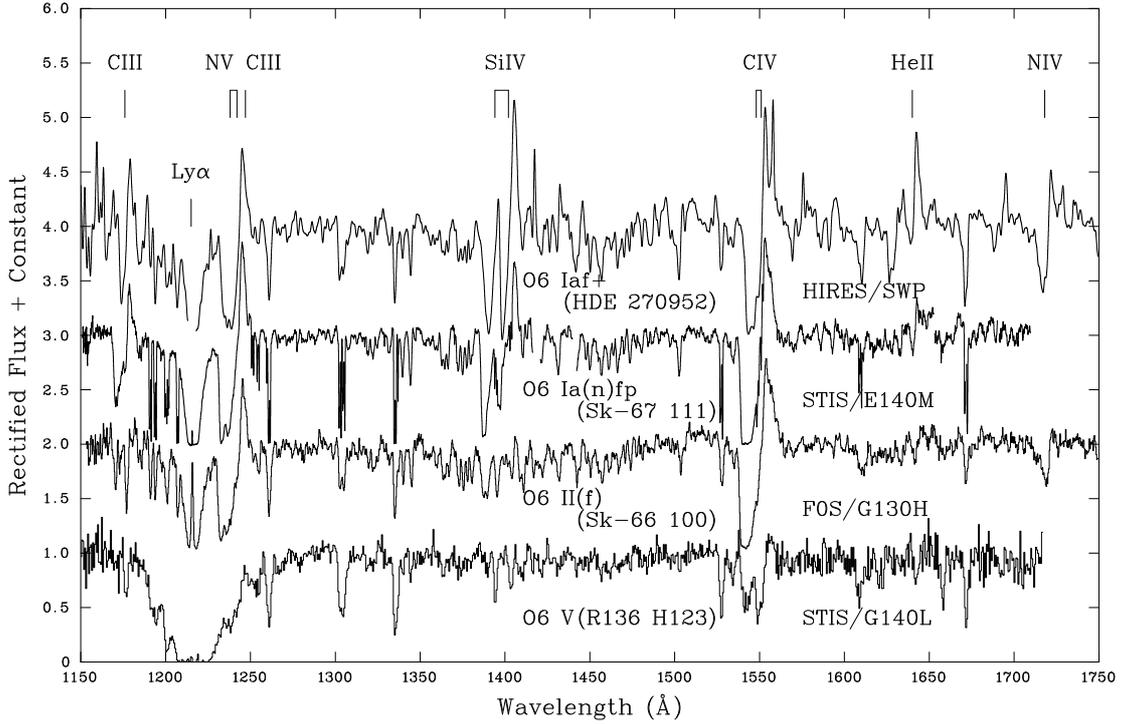} 
 \caption{Montage of O6 luminosity sequence. C\,{\sc iv} 
$\lambda\lambda$1548--51 and N\,{\sc v} $\lambda\lambda$1238--42 exhibit 
a weak P Cygni profile for dwarfs, in contrast to strong P Cygni profiles
at class II--I, with Si\,{\sc iv} $\lambda\lambda$1393--1402 also exhibiting 
saturated P Cygni profiles for supergiants. Narrow He\,{\sc ii} 1640 emission
is also observed in supergiants.}
\label{o6uv}
\end{center}
\end{figure*}

\begin{figure*}
\begin{center}
 \includegraphics[bb=15 45 500 
785,height=1.75\columnwidth,angle=-90]{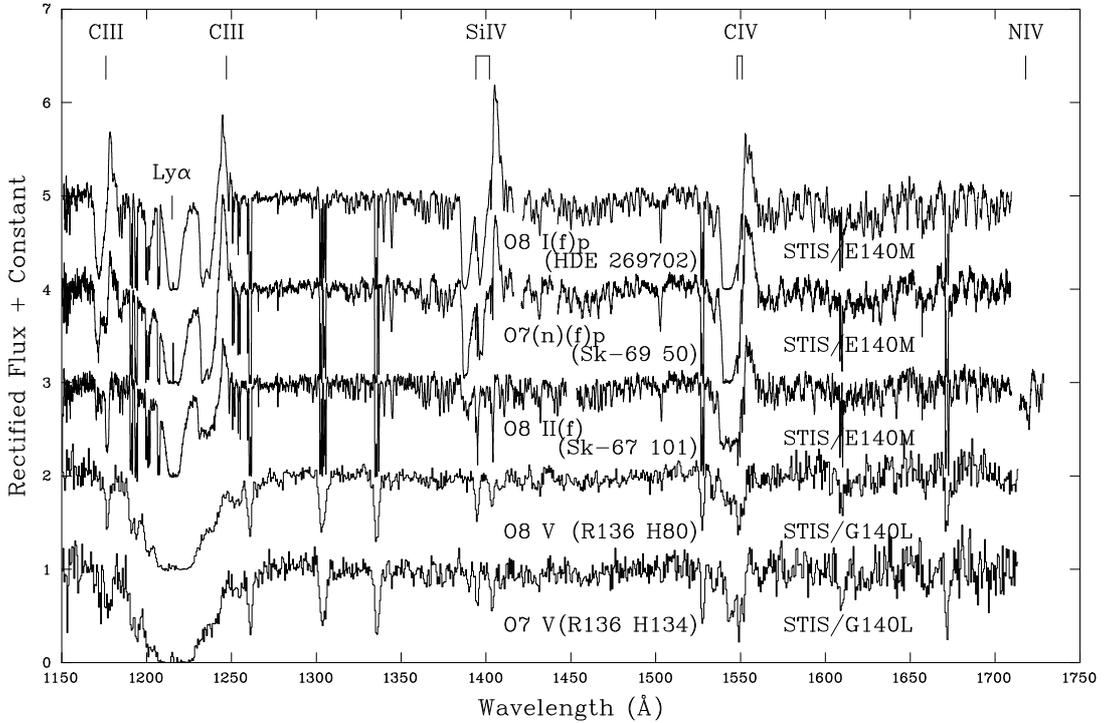} 
 \caption{Montage of O7-8 luminosity sequence. C\,{\sc iv} 
$\lambda\lambda$1548--51 exhibits a narrow, weak P Cygni profile
for dwarfs, with N\,{\sc v} $\lambda\lambda$1238--42 barely present.
Both lines display a prominent P Cygni profile at Class II (unsaturated
for N\,{\sc v}), while Si\,{\sc iv} 
$\lambda\lambda$1393--1402 exhibit saturated P Cygni profiles amongst 
supergiants.}
\label{o7-8uv}
\end{center}
\end{figure*}

\begin{figure*}
\begin{center}
 \includegraphics[bb=15 45 500 
785,height=1.75\columnwidth,angle=-90]{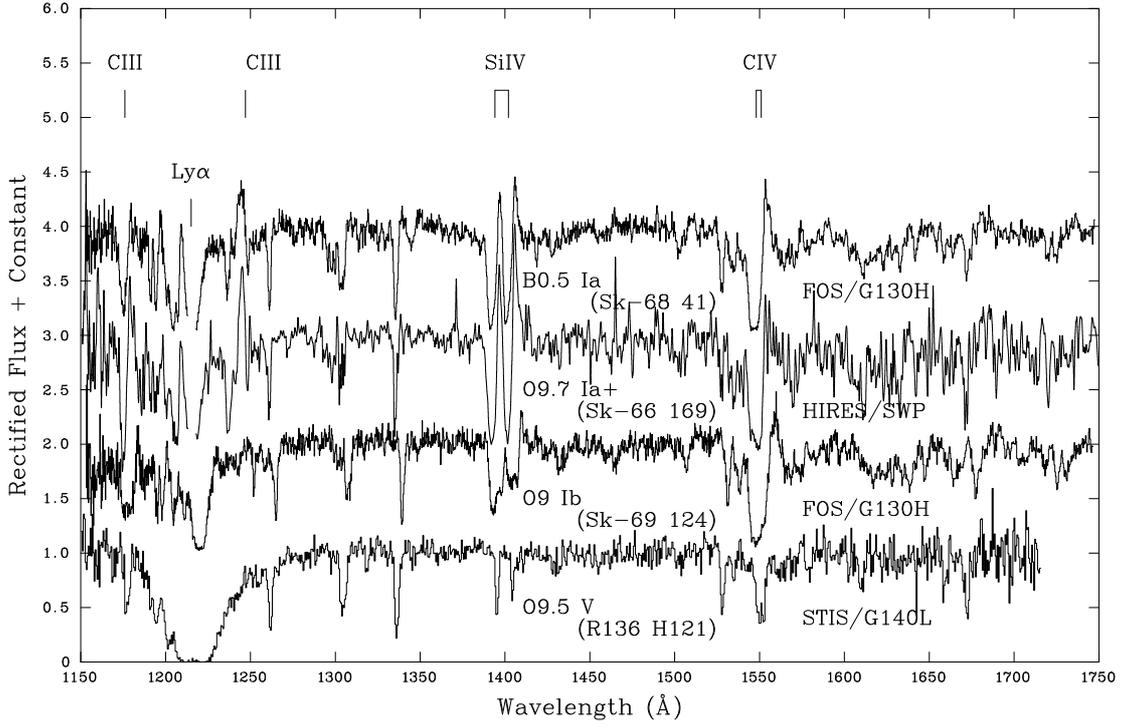} 
 \caption{Montage of O9--B0.5 luminosity sequence. C\,{\sc iv} 
$\lambda\lambda$1548--51 is photospheric in dwarfs, 
N\,{\sc v} $\lambda\lambda$1238--42 absent, while C\,{\sc iv} and
Si\,{\sc iv} $\lambda\lambda$1393--1402 are saturated P Cygni profiles in luminous supergiants.
C\,{\sc iii} $\lambda$1247 is also prominent in luminous supergiants.}
\label{o9+uv}
\end{center}
\end{figure*}

In this appendix we present an ultraviolet atlas of LMC O stars, primarily
obtained with the FOS or STIS instruments of Hubble Space Telescope. A log
of observations is presented in Table~\ref{lmc_uv_atlas}, with
morphological sequences for O dwarfs, giants and supergiants 
presented in Figs.~\ref{oearlydwarf-uv}--\ref{osuper-uv}. Luminosity sequences
for O2 -- O9+ stars are  presented in Figs~\ref{o2uv}-\ref{o9+uv}. More 
extensive IUE SWP/LORES atlases for LMC OB stars are presented by Smith 
Neubig \& Bruhweiler (1999).

\clearpage

\clearpage

\clearpage

\clearpage

\clearpage



\clearpage

\clearpage

\section{Ultraviolet atlas of early-type stars in R136}

\begin{figure*}
\begin{center}
 \includegraphics[bb=15 45 500 785, 
height=1.75\columnwidth,angle=-90]{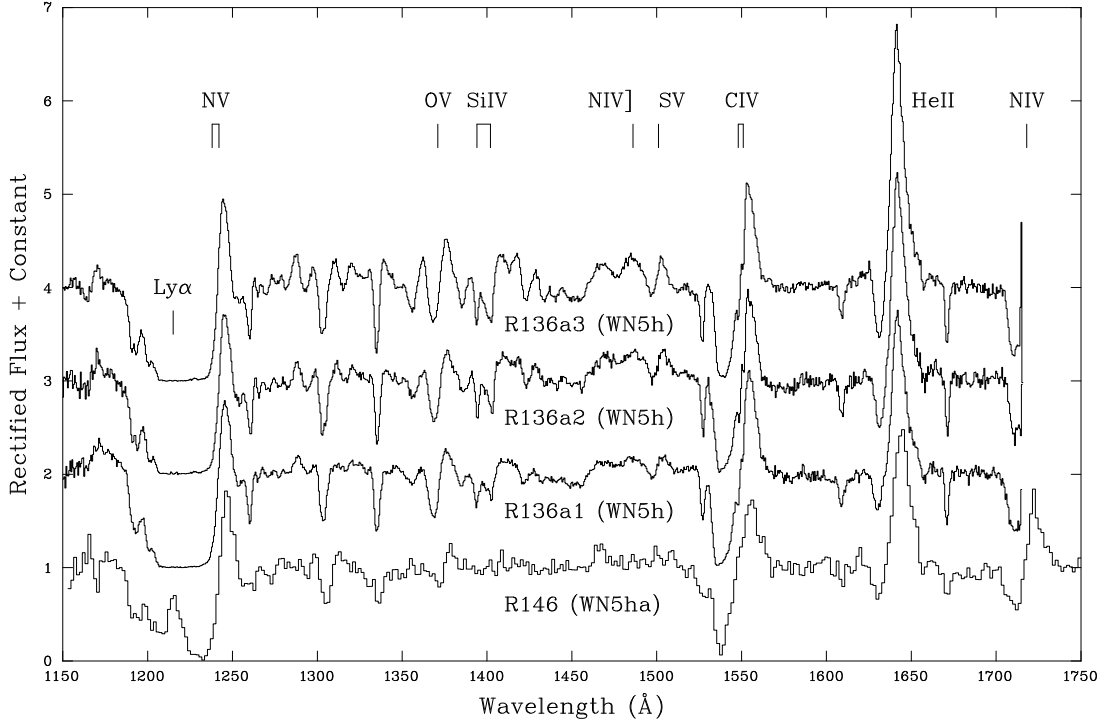} 
 \caption{HST/STIS ultraviolet spectroscopy of WN stars in R136, together
with an LMC template WN5 star R146 (Brey 88=BAT99-117, IUE SWP/LORES)}
\label{r136-wn-uv}
\end{center}
\end{figure*}

\begin{figure*}
\begin{center}
 \includegraphics[bb=15 45 500 
785,height=1.75\columnwidth,angle=-90]{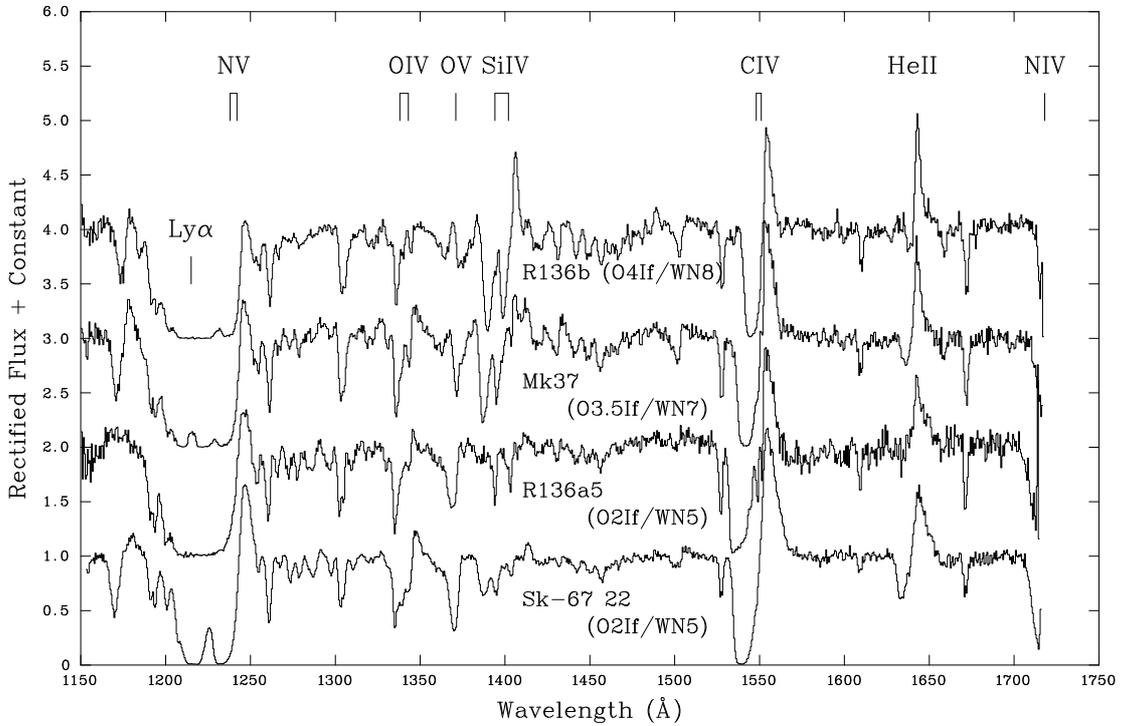} 
 \caption{HST/STIS ultraviolet spectroscopy of Of/WN stars in R136, 
together with LMC templates Sk--67$^{\circ}$ 22 and Melnick~37 (see 
Table~\ref{lmc_uv_atlas})}
\label{r136-osuper-uv}
\end{center}
\end{figure*}

\begin{figure*}
\begin{center}
 \includegraphics[bb=15 45 500 785,height=1.75\columnwidth,angle=-90]{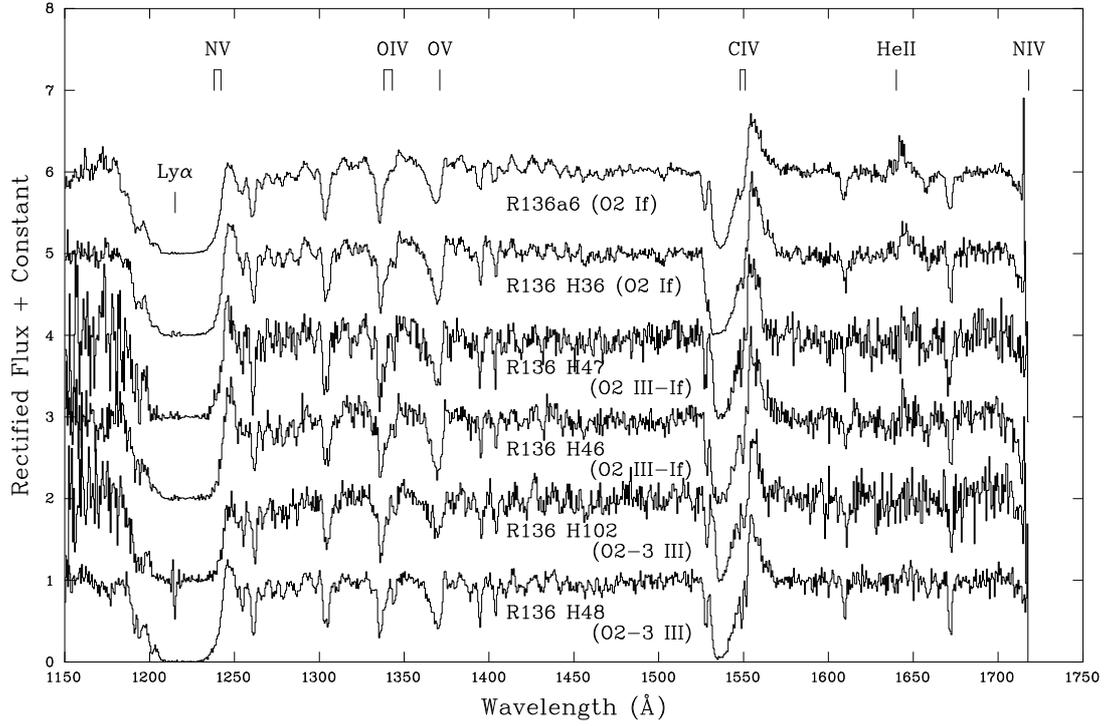} 
 \caption{HST/STIS ultraviolet spectroscopy of O2--3 (super)giants in R136, 
together with LMC templates R136 H48 (O2--3\,III) and H36 (O2\,If), see Table~\ref{lmc_uv_atlas}.}
\label{r136-o2super-uv}
\end{center}
\end{figure*}

\begin{figure*}
\begin{center}
 \includegraphics[bb=35 155 530 
680,width=1.75\columnwidth]{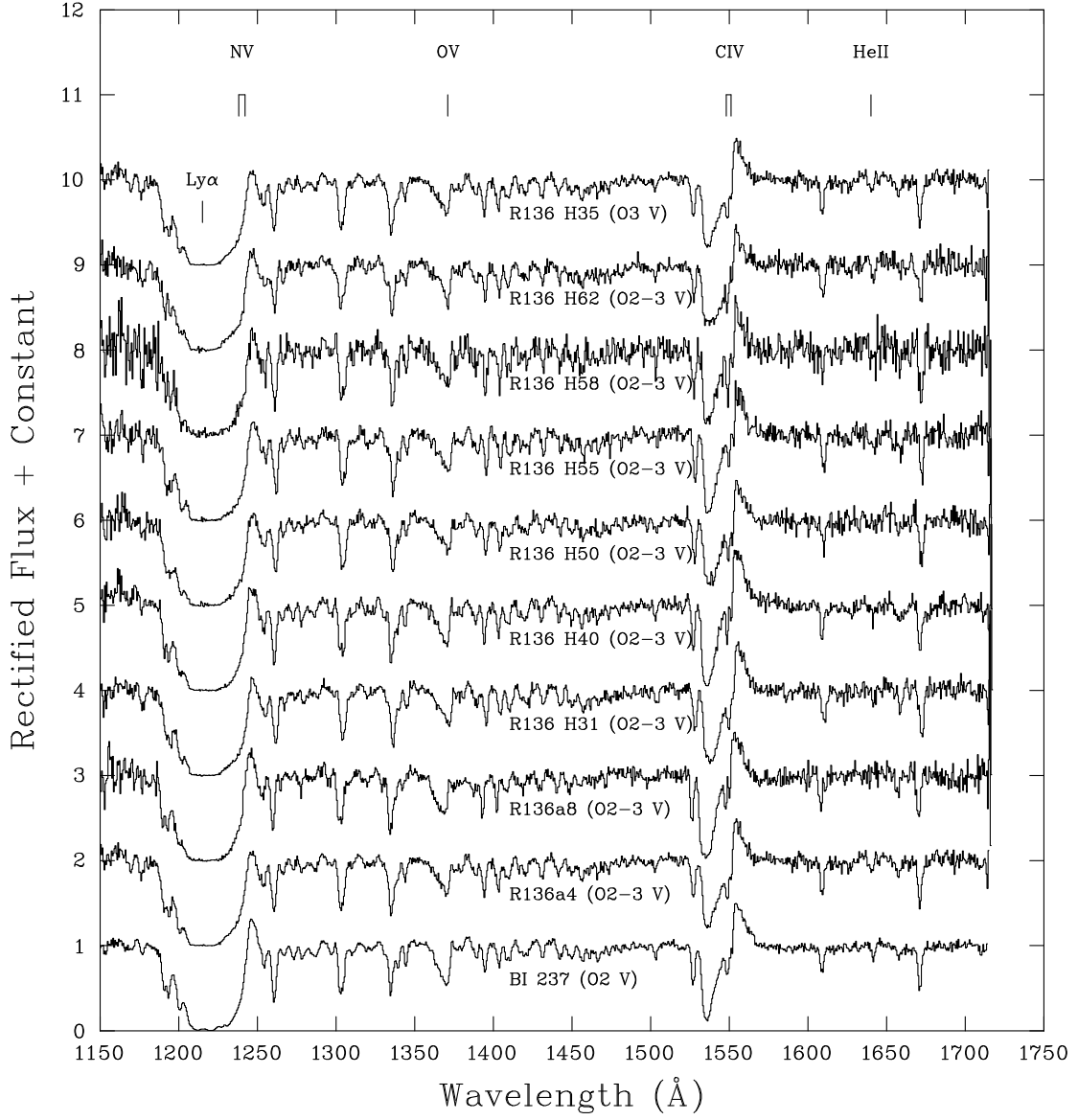} 
 \caption{HST/STIS ultraviolet spectroscopy of O2--3 dwarfs in 
R136, together with templates BI 237 (O2\,V) and R136 H35 (O3\,V), see Table~\ref{lmc_uv_atlas}.}
\label{r136-o2-3uv}
\end{center}
\end{figure*}

\begin{figure*}
\begin{center}
 \includegraphics[bb=35 165 530 680,width=1.75\columnwidth]{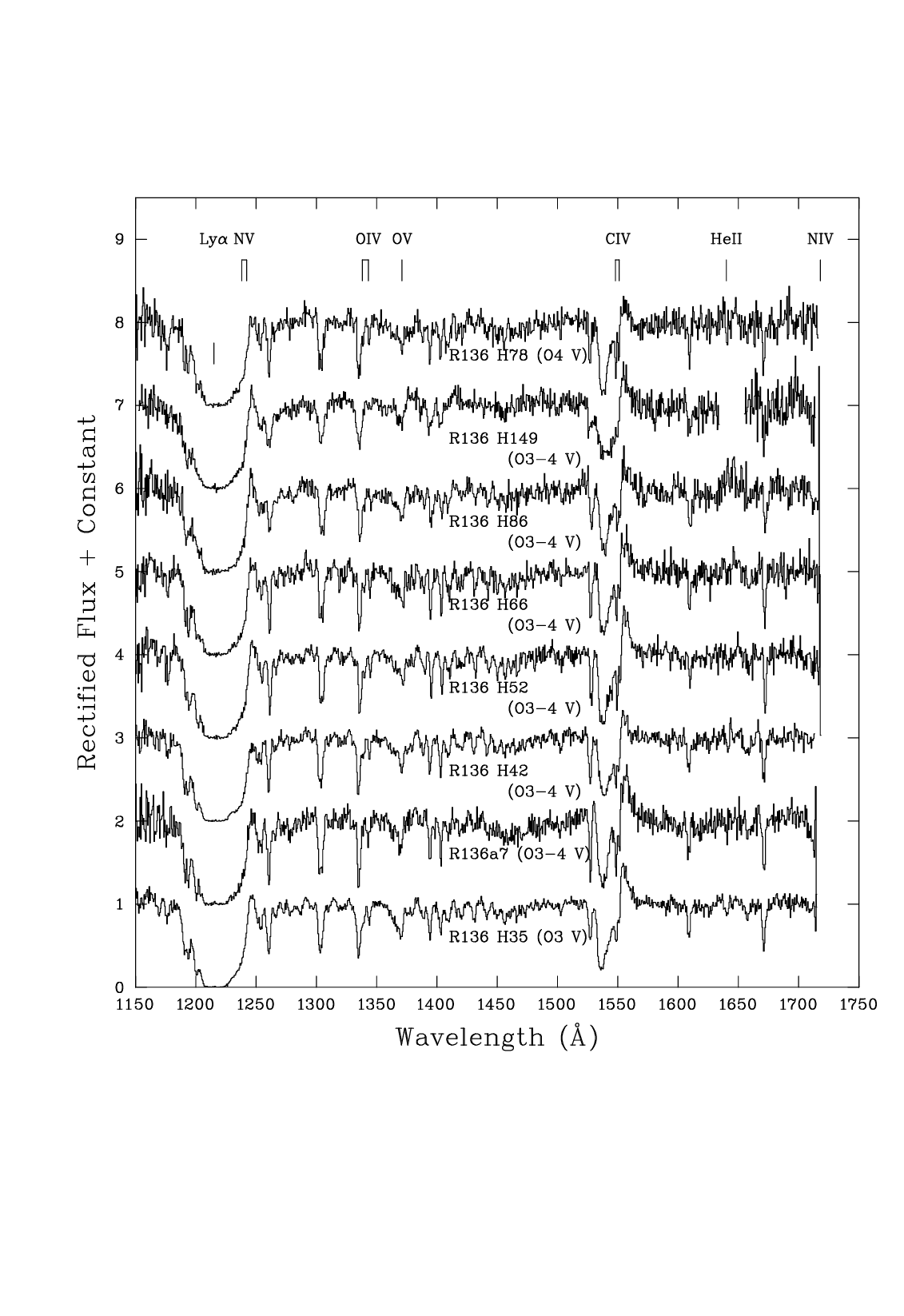} 
 \caption{HST/STIS ultraviolet spectroscopy of O3--4 dwarfs in 
R136, together with LMC templates R136 H35 (O3\,V) and H78 (O4\,V); see Table~\ref{lmc_uv_atlas}.}
\label{r136-o3-4uv}
\end{center}
\end{figure*}

\begin{figure*}
\begin{center}
 \includegraphics[bb=35 105 530 680,width=1.75\columnwidth]{r136-o4-5uv_p.eps} 
 \caption{HST/STIS ultraviolet spectroscopy of O4--5 dwarfs in R136,
together with LMC templates R136 H78 (O4\,V) and H70 (O5\,Vz); see 
Table~\ref{lmc_uv_atlas}.}
\label{r136-o4-5uv}
\end{center}
\end{figure*}

\begin{figure*}
\begin{center}
 \includegraphics[bb=35 105 530 680,height=1.75\columnwidth]{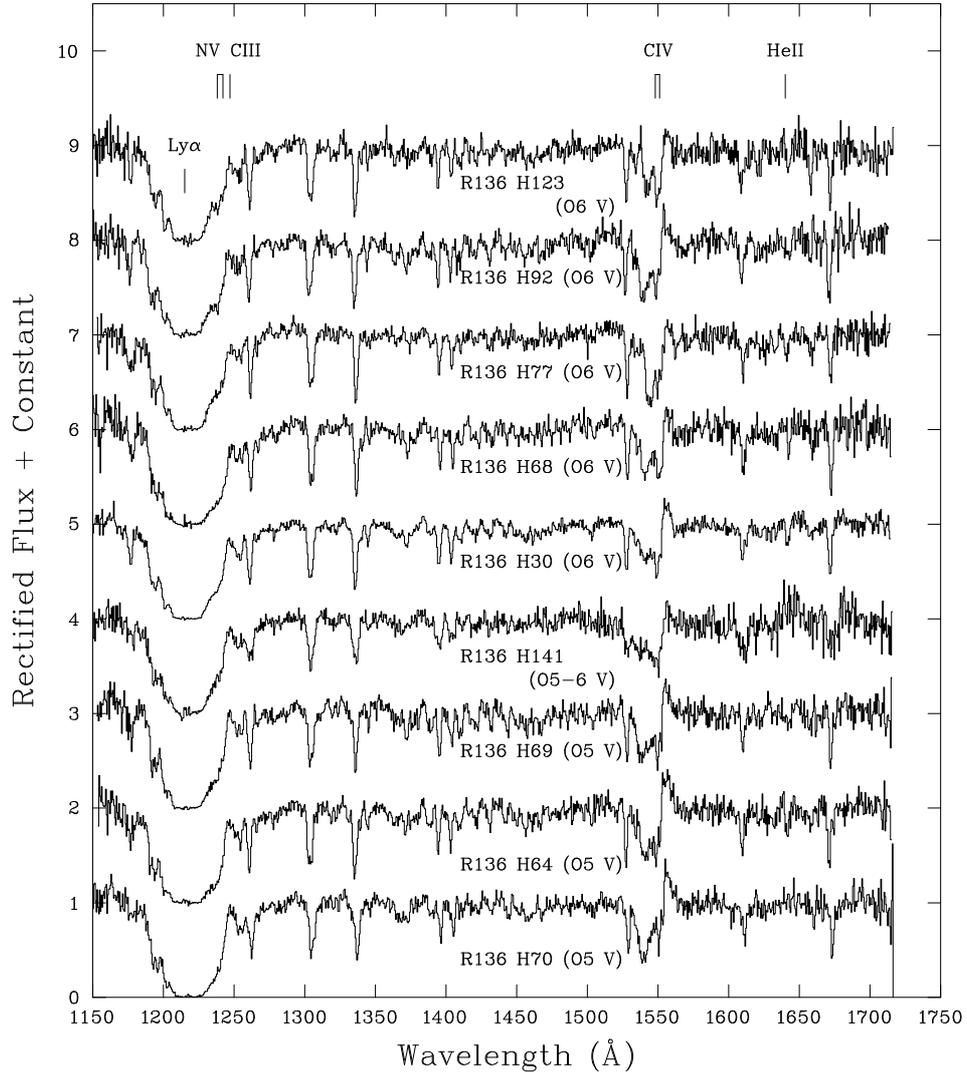} 
 \caption{HST/STIS ultraviolet spectroscopy of O5--6 dwarfs in 
R136, together with LMC templates R136 H70 (O5\,Vz) and H123 (O6\,V); see 
Table~\ref{lmc_uv_atlas}.}
\label{r136-o5-6uv}
\end{center}
\end{figure*}

\begin{figure*}
\begin{center}
 \includegraphics[bb=35 165 530 680,width=1.75\columnwidth]{r136-o7-8uv_p.eps} 
 \caption{HST/STIS ultraviolet spectroscopy of O7--8 dwarfs in R136, 
together with LMC templates R136 H134 (O7\,V) and H80 (O8\,V); see Table~\ref{lmc_uv_atlas}.}
\label{r136-o7-8uv}
\end{center}
\end{figure*}

\begin{figure*}
\begin{center}
 \includegraphics[bb=35 165 530 680,width=1.75\columnwidth]{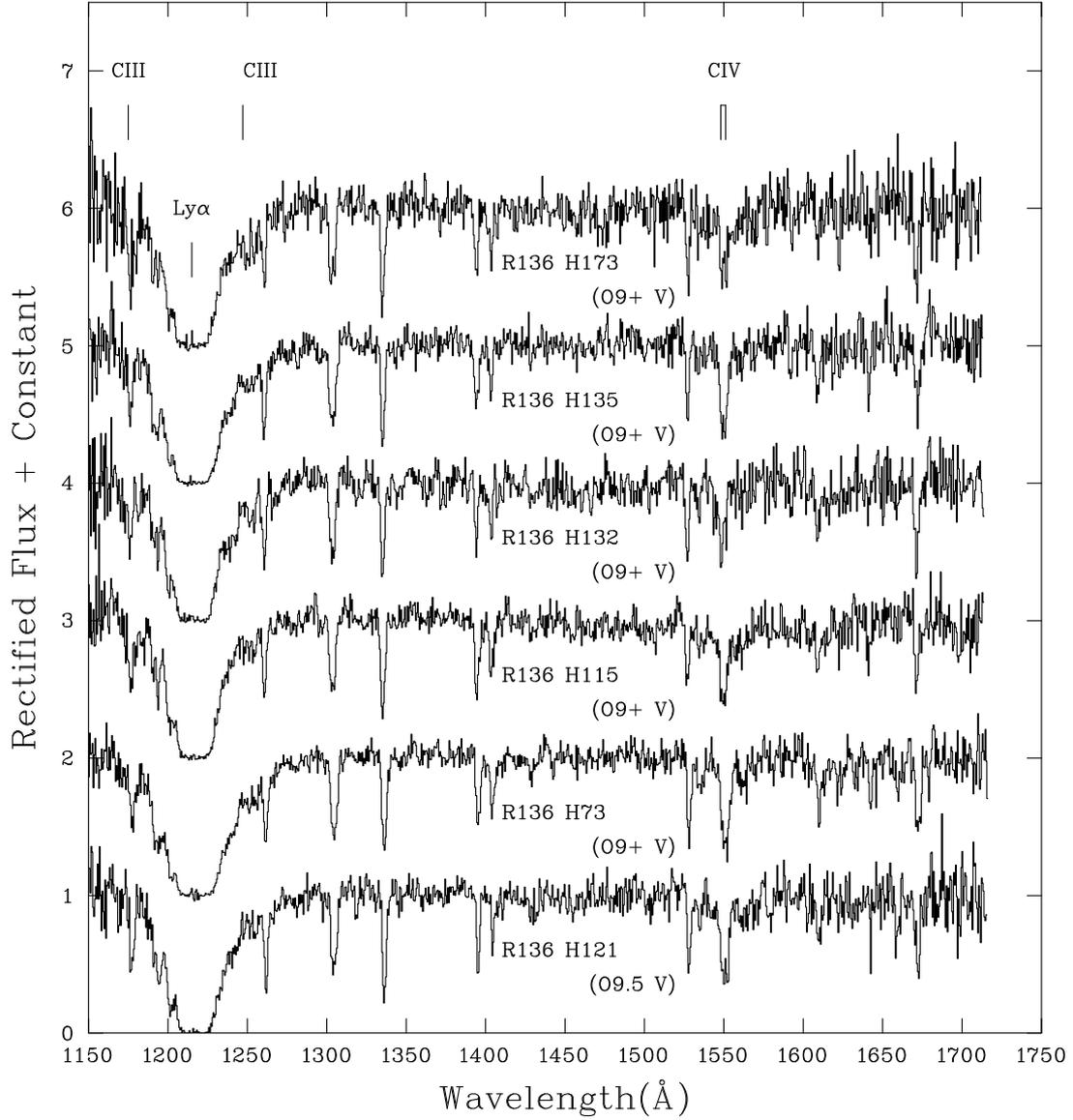} 
 \caption{HST/STIS ultraviolet spectroscopy of O9+ dwarfs in R136, 
together with LMC template R136 H121 (O9.5\,V); see Table~\ref{lmc_uv_atlas}.}
\label{r136-o9+uv}
\end{center}
\end{figure*}

\clearpage

\section{Wind velocities of LMC O-type stars}

\begin{table*}
\begin{center}
  \caption{Comparison of terminal wind velocities (km\,s$^{-1}$) of LMC O stars in the present study (indicated with a colon if via $v_{\rm edge}$)
    with literature values from far-UV spectroscopy (direct measurement or Sobolov with Exact Integral SEI approach).}
\label{winds}
\begin{tabular}{l@{\hspace{1.5mm}}l@{\hspace{1.5mm}}l@{\hspace{1.5mm}}l
@{\hspace{2mm}}l@{\hspace{2mm}}l
@{\hspace{2mm}}l@{\hspace{2mm}}l}
\hline
Star & Alias & Sp Type & $v_{\infty}$ & Dataset/diagnostic & $v_{\infty}$ & Dataset/diagnostic & Reference \\
     &       &         & km\,s$^{-1}$ & (this study)      & km\,s$^{-1}$ & (literature)       &           \\
\hline
BI 237             && O2\,V        &        &                    & 3400   & HST STIS/C\,{\sc iv} & f \\
BI 253             &VFTS 72& O2\,V        &        &                    & 3180   & HST STIS/C\,{\sc iv} & f \\
R136 H40           && O2--3\,V     & 2750   & HST STIS/C\,{\sc iv}& 3400, 3000, 3400 & HST GHRS/C\,{\sc iv} & b, c, e \\
R136 H50           && O2--3\,V     & 2620   & HST STIS/C\,{\sc iv}& 3125  & HST GHRS/C\,{\sc iv} & b \\
R136 H55           && O2--3\,V     & 2880   & HST STIS/C\,{\sc iv}& 3150, 2955, 3250 & HST GHRS/C\,{\sc iv} & b, c, e\\
R136 H58           && O2--3\,V     & 2980   & HST STIS/C\,{\sc iv}& 3050, 2840 & HST GHRS/C\,{\sc iv} & b, c \\
LH 101/W3-14       && O3\,V        &        &                    & 3100   & HST STIS/C\,{\sc iv} & f \\ 
R136 H33           && O3\,V        &        &                    & 3250   & HST STIS/C\,{\sc iv} & f \\
LH 101/W3-24       &&O3\,V         &        &                    & 2400   & HST STIS/C\,{\sc iv} & e \\
R136 H39           && O3\,V+O5.5\,V&        &                    & 2900   & HST GHRS/C\,{\sc iv} & b \\
LH 81/W28-23       && O3.5\,V((f+))&        &                    & 3050   & HST STIS/C\,{\sc iv} & f \\
R136a7   & R136 H24 & O3--4\,V     & 2710: & HST STIS/C\,{\sc iv}& 2900, 3100 & HST GHRS/C\,{\sc iv} & b, e\\
R136 H42  &         & O3--4\,V     & 2245   & HST STIS/C\,{\sc iv}& 2850  & HST GHRS/C\,{\sc iv} & b \\ 
LH 81/W28-5 &       & O4\,V((f+))  &        &                    & 2700   & HST STIS/C\,{\sc iv} & e \\
Sk--70$^{\circ}$ 60 && O4--5\,V((f))&        &                    & 2300   & FUSE/SEI             & d \\
LMC 054383        && O4--5\,V      &        &                    & 2380   & HST STIS/C\,{\sc iv} & g \\
LH 58-496    &     & O5\,V(f)      &        &                    & 2400   & HST STIS/C\,{\sc iv} & f \\
Sk--70$^{\circ}$ 69 && O5.5\,V((f)) &        &                    & 2750   & FUSE/SEI             & d \\
BI 229 &       & O7\,V--III        &        &                    & 1950   & FUSE/SEI             & d \\
Sk--67$^{\circ}$ 191 & &  O8\,V     &        &                    & 1950   & FUSE/SEI             & d \\ 
\\
HDE\,269810 & Sk--67$^{\circ}$ 211 & O2\,III(f) &        &        & 3600   & FUSE/SEI             & d \\
Sk--66$^{\circ}$ 172 &             & O2\,III    &        &        & 3100   & FUSE/SEI             & d \\
Sk--70$^{\circ}$ 91    &LH 62-1           & O2\,III    &        &        & 3150   & FUSE/SEI             & d \\
Sk---68$^{\circ}$ 137  &           &  O2\,III   &        &        & 3400   & HST FOS/SEI          & a \\
LH 64--16           &             & ON2\,III   &        &        & 3250   & HST STIS/C\,{\sc iv} & g \\
VFTS 16   &                       & O2\,III-If &        &        & 3250   & HST COS/SEI          & h \\
R136 H46  &                       & O2\,III-If & 3440   & HST STIS/C\,{\sc iv} & 3550, 3355 & HST GHRS/C\,{\sc iv} & b, c\\
R136 H47  &                       & O2\,III-If & 3045   & HST STIS/C\,{\sc iv} & 3625, 3305, 3500 & HST GHRS/C\,{\sc iv} & b, c, e\\
R136 H18  &     Melnick 33S    &  O3\,III   &        &        & 3200   & HST/STIS/C\,{\sc iv} & f \\
LH 90 ST2-22 &                    & O3.5\,III(f+) &     &        & 2560   & HST STIS/C\,{\sc iv} & f \\
Sk--67$^{\circ}$ 69 &              & O4\,III   &         &        & 2500   & FUSE/SEI             & d \\
HDE 269676  &  Sk--71$^{\circ}$ 45 & O4--5 III &         &        & 2500   & FUSE/SEI             & d \\
Sk--66$^{\circ}$ 100 &             & O6\,II(f) &         &        & 2075   & FUSE/SEI             & d \\ 
BI 272  &                         &  O7:\,III--II &     &        & 3400   & FUSE/SEI             & d \\
Sk--67$^{\circ}$ 101 &   LH 54-21        & O8\,II(f) &         &        & 2300   & FUSE/SEI             & d \\
BI 173 &                          & O8\,II    &         &        & 2850   & FUSE/SEI             & d \\
\\
LH 101/W3-19 &                    & O2\,If    &         &        & 2850   & HST STIS/C\,{\sc iv} & f \\ 
Melnick 42 &  R136 H2             & O2\,If    &         &        & 2800   & HST STIS/C\,{\sc iv} & i \\
R136 H36   &                      & O2\,If    & 3500    & HST STIS/C\,{\sc iv} & 3640, 3700, 3750 & HST GHRS/C\,{\sc iv} & b, c, e\\
Sk--65$^{\circ}$ 47 & LH 43--18    & O4\,I(n)f+p &       &        & 2100   & HST STIS/C\,{\sc iv} & f \\
HDE 269698 & Sk--67$^{\circ}$ 166 & O4\,If+   &         &        & 1900   & HST FOS/SEI          & a \\
Sk--67$^{\circ}$ 167 & LH 76-21   & O4\,Inf+  &         &        & 2150   & HST FOS/SEI          & a \\
Sk--67$^{\circ}$ 111 & LH 60-53    & O6\,Ia(n)fp var&    &        & 2000   & FUSE/SEI             & d \\
HDE 270952   & Sk--65$^{\circ}$ 22 & O6\,Iaf+ &          &        & 1350   & FUSE/SEI             & d \\
HDE 270145 & Sk--70$^{\circ}$ 115 & O6.5\,Iaf &          &        & 2200   & FUSE/SEI             & d \\
BI 170 &                         & O9.5\,Ib  &          &        & 1700   & FUSE/SEI             & d \\
Sk--69$^{\circ}$ 124 &            & O9\,Ib    &          &        & 1600   & FUSE/SEI             & d \\
Sk--65$^{\circ}$ 21  &            & O9.7\,Iab &          &        & 1700   & FUSE/SEI             & d \\
Sk--66$^{\circ}$ 169 &            & O9.7\,Ia+ &          &        &  800   & FUSE/SEI             & d \\
HDE\,269896 & Sk--68$^{\circ}$ 135 & ON9.7\,Ia+ &         &       & 1050   & FUSE/SEI             & d \\
HDE\,268605 & Sk--67$^{\circ}$ 05 & O9.7\,Ib &            &       & 1400   & FUSE/SEI       &       d \\   
\\
Sk--67$^{\circ}$ 22 &       & O2\,If/WN5     &            &       & 2650   & HST STIS/C\,{\sc iv} & f \\
R136a5     & R136 H20      & O2\,If/WN5     &  3045      & HST STIS/C\,{\sc iv} & 3000, 3400 & HST/GHRS C\,{\sc iv} & b, e\\
Melnick 39 & R136 H7       & O2.5\,If/WN6   &            &       & 2100   & HST STIS/C\,{\sc iv}+Si\,{\sc iv} & f \\
Melnick 37 & R136 H14      & O3.5\,If/WN7   &            &       & 2000   & HST STIS/C\,{\sc iv}+Si\,{\sc iv} & f \\
LH 90 Brey 58 & AB4        & O3.5\,If/WN7   &            &       & 1900   & HST STIS/Si\,{\sc iv} & f \\ 
\hline
\hline
\multicolumn{8}{l}{
  \begin{minipage}{1.75\columnwidth}~\\
(a) Walborn et al. (1995);
(b) de Koter et al. (1998); (c) Prinja \& Crowther (1998); (d) Massa et al. (2003); (e) Massey et al. (2004); (f) Massey et al. (2005); (g)
Massey et al. (2009); (h) Evans et al. (2010); (i) Bestenlehner et al. (2014)
  \end{minipage}
    }\\
\end{tabular}
\end{center}
\end{table*}

\clearpage

\section{Additional R136 sources}

\begin{table*}
\begin{center}
\caption{HST/WFC3 photometry (de Marchi et al. 2011) for additional R136 sources brighter than $m_{\rm 
F555W}$ = 17.0  mag within a projected distance of 4.1 arcsec (1 
parsec) from R136a1. Identifications  are from Weigelt \& Baier (1985, WB85) or 
Hunter et al. (1995, HSH95). $A_{\rm 
F555W}$ is calculated from observed  colours for an adopted reddening law 
of $R_{\rm 5495}$ = 4.1, and DM = 18.49 (Pietrzynski et 
al. 2013). F555W-band photometry shown in italics is from Hunter et al. 
(1995) offset by --0.17 mag (see Sect.~\ref{census}). R136 H351 is included since it is one of the 70 far-UV brightest sources in R136a1.}
\label{faint_targets}
\begin{tabular}{
c@{\hspace{1mm}}
c@{\hspace{1mm}}
c@{\hspace{1mm}}l@{\hspace{1mm}}c
@{\hspace{1mm}}c@{\hspace{1.mm}}c
@{\hspace{1mm}}c@{\hspace{1mm}}r
@{\hspace{1mm}}r@{\hspace{1mm}}c@{\hspace{1mm}}c}
\hline
HSH95 & Sp & Ref & $r$ & $m_{\rm F555W}$ & $m_{\rm F336W}$--$m_{\rm 
F438W}$ & $m_{\rm F438W}$--$m_{\rm F555W}$ & 
$m_{\rm F555W}$--$m_{\rm F814W}$ & $A_{\rm F555W}$ & 
$M_{\rm F555W}$ & $10^{14} F_{\rm 1500}$ & Slit\\
(WB85) & Type &  & arcsec & mag& mag& mag & mag & mag &  mag  
&erg\,s$^{-1}$\,cm$^{-2}$\,\AA\ & \\
\hline
379 &  &  & 0.41 & 16.51$\pm$0.05 & --0.97$\pm$0.06 & 0.01$\pm$0.06 & 
0.60$\pm$0.08 & 1.87 & --3.85 & 0.5$\pm$0.1 & SE2 \\
235 &  &  & 0.50 & {\it 16.63\phantom{$\pm$0.01} } &  &      &               
& 1.72 & --3.58 & 1.2$\pm$0.1 & SE2 \\ 
%
210 &&& 0.54 & {\it 16.47\phantom{$\pm$0.01} }  &      &         &        
&  1.72 & --3.74 & 
1.2$\pm$0.1 & NW3 \\
378&& & 0.56 & 16.06$\pm$0.03 &        &       &  &  1.72 & --4.15 & 1.0$\pm$0.1 & SE2 \\
306&& & 0.70 & 16.42$\pm$0.05 &        &             & 0.33$\pm$0.08 & 1.81& --3.88 & 0.5$\pm$0.1 & NW4 \\
196&& & 0.74 & {\it 16.36\phantom{$\pm$0.01} } &      &          &        
&  1.72 & --3.85 & & NW5 \\  
231&& & 0.88 & 16.66$\pm$0.06 & --1.23$\pm$0.08 & 0.11$\pm$0.08 & 
0.34$\pm$0.11 & 1.58 & --3.41 & 0.4$\pm$0.1 & NW2 \\
305&& & 0.91 & {\it 16.96\phantom{$\pm$0.01} } &  &  &
              & 1.72 & --3.25 &             & NW4 \\ 
291&& & 0.94 & 16.67 $\pm$0.05& --1.20$\pm$0.07 & 0.16$\pm$0.07 & 
0.42$\pm$0.08 & 1.72 & --3.54 & 0.4$\pm$0.1 & SE4 \\
259&& & 1.00 & {\it 16.75\phantom{$\pm$0.01} } &      &          &    &   
1.72 & --3.46 & 
0.4$\pm$0.1 & NW5 \\
217&& & 1.01 & 16.52$\pm$0.04 & --1.36$\pm$0.06 & 0.18$\pm$0.06 &         
& 1.48 & --3.45 & 0.5$\pm$0.1 & NW1 \\
274&& & 1.06 & 16.87$\pm$0.07 &        &         &  0.29$\pm$0.12 & 1.72 & 
--3.34 & 0.4$\pm$0.1 & NW4 \\
187&& & 1.07 & 16.16$\pm$0.03 & --1.31$\pm$0.04 & 0.11$\pm$0.05 & 
0.27$\pm$0.05 
& 1.44 & --3.77 & 1.0$\pm$0.1 & SE4 \\
203&& & 1.21 & 16.33$\pm$0.04 & --1.27$\pm$0.06 & 0.17$\pm$0.06 & 
0.32$\pm$0.07 
& 1.61 & --3.77 & 0.6$\pm$0.1 & NW3 \\
233&& & 1.24 & 16.76$\pm$0.05 & --1.25$\pm$0.08 & 0.15$\pm$0.08 & 
0.32$\pm$0.09 
& 1.61 & --3.34 & 0.3$\pm$0.1 & SE2 \\
293&& & 1.26 & 16.99$\pm$0.07 & --1.03$\pm$0.11 & 0.17$\pm$0.10 & 
0.30$\pm$0.11 
& 2.05 & --3.55 & 0.3$\pm$0.1 & NW7 \\    
159&& & 1.34 & 16.14$\pm$0.03 & --1.24$\pm$0.04 & 0.10$\pm$0.04 & 
0.33$\pm$0.04 
&1.65 &--4.00 & 0.7$\pm$0.1 & SE6 \\
176&& & 1.35 & 16.14$\pm$0.03 & --1.36$\pm$0.04 & 0.13$\pm$0.04 & 
0.22$\pm$0.05 
& 1.36 & --3.71 & 0.9$\pm$0.1 & NW7 \\
179&& & 1.40 & 16.23$\pm$0.03 & --1.31$\pm$0.05 & 0.13$\pm$0.05 &  & 1.47 
&  --3.73 & 0.9$\pm$0.1 & NW6 \\
254&& & 1.43 & 16.87$\pm$0.05 & --1.25$\pm$0.08 & 0.10$\pm$0.08 & 
0.32$\pm$0.09 
& 1.51 & --3.13 & 0.5$\pm$0.1 & NW3 \\
302&& & 1.48 & 16.98$\pm$0.06 &--1.28$\pm$0.09 & 0.10$\pm$0.09 & 
0.18$\pm$0.12 & 1.47 & 
--2.98 & 0.9$\pm$0.1 & NW8 \\
223&& & 1.53 & 16.61$\pm$0.04 & --1.30$\pm$0.06 & 0.15$\pm$0.06 & 
0.26$\pm$0.07 
& 1.51 & --3.39 & 0.6$\pm$0.1 & NW2 \\
285&& & 1.58 & 16.82$\pm$0.05 & --1.23$\pm$0.08 & 0.17$\pm$0.07 & 
0.53$\pm$0.08 
&  1.72 & --3.39 & 0.4$\pm$0.1 & NW2 \\
288&& & 1.62 & 16.80$\pm$0.04 & --1.18$\pm$0.06 & 0.13$\pm$0.06 & 0.29$\pm$0.07 & 1.69 & --3.38 & &\ldots \\ 
281&& & 1.63 & 16.71$\pm$0.04 & --1.21$\pm$0.06 & 0.14$\pm$0.05 & 
0.39$\pm$0.06 &   1.65 & --3.43 & 0.5$\pm$0.1 & SE1 \\
139&& & 1.65 & 16.02$\pm$0.02 &        &       & 0.22$\pm$0.04 & 1.72 & 
--4.19 & 1.1$\pm$0.1 & SE3 \\
162&& & 1.72 & {\it 16.06\phantom{$\pm$0.01} } &        &       &       & 
1.72 & --4.15 & 
0.8$\pm$0.1 & SE2 \\
386&& & 1.75 & 16.87$\pm$0.05 & --0.97$\pm$0.09 & 0.12$\pm$0.07 & 0.36$\pm$0.08 & 2.07 & --3.69 & 0.2$\pm$0.1 & NW4 \\ %
311&& & 1.78 & 16.87$\pm$0.05 & --0.97$\pm$0.08 &  0.12$\pm$0.07 &  
0.36$\pm$0.08  
& 2.07 & --3.69 & 0.3$\pm$0.1 & NW4 \\
211&& & 1.78 & 16.52$\pm$0.03 & --1.31$\pm$0.04 & 0.07$\pm$0.04 & 
0.22$\pm$0.06 & 
 1.72 & --3.69 & &\ldots \\ 
181&& & 1.87 & {\it 16.17\phantom{$\pm$0.01} } &     &        &     & 1.72 
& --4.04 & 
0.7$\pm$0.1 & NW1 \\
240&& & 2.11 & {\it 16.66\phantom{$\pm$0.01} } &       &          &           
&  1.72 & --3.55 
& & \ldots\\
237&& & 2.18 & 16.66$\pm$0.03 & --1.04$\pm$0.04 & 0.09$\pm$0.04 & 
0.30$\pm$0.05 
& 1.90 & --3.73 & &\ldots \\
416&& & 2.20 & 16.95$\pm$0.04 &  --0.90$\pm$0.06 & 0.11$\pm$0.05 & 
0.44$\pm$0.06 & 2.18 & --3.72 
& &\ldots \\
213&& & 2.23 & 16.68$\pm$0.03 &--1.10$\pm$0.05 & 0.12$\pm$0.05 & 
0.36$\pm$0.05 
&1.83 & --3.64 & &\ldots \\
253&& & 2.30 & 16.89$\pm$0.04 & --1.27$\pm$0.05 & 0.14$\pm$0.05 & 
0.27$\pm$0.06 
& 1.56 & --3.16 & & \ldots \\
392&& & 2.33 & 16.73$\pm$0.03 & --1.11$\pm$0.07 & 0.07$\pm$0.07 & 
0.46$\pm$0.06 
& 1.72 & --3.48 & 0.3$\pm$0.1 & SE8 \\
205&& & 2.37 & 16.37$\pm$0.02 & --1.22$\pm$0.03 & 0.19$\pm$0.03 
& 0.46$\pm$0.03 & 1.75 & --3.87 &  & \ldots \\ %
258&& & 2.39 & 16.74$\pm$0.03 &          & 0.17$\pm$0.05 & 0.44$\pm$0.05 & 
1.72 & --3.47 & & \ldots \\
189&& & 2.42 & 16.31$\pm$0.02 & --1.23$\pm$0.03 & 0.18$\pm$0.03 
& 0.32$\pm$0.03 & 1.72 & --3.89 & 1.0$\pm$0.2 & SE8 \\
97    &     &   & 2.42 & 15.46$\pm$0.01 & --1.31$\pm$0.02 & 0.18$\pm$0.02 & 0.34$\pm$0.02 & 1.76 & --4.79 &  & \ldots \\ 
87    &     &   & 2.47 & 15.25$\pm$0.01 & --1.32$\pm$0.01 & 0.13$\pm$0.01 & 0.32$\pm$0.02 & 1.65 & --4.89 &  & \ldots \\ 
145   &     &   & 2.47 & 16.10$\pm$0.01 & --1.30$\pm$0.02 & 0.10$\pm$0.02 & 0.28$\pm$0.03 & 1.42 & --3.81 & 0.8$\pm$0.1 & NW2 \\ 
547   &     &   & 2.51 & 16.75$\pm$0.03 &                 &               & 0.64$\pm$0.04 & 1.72 & --3.46 &  & \ldots \\
255   &     &   & 2.55 & 16.77$\pm$0.03 & --1.17$\pm$0.04 & 0.18$\pm$0.04 & 0.48$\pm$0.04 & 1.81 & --3.53 &  & \ldots \\
267   &     &   & 2.60 & 16.90$\pm$0.02 & --1.19$\pm$0.03 & 0.12$\pm$0.03 & 0.27$\pm$0.04 & 1.68 & --3.27 &  & \ldots \\
206   &     &   & 2.63 & 16.53$\pm$0.02 &                 & 0.16$\pm$0.03 & 0.23$\pm$0.04 & 1.72 & --3.68 &  & \ldots \\
127   &     &   & 2.67 & 15.94$\pm$0.02 & --1.33$\pm$0.02 & 0.11$\pm$0.02 & 0.30$\pm$0.03 & 1.59 & --4.15 &  & \ldots \\
177   &     &   & 2.80 & 16.26$\pm$0.02 & --1.18$\pm$0.02 & 0.26$\pm$0.02 & 0.52$\pm$0.03 & 1.94 & --4.17 & 0.4$\pm$0.1 & NW5 \\ 
33  & O3\,V & b & 2.80 & 14.36$\pm$0.01 & --1.36$\pm$0.01 & 0.06$\pm$0.01 & 0.28$\pm$0.01 & 1.45 & --5.58 &  & \ldots \\
251   &     &   & 2.81 & 16.83$\pm$0.02 &                 & 0.13$\pm$0.03 &               & 1.72 & --3.38 &  & \ldots \\
184   &     &   & 2.82 & 16.40$\pm$0.02 & --1.32$\pm$0.02 & 0.17$\pm$0.02 & 0.29$\pm$0.03 & 1.52 & --3.61 &  & \ldots \\
294   &     &   & 2.86 & 17.00$\pm$0.03 & --1.24$\pm$0.04 & 0.17$\pm$0.04 & 0.46$\pm$0.04 & 1.67 & --3.16 &  & \ldots \\
236   &     &   & 2.90 & 16.73$\pm$0.02 & --1.24$\pm$0.03 & 0.13$\pm$0.03 & 0.35$\pm$0.04 & 1.59 & --3.35 &  & \ldots \\
241   &     &   & 2.92 & 16.80$\pm$0.02 &                 &               & 0.17$\pm$0.04 & 1.72 & --3.41 &  & \ldots \\
272   &     &   & 2.93 & 16.95$\pm$0.02 & --1.08$\pm$0.04 & 0.21$\pm$0.04 & 0.54$\pm$0.04 & 2.03 & --3.57 &  & \ldots \\
170   &     &   & 2.97 & 16.25$\pm$0.01 & --1.23$\pm$0.02 & 0.07$\pm$0.02 & 0.33$\pm$0.03 & 1.50 & --3.74 &  & \ldots \\
216   &     &   & 2.99 & 16.60$\pm$0.01 & --1.23$\pm$0.02 & 0.10$\pm$0.02 & 0.27$\pm$0.03 & 1.55 & --3.44 &  & SE7 \\
\hline
\hline
\end{tabular}
\end{center}
(a) Crowther \& Dessart (1998); (b) Massey et al. (2005)
\end{table*}

\addtocounter{table}{-1}

\begin{table*}
\begin{center}
\caption{(continued)}
\begin{tabular}{
c@{\hspace{1mm}}
c@{\hspace{1mm}}
c@{\hspace{1mm}}l@{\hspace{1mm}}c
@{\hspace{1mm}}c@{\hspace{1.mm}}c
@{\hspace{1mm}}c@{\hspace{1mm}}r
@{\hspace{1mm}}r@{\hspace{1mm}}c@{\hspace{1mm}}c}
\hline
HSH95 & Sp & Ref & $r$ & $m_{\rm F555W}$ & $m_{\rm F336W}$--$m_{\rm 
F438W}$ & $m_{\rm F438W}$--$m_{\rm F555W}$ & 
$m_{\rm F555W}$--$m_{\rm F814W}$ & $A_{\rm F555W}$ & 
$M_{\rm F555W}$ & $10^{14} F_{\rm 1500}$ & Slit\\
(WB85) & Type &  & arcsec & mag& mag& mag & mag & mag &  mag  
&erg\,s$^{-1}$\,cm$^{-2}$\,\AA\ & \\
\hline
222   &     &   & 3.06 & 16.71$\pm$0.02 & --1.15$\pm$0.03 & 0.22$\pm$0.03 & 0.55$\pm$0.03 & 1.92 & --3.70 & 0.1: & SE8 \\
250   &     &   & 3.09 & 16.95$\pm$0.02 & --1.09$\pm$0.02 & 0.22$\pm$0.03 & 0.46$\pm$0.03 & 2.04 & --3.58 & 0.2$\pm$0.1  & NW1 \\
277   &     &   & 3.22 & 16.96$\pm$0.02 & --1.13$\pm$0.03 & 0.25$\pm$0.03 & 0.62$\pm$0.03 & 2.01 & --3.54 & 0.2$\pm$0.1  & SE6 \\
317   &     &   & 3.27 & 16.25$\pm$0.02 & --1.50$\pm$0.02 & 0.63$\pm$0.02 &               & 2.03 & --4.27 &  & \ldots \\ 
120   &     &   & 3.29 & 15.88$\pm$0.01 & --1.23$\pm$0.02 & 0.05$\pm$0.02 & 0.32$\pm$0.02 & 1.65 & --4.26 & 1.1$\pm$0.1 & SE4 \\
10 (c)&WN5h & a & 3.44 & 13.43$\pm$0.01 & --1.20$\pm$0.02 & 0.31$\pm$0.01 & 0.82$\pm$0.01 & 2.33 & --7.39 &  & \ldots \\ 
128   &     &   & 3.52 & 16.00$\pm$0.01 & --1.28$\pm$0.01 & 0.07$\pm$0.02 & 0.28$\pm$0.02 & 1.41 & --3.90 &  & \ldots \\
136   &     &   & 3.64 & 15.96$\pm$0.01 & --1.15$\pm$0.02 & 0.23$\pm$0.02 & 0.62$\pm$0.02 & 1.94 & --4.47 &  & \ldots \\
150   &     &   & 3.74 & 16.11$\pm$0.01 & --1.30$\pm$0.02 & 0.16$\pm$0.02 & 0.32$\pm$0.02 & 1.54 & --3.92 &  & \ldots \\
165   &     &   & 3.81 & 16.16$\pm$0.01 & --1.27$\pm$0.02 & 0.18$\pm$0.02 & 0.35$\pm$0.02 & 1.63 & --3.98 &  & \ldots \\ 
57   &     &   & 3.87 & 14.80$\pm$0.01 & --1.16$\pm$0.01 & 0.26$\pm$0.01 & 0.66$\pm$0.01 & 2.17 & --5.86 &  & \ldots \\ 
197  &      &   & 4.00 & 16.38$\pm$0.01 & --1.17$\pm$0.02 & 0.13$\pm$0.02 & 0.45$\pm$0.02 & 1.72 & --3.83 & & \ldots \\
183  &      &   & 4.07 & 16.33$\pm$0.01 & --1.26$\pm$0.02 & 0.24$\pm$0.02 & 0.49$\pm$0.02 & 1.76 & --3.92 & & \ldots \\
140  &      &   & 4.09 & 15.99$\pm$0.01 & --1.29$\pm$0.02 & 0.11$\pm$0.02 & 0.26$\pm$0.02 & 1.47 & --3.97 & & \ldots \\
246  &      &   & 4.11 & 16.75$\pm$0.01 & --1.36$\pm$0.02 & 0.12$\pm$0.02 & 0.15$\pm$0.02 & 1.36 & --3.10 & & \ldots \\
%
%
%
%
%
%
%
%
%
%
\hline
351   &     &   & 0.55 & {\it 17.18\phantom{$\pm$0.01} }    &                 
&               &               & 1.72 & --3.03 & 0.5$\pm$0.1 & SE2 \\
\hline
\hline
\end{tabular}
\end{center}
(a) Crowther \& Dessart (1998); (b) Massey et al. (2005)
\end{table*}

\label{lastpage}
\end{document}